\documentclass{aa}

\usepackage{graphicx}
\usepackage{gensymb}
\usepackage{float}
\usepackage{hyperref}
\usepackage{caption}
\usepackage{subcaption}
\hypersetup{colorlinks=true,linkcolor=blue,citecolor=blue,filecolor=blue,urlcolor=blue,}
\usepackage{siunitx}
\usepackage{placeins}
\newcommand{\orcid}[1]{\protect\href{https://orcid.org/#1}{\protect\includegraphics[width=10pt]{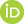}}}
\usepackage{txfonts,textcomp}

\newcommand{\A}{\r{A}\,} 
\newcommand{\ha}{H$\alpha$\,} 
 
\newcommand{\hb}{H$\beta$\,}

\newcommand{\hei}{He\,{\footnotesize I}\,}
\newcommand{\heii}{He\,{\footnotesize II}\,}
\newcommand{\nii}{[N\,{\footnotesize II}]\,}
\newcommand{\sii}{[S\,{\footnotesize II}]\,}
\newcommand{\siii}{[S\,{\footnotesize III}]\,}
\newcommand{\sitwo}{[Si\,{\footnotesize II}]\,}
\newcommand{\sithree}{[Si\,{\footnotesize III}]\,}

\newcommand{\civ}{C\,{\footnotesize IV}\,}
\newcommand{\oii}{[O\,{\footnotesize II}]\,}
\newcommand{\oiii}{[O\,{\footnotesize III}]\,}
\newcommand{\ciii}{[C\,{\footnotesize III}]\,}

\defcitealias{2024A&A...687A..20G}{BETIS I}
\usepackage{color}

\begin{document}

   \title{He II emitters at cosmic noon and beyond}

   \subtitle{Characterising the He II $\lambda$1640 emission with MUSE and JWST/NIRSpec}

   \author{R. Gonz\'alez-D\'iaz
          \inst{1}\orcid{0000-0002-0911-5141}
          \and
          J. M. Vílchez
          \inst{1}\orcid{0000-0001-7299-8373}
          \and
          C. Kehrig
          \inst{1,2}\orcid{0000-0003-1231-1482}
          \and
          I. del Moral-Castro
          \inst{3}\orcid{0000-0001-8931-1152}
          \and
          J. Iglesias-Páramo
          \inst{1,4}\orcid{0000-0003-2726-6370}
          }

   \institute{Instituto de Astrofísica de Andalucía (IAA-CSIC), Glorieta de la Astronomía s/n, E-18008 Granada, Spain\\ 
              \email{ragonzalez@iaa.es}
         \and
         Observat\'orio Nacional/MCTIC, R. Gen. Jos\'e Cristino, 77, 20921-400, Rio de Janeiro, Brazil\label{newinst}
         \and
         Instituto de Astrofísica, Facultad de Física, Pontificia Universidad Católica de Chile, Av. Vicuña Mackenna 4860, Santiago, Chile 
         \and
         Centro Astronómico Hispano en Andalucía, Observatorio de Calar Alto, Sierra de los Filabres, 04550 Gérgal, Spain
         }

   \date{Received ... ; accepted ... }

 
\abstract{The study of high-redshift galaxies provides critical insights into the early stages of cosmic evolution, particularly during what is known as cosmic noon, when star formation activity reached its peak. Within this context, the origin of the nebular \heii emission remains an open question. For this work, we conducted a systematic multi-wavelength investigation of a sample of z $\sim$ 2-4 \heii$\lambda$1640 \A emitters from the MUSE Hubble Ultra Deep Field surveys, utilising both MUSE and JWST/NIRSpec data and extending the sample presented by previous studies. We derived gas-phase metallicities and key physical properties, including electron densities, temperatures, and the production rates of hydrogen- and He$^+$-ionising photons. Our results suggest that a combination of factors, such as stellar mass, initial mass function, stellar metallicity, and stellar multiplicity, likely contributes to the origin of the observed nebular \heii emission. Specifically, for our galaxies with higher gas-phase metallicity (12 + log(O/H) $\gtrsim$ 7.55), we find that models for binary population with Salpeter IMF (M$_{up}$=100 M$_{\odot}$) and stellar metallicity $Z_{\star}\approx10^{-3}$ (i.e. similar to that of the gas) can reproduce the observed \heii ionising conditions.  However, at lower metallicities, models for binary populations with a `top-heavy' initial mass function (M$_{up}$ = 300 M$_{\odot}$) and $Z_{\star}$ much lower than that of the gas ($10^{-4} < Z_{\star} < 10^{-5}$)  are required to fully account for the observed \heii ionising photon production. These results reinforce that the \heii ionisation keeps challenging current stellar populations, and the \heii ionisation problem persists in the very low-metallicity regime.

}


   \maketitle
%

\section{Introduction}

The study of high-redshift galaxies offers critical insights into the early stages of cosmic evolution, particularly during what is called cosmic noon, a period around z $\approx$ 2–3 when star formation (SF) activity peaked \citep[e.g. ][]{LillyCosmo95,MadauCosmo96}.  Understanding the physical processes and stellar populations present during this epoch, and extending into the cosmic dawn at $6 \lesssim z \lesssim 10$ when the first ionising sources emerged, is essential to fully understand the processes of cosmic reionisation and the subsequent evolution of the universe \citep[e.g. ][]{TumlinsonZeroAge00,Wise12,BrommReview13,Wise14,Finkelstein19}. In this context,  the budget of the double ionised helium (He$^{++}$) is still a mystery. This ion produces recombination emission lines (\heii);  the most prominent features are the optical $\lambda$4686 \A and the ultraviolet (UV) $\lambda$1640 \A narrow lines. This implies  the existence of very strong ionising radiation with photon energies higher than 54.4 eV (or $\lambda\leq228$ \A). In the far Universe, the presence of these nebular lines is observed to be more common in high-redshift galaxies than in local galaxies \citep[e.g. ][]{KehrigM3311,CassataPopIII13,KehrigIZw15,KehrigSBS18}, making the study of galaxies exhibiting nebular \heii lines in their spectra (hereafter referred to as \heii emitters) important in understanding the nature of the ionising sources, the physical conditions of the interstellar medium (ISM), and the stellar populations driving the early phases of cosmic evolution. In addition, the intensity of the nebular \heii line is observed to be higher in lower metallicity galaxies \citep{Guseva00, BrommReview13}. In consequence, the presence of nebular \heii lines in high-redshift galaxies is suggested as a tracer of Population III (PopIII; \citealt{SchaererWRPopII02,SchaererPOPIII08}) stars, which are extremely hot and metal-poor (often considered essentially metal-free, Z~$\approx$~0) as they formed during the first million years of the universe from primordial hydrogen and helium. These stars would emit a significant hard UV ionising continuum, potentially making them major contributors to cosmic reionisation \citep[e.g. ][]{YoonPopIIImodels12,CassataPopIII13,NakajimaPopIII22}.
Although a prominent PopIII ionising continuum meets the conditions required to produce strong \heii and Ly$\alpha$ lines in high-redshift star-forming galaxies as observed at z > 6 with HST (e.g. \citealt{CaiHST11}) and JWST (e.g. \citealt{ToppingJWSTz624,VendittiJWSTz624, MondalJWST2025}), this becomes a problem at lower redshifts, where the presence of PopIII stars becomes increasingly unlikely.

Wolf--Rayet (WR) stars were proposed as a candidate to solve this problem, \citep{GarnettWR91,SchaererWR96}. WR stars are massive evolved stars in a late evolutionary stage, where core helium burning occurs and the hydrogen envelope is stripped away \citep{AllenWR76,IzotovWR04,RoyWR20,Roy25}. In this phase, some early-type WR stars are characterised by intense mass loss driven by dense and powerful stellar winds, while the stripped stellar surface reaches extremely high temperatures (T$\sim$100 kK) capable of doubly ionising the helium. WR winds can explain the origin of the nebular \heii in high-metallicity galaxies \citep{ShiraziWR12,Roy25}. In particular, systematic studies using integral field spectrographs (IFS) such as VIMOS \citep{LeFevreVIMOS03,CassataPopIII13} and MUSE \citep{BaconMuse10,NanayakkaraHeII19}, as well as observations in individual galaxies such as NGC1569 with MEGARA \citep{Mayya20} or the Cartwheel Galaxy with MUSE \citep{MayyaWR23}, have found that nebular \heii emission is most likely powered by WR stars. However, the problem persists in HeII-emitting metal-deficient star-forming galaxies, where WR features are not prominent \citep[e.g. ][]{IzotovWR04,KehrigSBS18,NanayakkaraHeII19,Roy25}. WR stars are not expected to be seen in the spectra of extremely metal-poor galaxies (Z$\sim$0.03Z$_{\odot}$) such as
SBS 0335-052E \citep{IzotovWR99} and IZw18 \citep{ThuanIZw1804}. In these cases and in similar extremely metal-poor galaxies \citep{EnriqueHeII20}, the ionising power of the WR populations has been found insufficient to supply the UV radiation required to fully ionise He$^+$, leaving this scenario still open to debate. On the other hand, interacting binary stellar populations composed of young massive stars have also been shown to reproduce similar ionising conditions, particularly in metal-poor environments (Z$\sim$0.03Z$_{\odot}$) with a top-heavy initial mass function (IMF). These conditions favour the formation of stars with higher surface temperatures, resulting in a harder ionising spectrum and a greater production of photons with energies exceeding 54.4 eV \citep[e.g. ][]{EldridgeBinary12,EldridgeBinaries17,GotbergBinary17,SmithBinary18,GotbergBinary1719,GotbergStrip23}.  According to binary stellar evolution models such as BPASS \citep{EldridgeBinaries17,XiaoBPASS18}, the predicted ionising flux from systems that include metal-poor binary interactions, stellar rotation, and a top-heavy IMF can account for the overall \heii photon budget \citep[e.g. ][]{KehrigSBS18,NanayakkaraHeII19,HawcroftSTARBURST25}. Nevertheless, in extremely metal-poor galaxies, these models cannot fully account for the \heii photon production when the stellar Z is the same as that of the gas, even when including contributions from X-ray photons \citep[e.g. ][]{SenchynaXrays20,KehrigXRays21}, shocks \citep{GotbergStrip23} or gas-stripping processes \citep[e.g. ][]{GotbergBinary17,GotbergStrip23}. This limitation necessitates invoking nearly metal-free stars, suggesting a decoupling between stellar and gas-phase metallicities \citep[e.g. ][]{KehrigSBS18,EldridgeRev22}. 

In general, the origin of the nebular \heii emission is often attributed to WR stars or metal-poor binary systems with a top-heavy IMF in high-metallicity and metal-poor galaxies (up to 12 + log(O/H) $\approx$ 8; \citealt{KehrigSBS18,NanayakkaraHeII19,EnriqueHeII20}). However, an apparent anti-correlation between \heii photon production and gas-phase metallicity has been observed, which remains largely unexplained \citep[e.g. ][]{ShiraziWR12,SchaererXray19,EnriqueHeII20,Roy25}. Commonly invoked ionising sources, such as those mentioned, appear insufficient to account for the observed \heii photon production as the gas-phase metallicity of the emitters decreases, indicating that additional ionising mechanisms may be required, especially in extremely metal-poor galaxies (7.2 $\lesssim$ 12 + log(O/H) $\lesssim$ 7.7; \citealt{KehrigIZw15,KehrigIZw16,KehrigSBS18}). This is one of the primary reasons why PopIII-like stars are often invoked in such scenarios, as their extremely hot metal-free nature enables them to produce the hard ionising continuum required to explain the observed \heii emission in the most metal-poor galaxies \citep[e.g. ][]{SchaererPOPIII08,SchaererWRPopII02,CassataPopIII13,NanayakkaraHeII19,NakajimaPopIII22}. However, most studies that explore the \heii ionisation budget in detail tend to focus on individual galaxies—such as Cartwheel or SBS 0335-052E with MUSE \citep[e.g. ][]{KehrigIZw16,KehrigSBS18,Mayya20,MayyaWR23}—without testing whether the observed anti-correlation between \heii emission and gas-phase metallicity is also present in other galaxies. On the other hand, larger systematic studies using integral field units (IFUs), such as \citet{CassataPopIII13} with VIMOS and \citet{NanayakkaraHeII19} with MUSE, often do not fully exploit the rich two-dimensional spatial and spectral information that these instruments can provide. 

\noindent
The aim of this study was to carry out a systematic multi-wavelength investigation of the \heii ionisation budget by analysing a sample of \heii-emitting star-forming galaxies observed with the MUSE instrument, extending the dataset originally presented by \citet{NanayakkaraHeII19}. We made use of all available emission lines with a sufficiently high and reliable signal-to-noise ratio (S/N)  in the MUSE Hubble Ultra Deep Field surveys \citep{MHXDFBacon17, MHXDFBacon23} and complemented the MUSE data with observations from the James Webb Space Telescope's Near-Infrared Spectrograph (JWST/NIRSpec). This combined dataset allowed us to derive the gas-phase abundances of our \heii emitters and to perform a series of diagnostic analyses involving the \heii$\lambda$1640 line in conjunction with other key UV features, such as \oiii$\lambda$1661,1666, \ciii$\lambda$1907,1909, and \civ$\lambda\lambda$1548,1551, in order to constrain the origin of the \heii ionisation. 

The structure of this paper is as follows. Section \ref{sec:2} presents the MUSE and JWST sample, outlining the selection criteria and characterisation of the galaxies. In Section \ref{sec:3} we derive the gas-phase metallicities and determine key physical properties, including the electron density and the production rates of hydrogen- and He$^+$-ionising photons. Section \ref{sec:4} compares our results with the photoionisation models of \citet{GutkinModels16} and the BPASS binary stellar population models in order to constrain the possible origins of the \heii emission. Additionally, we discuss potential discrepancies between metallicity estimates based on UV calibrators and those obtained via direct methods using electron temperature measurements from JWST data. We also examine the impact of dust reddening across different wavelengths.

We adopt the standard $\Lambda$CDM cosmology with H$_0$ = 70 km/s/Mpc,\, $\Omega_{\Lambda}$ = 0.7, and\,$\Omega_M$ = 0.3.

\section{Sample characterisation}\label{sec:2}

\subsection{MUSE sample}\label{sec:data_UV}

\begin{figure*}[ht!]
\centering
    \includegraphics[width=\textwidth]{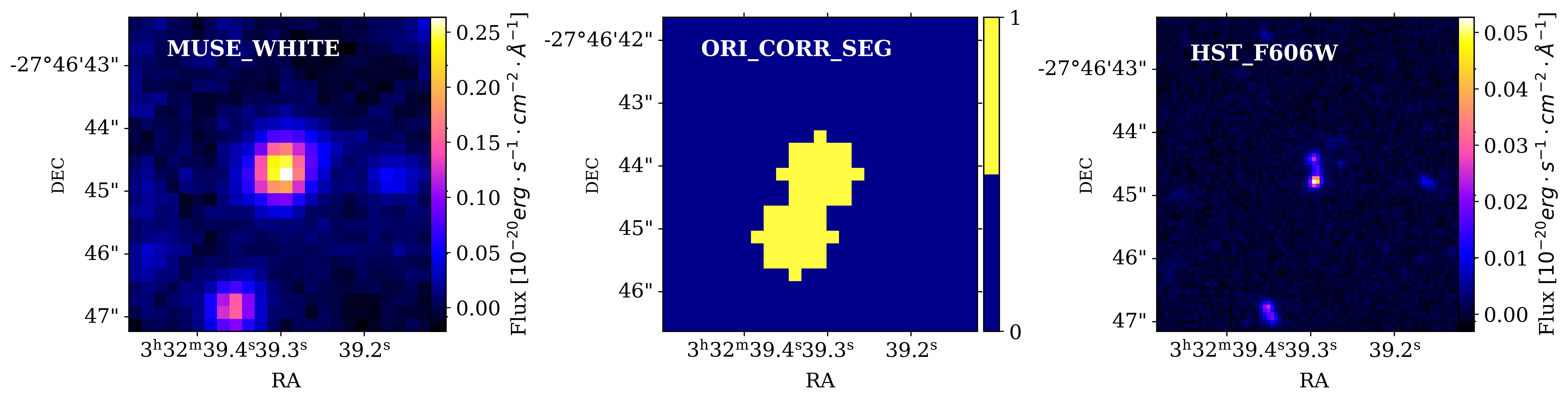}
    \includegraphics[width=\textwidth]{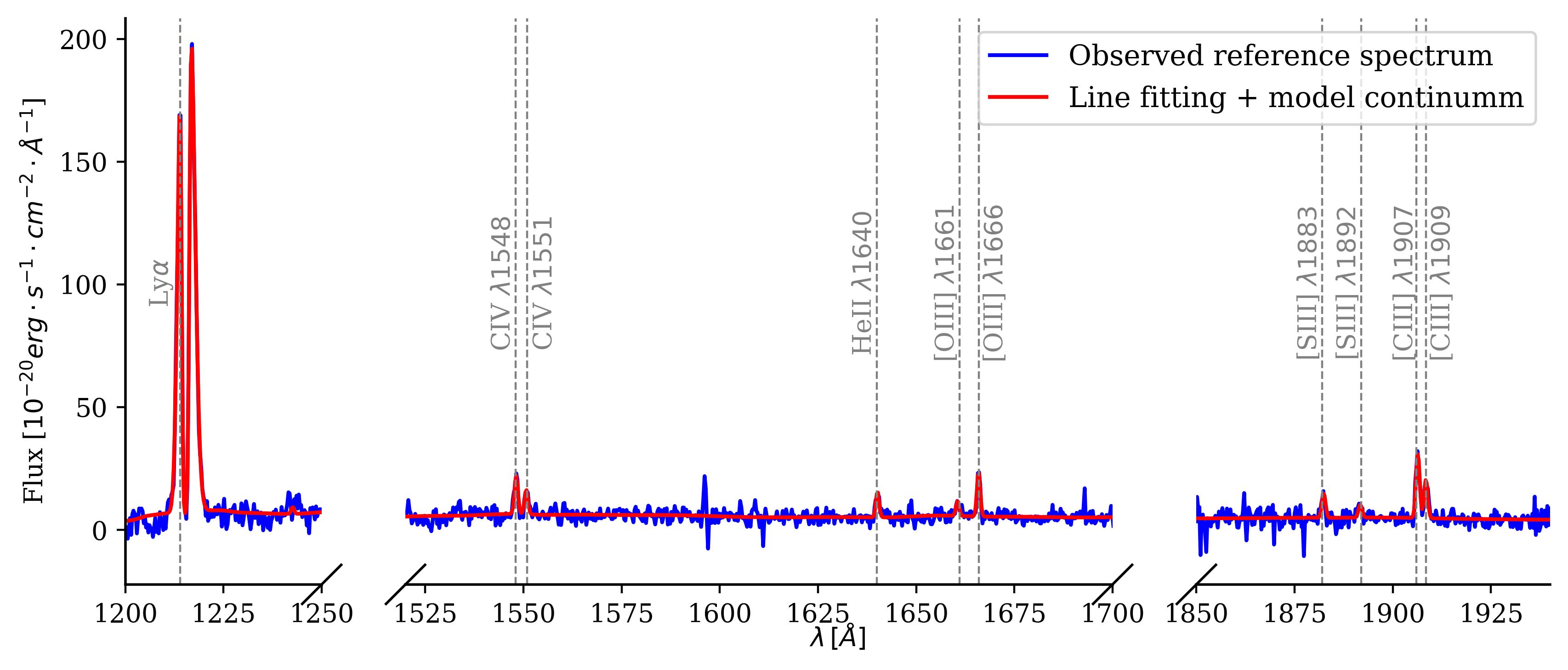}
\caption{Example of a detected source from MHUDF, representing the galaxy with ID 106 in the MXDF. Upper panels: MUSE white-light image (left), ORIGIN segmentation map of the source (centre), and HST F606W filter image (right). Lower pannel: Observed reference spectrum of the source extracted with ORIGIN (blue) and its respective continuum + emission line fit (red).}
\label{fig:maps_spec}     
\end{figure*}

The physical characterisation of \heii emitters requires medium  to high spectral resolution (R$\gtrsim$1000) in order to discriminate the nebular \heii line, along with broad spectral coverage to discern key UV features such as the \oiii $\lambda$1661,6 and \ciii $\lambda$1907,9, as well as Ly$\alpha$ emission. The Multi-Unit Spectrographic Explorer (MUSE; \citealt{BaconMuse10}) IFS at the ESO-VLT 8.2m telescope offers a rest-frame spectral range of $4650-9300$ \r{A}, with a resolution varying from 1750 at 4650 \r{A} to 3750 at 9300 \r{A}, and a spectral sampling of 1.25 \r{A}. It provides a field of view (FoV) of 1 arcmin$^2$ with a spatial sampling of $0.2\times0.2$ arcsec. These characteristics enable the detection of the \heii line in galaxies from the cosmic noon to earlier epochs (1.9 < z < 4.7).

Our sample was selected from the MUSE Hubble Ultra Deep Field surveys (MHUDF; \citealt{MHXDFBacon17}), comprising three MUSE fields: the MUSE eXtremly Deep Field (MXDF, 141-h depth), the Ultra Deep Field survey (UDF-10, 1 arcmin$^2$ 31-h depth field), and the 3$\times$3 arcmin mosaic of nine MUSE fields with 10-h depth (MOSAIC). The  MHUDF data release includes 2221 different spectroscopic sources detected in those surveys \citep{MHXDFBacon23}, all of them published in The Advanced MUSE Data products (AMUSED) web interface.\footnote{\href{https://amused.univ-lyon1.fr/}{AMUSED} web interface.} The available data products include datacubes that have already been reduced using a scheme to remove systematics and a self-calibration process to mask instrumental artefacts (see \citealt{MHXDFBacon17} for more details regarding the data reduction). Additionally, the Zurich Atmospheric Purge (ZAP) software \citep{SotoZap16} was applied to remove sky residuals and improve the overall datacube quality. 
The database includes a reference observed spectrum for each source, along with the corresponding continuum model and line fittings. The reference spectrum represents the observed integrated spectrum of the source, where the detected sources within the datacube and subsequent extraction were performed using the ORIGIN\footnote{In addition to ORIGIN, another algorithm developed by \citet{MHXDFBacon23}, called ODHIN, was used to extract spectra from deep-exposure sources. ODHIN is optimised for extracting spectra from blended sources, a common issue in MOSAIC and MXDF, where exposure times exceed 10 hours.} software \citep{MaryORIGIN20}, which is optimised for detecting faint emission sources in MUSE datacubes. Figure \ref{fig:maps_spec} presents an example of a source detected in this work and included in the AMUSED database. The left panel displays a white-light continuum image extracted directly from the MUSE datacube, while the central panel shows the segmentation map generated by the ORIGIN software, from which the reference spectrum was extracted, and the right panel shows the corresponding HST F606W filter image. The lower panel presents the spectrum extracted using the ORIGIN segmentation map (in blue), along with the continuum model and line fitting (in red), measured using the {\sc pyPlatefit} tool. This tool was specifically developed to analyse the line features present in this dataset \citep{MHXDFBacon23}, measuring the fluxes, errors, S/N, and FWHM for $\sim$65 lines that are included in the dataset. Additionally, the spectral energy distribution (SED), stellar mass, and star formation rate (SFR) for each detection, computed with {\sc Prospector} \citep{prospectorJohnson21} using Hubble Space Telescope (HST) photometry, are also provided. 

This dataset has  already been used in a wide variety of works such as those involving the Ly$\alpha$ luminosity function \citep{DrakeMHUDF17}, the MgII emission \citep{FeltreMHUDF18}, or extreme emission line galaxies \citep{IgnacioMHUDF24}. From this dataset we selected only sources with a S/N in the \heii line greater than 3. Only 25 out of the 2221 sources meet this criterion, two of which are confirmed active galactic nuclei (AGNs; ID:1051 and ID:1056; \citealt{NanayakkaraHeII19}), which present broad \heii, \ciii, and \civ features. We identified a greater number of \heii emitters in the MHUDF compared to previous studies (e.g. \citealt{NanayakkaraHeII19});  five of these sources are present in the \citet{NanayakkaraHeII19} dataset (without taking into account the two AGNs; see \ref{A2:fluxes_compar} for more details). In addition, 12 of them also present a prominent Ly$\alpha$ emission line.  We note that we do not know whether the rest of the galaxies present  Ly$\alpha$ emission since the 1216 \r{A} wavelength is outside the spectral range at that redshift;\footnote{Ly$\alpha$ falls within the MUSE wavelength range for 2.82 $\lesssim z \lesssim$ 6.64.} however, it is important to note that all the galaxies in this sample that show \heii emission also show Ly$\alpha$ emission when the $\lambda=1216$ \r{A} is in the observed rest-frame spectrum, when, in general, not necessary a galaxy with \heii emission present Ly$\alpha$ \citep{CassataPopIII13}. Figure \ref{fig:collage} shows the \heii emission detected in the 25 sources; the observed and fitted spectra (blue and red, respectively) are the reference and the continuum + fit spectrum as in Figure \ref{fig:maps_spec}. Additionally, the figure shows the \heii S/N and Ly$\alpha$ S/N for galaxies where Ly$\alpha$ is detected. All galaxies present a clearly visible and prominent \heii line;  furthermore, the Ly$\alpha$ emission is also prominent (except for ID:103), with S/N > 100 in some cases. A detailed spatially resolved analysis of Ly$\alpha$ profiles will be done in a forthcoming paper as half of the Ly$\alpha$ emitters in this sample exhibit a double-peaked profile, and, in general, all are partially absorbed.

The main characteristics of the sample are listed in Table \ref{tab1:Sample} (all tables can be found in Appendix \ref{A3:tables}).  
Figure \ref{fig:SFMS} shows the classical star formation main sequence (SFMS) for our sample (excluding the two AGNs), covering a stellar mass ($M_{\star}$) distribution in the range $7.98^{+0.68}_{-0.37}$ < $log(M_{\star}/M_{\odot})$ < $10.41^{+0.21}_{-0.24}$ with a median of $9.32^{+0.53}_{-0.49}$ $log(M_{\star}/M_{\odot})$, as well as a SFR distribution of $-1.30^{+2.40}_{-0.33}$ < $log(M_{\odot}\cdot yr^{-1})$ < $2.52^{+0.26}_{-0.35}$ with a median value of $0.36^{+0.59}_{-0.25}$ $log(M_{\odot}\cdot yr^{-1})$. These galaxies span  1.907 < z < 4.41, with a median z = 2.974, thus one-half of the sample lies beyond the cosmic noon (z > 3). The z = 3 theoretical SFMS derived by \citet{BoogaardSFMS18} is also shown, exhibiting a small offset with the data. This offset, also reported by \citet{MHXDFBacon23} for the entire HMUDF sample, is likely due to differences in the SFR calculation methods used by \citet{BoogaardSFMS18} and \citet{MHXDFBacon23} (see the latter for more details).

\begin{figure}[ht!]
\centering
    \includegraphics[width=\columnwidth]{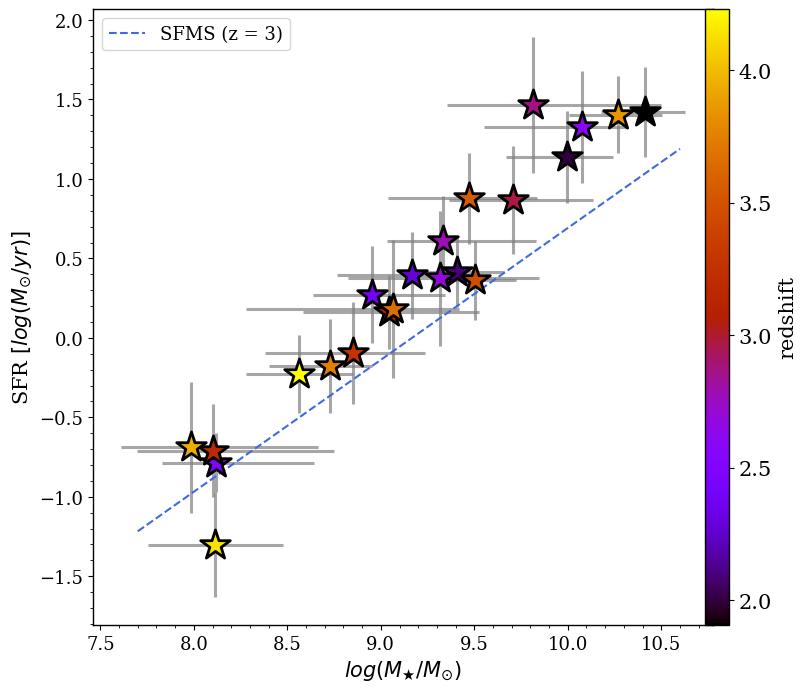}
\caption{Stellar mass vs. star formation rate for the sample, derived from the SED reported in the AMUSED database \citep{MHXDFBacon23}, colour-coded by redshift and excluding the two AGNs. The z = 3 SFMS, as derived by \citet{BoogaardSFMS18}, is represented by a dashed blue line. The sample shows a slight offset relative to the z = 3 SFMS, but follows the expected linear trend.}
\label{fig:SFMS}     
\end{figure}

A summary of the observed fluxes, errors, and FWHMs of all lines detected in the sample is given in Table \ref{tab:MUSE_lines} (the full table is available online). The table includes all emission and absorption lines detected in the galaxies; there are a  total of 54 lines ranging between the Ly$\alpha$ emission at 1216 \r{A} and the MgI absorption line at 2853 \r{A}. Only lines presenting a S/N greater than 3 are considerer in the analysis; the rest are excluded. Consequently, undetected lines (outside the wavelength range or having a low S/N) are represented as blank spaces in the table.

We derived the dust attenuation A$_V$ from the continuum UV slope $\beta$, parametrising the continuum as a power law \citep{CalzettiBeta94}: $f_{\lambda}\propto \lambda^{\beta}$, where $f_{\lambda}$ is the observed flux at a wavelength $\lambda$. The emission and absorption UV features within the fitting range (1300–1900 \r{A}) are masked to prevent them from altering the continuum shape; these features are listed in Table 2 of \citet{CalzettiBeta94}.
The $\beta$ slope can be related with the total extinction at $\lambda=1600$ \r{A} as $A_{1600} = 4.43 + 1.99\beta$ \citep{MeurerBeta99}, and  thus the $A_V$ derived from $\beta$ is
\begin{equation}\label{eq:Av_beta}
    A_V=R_V\frac{A_{1600}}{k_{1600}},
\end{equation}

\noindent
with $\rm k_{1600}=9.97$ derived from \citet{CalzettiExtCurve00} extinction curve at $\lambda=1600$ \r{A} and $\rm R_V=4.05$ the total V attenuation \citep{CalzettiExtCurve00}. Four of the galaxies present an insufficient continuum S/N to perform a reliable power-law fitting, giving a negative $\beta$ slope. For those cases (shown as blank spaces in the A$_V$ column of Table \ref{tab1:Sample}) we assume that $A_V=0$; for the rest, we correct all the fluxes using the \citet{CalzettiExtCurve00} extinction curve: $\rm F_{int}=F_{obs}\times10^{0.4A_Vk_{\lambda}/R_V}$.

Once the fluxes are corrected for extinction, we compute the UV magnitude, defined as the absolute magnitude at the 1500 \r{A} band (M$_{1500}$), i.e. the absolute AB magnitude derived from a simulated filter centred at $\lambda = 1500 \pm 100$ \r{A}. We obtain M$_{1500}$ by computing a running median, averaging the convolution of the corrected flux between 1400 and 1600 \r{A} with a window of size 20 \r{A}. M$_{1500}$ uncertainties are obtained as the standard deviation of 10000 Monte Carlo runs. 
\noindent
The magnitudes obtained are distributed between $-17 \gtrsim \rm M_{1500} \gtrsim -22$ with a median of $\sim-18$, which is comparable with the UV magnitudes of the \heii emitters reported by other authors (e.g. \citealt{NanayakkaraHeII19}). 
Both $M_{1500}$ and $A_V$ values are included in Table \ref{tab1:Sample}, along with the derived rest-frame \heii and Ly$\alpha$ equivalent widths (EWs). These EWs were calculated using the observed and continuum-fitted spectra available in the AMUSED database, following the definition: EW$=\int\rm (1-F_o/F_c)d\lambda$, where F$_o$ and F$_c$ represent the observed flux and the continuum-fitted flux, respectively. The EW is computed by centring the integral on the target line (1216 \r{A} for Ly$\alpha$ and 1640 \r{A} for \heii) within a 20 \r{A} window.

\subsection{JWST-NIRSpec optical sample}\label{sec:JWST_data}

\begin{figure*}[ht!]
    \centering
    \begin{subfigure}{0.79\textwidth}
        \includegraphics[width=\linewidth]{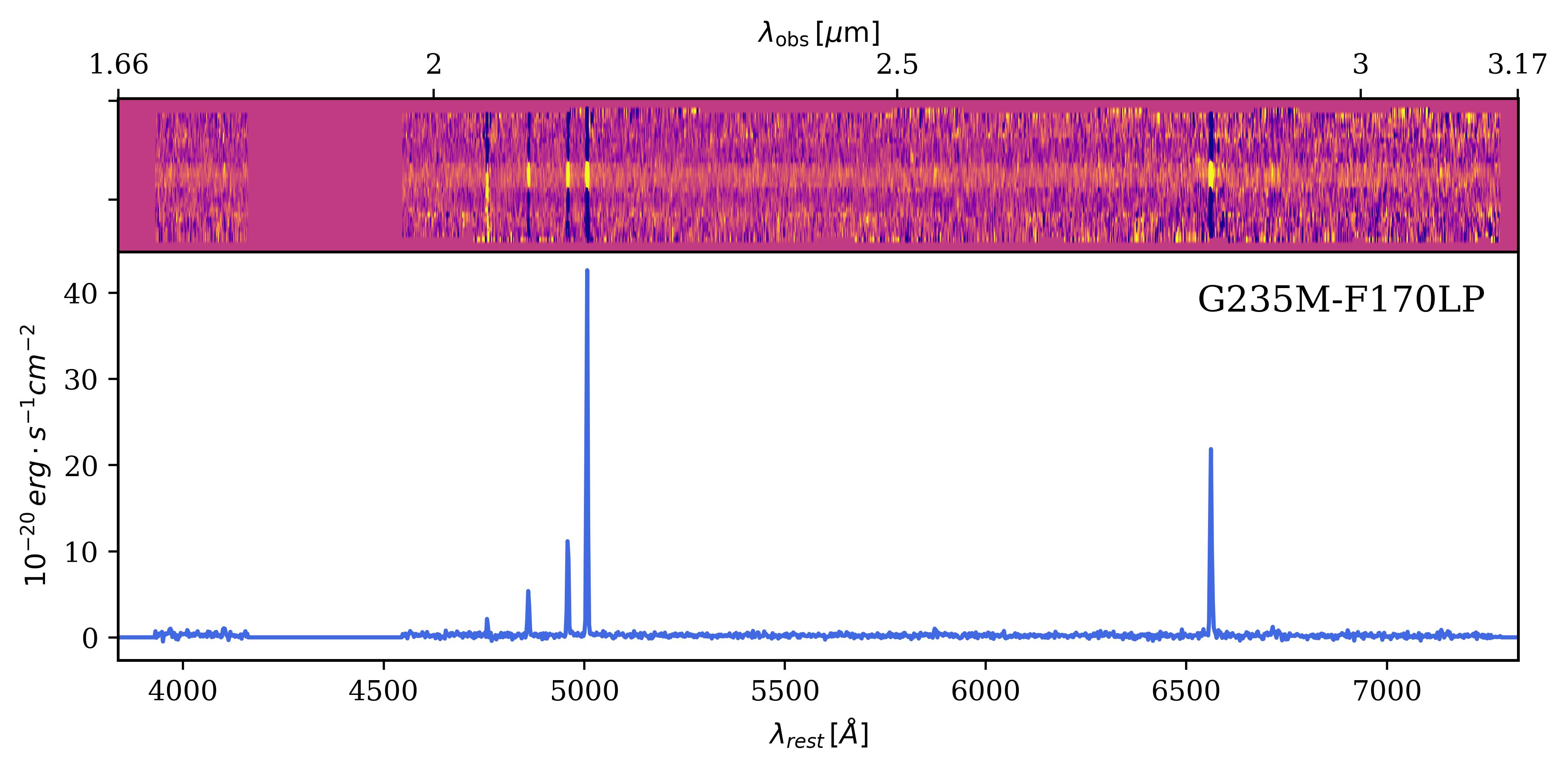}
        
    \end{subfigure}
    \begin{subfigure}{0.19\textwidth} 
        \centering
        \begin{subfigure}{\linewidth}
            \includegraphics[width=\linewidth]{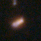}
            
        \end{subfigure}

        \begin{subfigure}{\linewidth}
            \includegraphics[width=\linewidth]{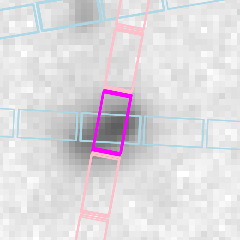}
            
        \end{subfigure}
    \end{subfigure}
    \caption{Reduced 2D and 1D NIRSpec spectrum of the galaxy with ID:50 at z = 3.325, obtained using the G235M/F170LP grating–filter combination. The right panels display a NIRCam RGB cutout of the galaxy, constructed using the F115W, F277W, and F444W filters (top), and a F444W cutout with the NIRSpec slit location overplotted in magenta (bottom). Both NIRCam cutouts were obtained from the {\sc grizli} public database.}
    \label{fig:NIRSpec_example}
\end{figure*}

\begin{figure}[ht!]
\centering
    \includegraphics[width=\columnwidth]{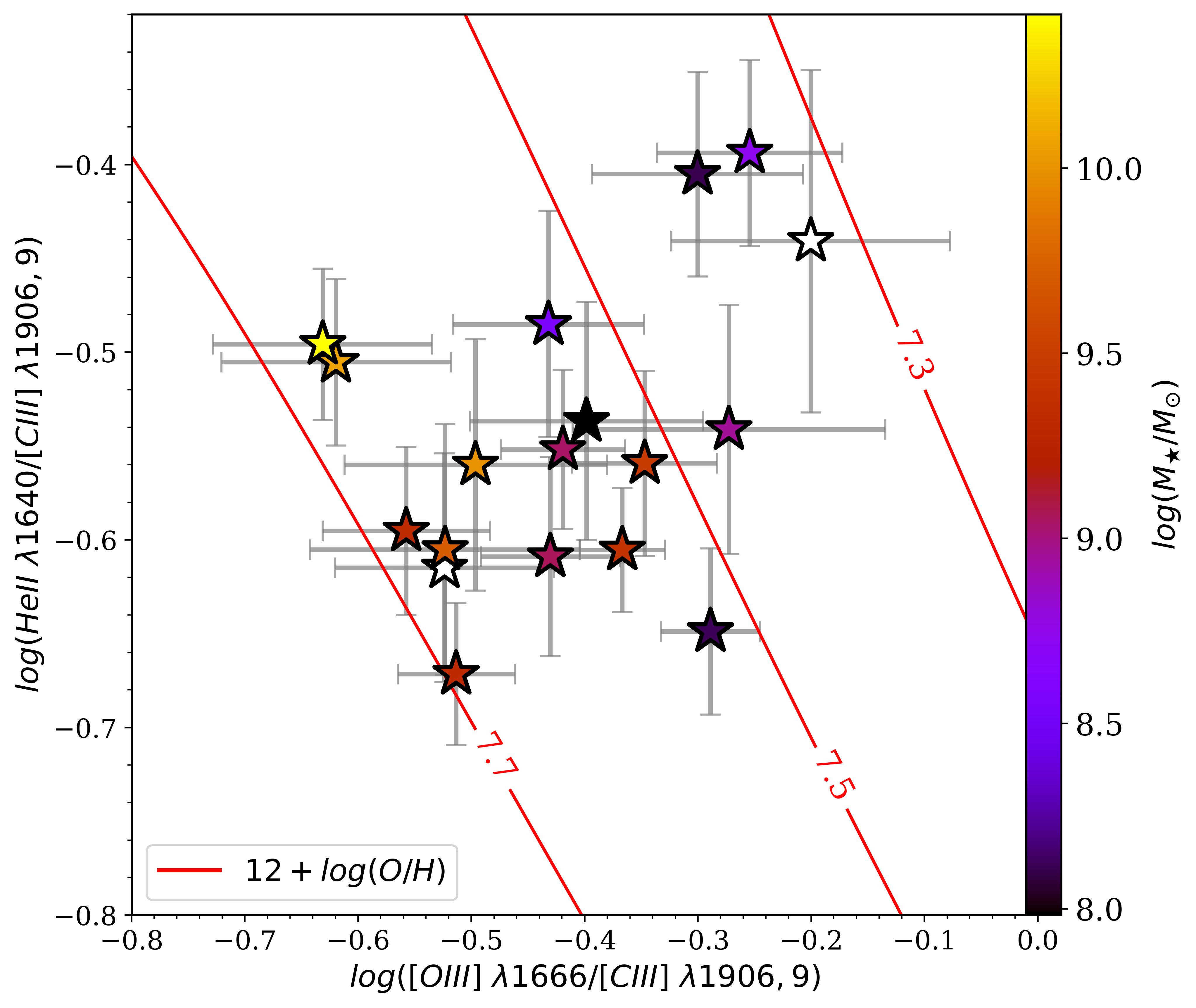}
\caption{He2-O3C3 diagnosis diagram. The red lines represent the polynomial from equation \ref{eq:He2-O3C3} at 12 + log$_{10}$(O/H) = 7.3, 7.5, and 7.7 respectively. The colours indicate stellar mass, while the open points correspond to galaxies without SED information, and consequently no measured $M_{\star}$. Only 18 points are shown as the rest present a S/N lower than 3 in at least one of the involved emission lines. All galaxies fall within a metal-poor regime;  three of them approach the extremely metal-poor regime (12 + log(O/H) $\approx$ 7.3), showing a trend with $M_{\star}$.}
\label{fig:He2-O3C3}     
\end{figure}

In addition to the MUSE UV rest frame data of the sample data, we included a complementary rest-frame optical dataset for the sample as having simultaneous optical and UV data enables a more comprehensive analysis of the physical and chemical properties of these galaxies.
The spectroscopic data from the Near-Infrared Spectrograph on the James Webb Space Telescope (NIRSpec-JWST) gives a wide range of emission and absorption features in the near-infrared (0.6–5.3 $\mu m$). At the redshift range of our galaxies (1.9 < z < 4.4), this spectral coverage corresponds to the optical regime, making NIRSpec spectra the ideal dataset for the mentioned task. NIRSpec gives spectra measured with different configurations, including a prism and several filter–grating combinations. 
The prism covers the entire NIRSpec spectral range, but at low resolution (R$\sim100$), causes a full blending of important emission lines, such as the \ha + \nii $\lambda\lambda$6548, 84 doublet or the \oiii $\lambda\lambda$4959, 5007 doublet. In consequence, we used spectra taken with the grating--filter combinations G140M-F070PL, G235M-F170LP, and G395-F290LP, which provide medium resolution (R$\sim1000$) and a coverage of 0.70–1.27 $\mu m$, 1.66–3.07 $\mu m$, and 2.87–5.10 $\mu m$, respectively.

In order to include a homogeneously reduced and comprehensive NIRSpec dataset, we conducted an extensive search in the DAWN JWST Archive (DJA), a public repository of JWST galactic photometric and spectroscopic data, reduced using the {\sc grizli}\footnote{\href{https://zenodo.org/records/8370018}{grizli} pipeline.} and {\sc msaexp} pipelines\footnote{\href{https://zenodo.org/records/8319596}{msaexp} pipeline.} \citep{ValentinoGrizli23, HeintzMSAEXP24, GraaffMSAEXP24}. We found in the DJA that only 5 of the 25 galaxies in our sample have reduced spectroscopic data for at least one of the mentioned filter--grating combinations, all of which were included in the JWST Advanced Deep Extragalactic Survey (JADES; \citealt{EugenioJADES25}). Figure \ref{fig:NIRSpec_example} presents an example of a 2D and a 1D NIRSpec spectrum for the galaxy with ID:50, obtained using the G235M/F170LP grating--filter combination. The figure also includes a NIRCam RGB cutout of the galaxy, created using the F115W, F277W, and F444W filters (top), along with a NIRCam F444W cutout displaying the NIRSpec slit location, marked by a magenta rectangle.\footnote{Both cutouts were obtained in the {\sc grizli} interactive data products interface: \href{https://dawn-cph.github.io/dja/general/mapview/}{https://dawn-cph.github.io/dja/general/mapview/}.} This galaxy exhibits the largest number of detected lines, ranging from Ly$\alpha$ at 1216 \r{A} to Pa$\gamma$ at 10938 \r{A}, offering an extensive dataset that spans from the UV to the near-IR.

We measured all the emission lines available within the spectral range of every galaxy, modelling the continuum as a power-law function, and using the Python {\sc MPFIT}\footnote{\href{ https://github.com/segasai/astrolibpy/tree/master/
mpfit}{MPFIT} documentation.} module to fit the lines. This module enables multi-peak Gaussian fitting for systems of multiplet lines, such as the \ha + \nii $\lambda\lambda$6548, 84 doublet or the \oiii $\lambda\lambda$4959, 5007 doublet (see \citealt{BETISI} for more details). Table \ref{tab:JWST_lines} shows the flux, error, and FWHM of the emission lines measured in the NIRSpec sample. As for the UV lines in section \ref{sec:data_UV}, we only considered those lines with a S/N greater than 3. 

This compilation of 1D MUSE and JWST data provided a diverse and heterogeneous sample, allowing the physical characterisation of \heii emitters and the measurement of chemical abundances, due to the large number of available emission lines. Furthermore, the complementary JWST/NIRSpec data allowed a more in-depth study of metallicity, electron temperature, and density. In addition, the MUSE 2D data available in the AMUSED database enabled the study of \heii\ extended emission. The 2D characterisation, along with the analysis of Ly$\alpha$ profiles, will be approached in a future paper.

\section{\heii emitters physical and chemical properties}\label{sec:3}

\subsection{Metallicity determination}\label{sec:metal}

Nebular emission-line models predict that gas-phase metallicity and the hardness of the ionising SEDs are sensitive to UV stellar absorption, where nebular lines make a significant contribution in both local and high-redshift galaxies and, consequently, UV fluxes correlate with gas-phase metallicity \citep{BylerUVZ17, BylerUVZ18}. In particular, since \ciii $\lambda$1907 is one of the  brightest UV lines, along with \oiii $\lambda$1661,6 \A, these lines, together with \heii 1640 \A, have proven to be promising UV metallicity calibrators \citep{FeltreUVZ16,JaskotUVZ16} yielding results similar to those obtained from direct optical determinations \citep{BylerUVZ20, IaniUVZ23}. 

We used the optimal 2D polynomial fitted to the model grid surface obtained in \citet{BylerUVZ20}, which relates the gas-phase metallicity with the He2-O3C3 diagnosis as 

\begin{equation} \label{eq:He2-O3C3}
    \begin{aligned}
    \rm 12 + log(O/H) = & \, \rm 6.88-1.13 x - 0.46 x^2 - 0.03 x ^3 \\
&  \rm - 0.61y + 0.02y^2 - 0.04 y ^3 \\
& \rm - 0.32 xy + 0.03xy^2 - 0.21 x^2y ,
    \end{aligned}
\end{equation}

\noindent
where x is log$_{10}$(\oiii$\lambda$1666/\ciii$\lambda$1906,9) and y is log$_{10}$(\heii $\lambda$1640/\ciii $\lambda$1906,9).

Out of the 25 galaxies, we calculated the metallicity for 18, as the remaining galaxies have a signal-to-noise ratio (S/N) lower than 3 in at least one of the involved emission lines. Figure \ref{fig:He2-O3C3} presents the He2-O3C3 diagnostic diagram for these 18 galaxies. Three red lines are plotted, representing the polynomial from Equation \ref{eq:He2-O3C3} at fixed values of 12 + log(O/H) (7.3, 7.5, and 7.7, respectively). The metallicity values for the galaxies can be found in Table \ref{tab:results}, in the range between 7.3 and 7.7 and with uncertainties of $\sim\pm$0.15, which places them within the low-Z regime; only two galaxies from our sample show metallicities ($\approx7.3$) slightly above  those (12 + log(O/H) $\approx$ 7.2 - 7.3) of the well-known extremely metal-deficient \heii emitters IZw18 and SBS0335-052E (e.g. \citealt{IzotovWR99,ThuanIZw1804,KehrigIZw16,KehrigSBS18}), which show relatively low $M_{\star}$: 10$^{6.63}M_{\odot}$ for IZw18 \citep{ZhouIWZ18mass21} and 10$^{7.43}M_{\odot}$ for SBS 0335-052E \citep{HuntSBSmass18}. In our galaxy sample, the stellar masses range from 10$^{7.98}M_{\odot}$ to 10$^{11.63}M_{\odot}$, with metallicity appearing to increase in more massive galaxies (see Figure \ref{fig:He2-O3C3}).

The C/O and N/O ratios were determined using the empirical relationship with metallicity derived by \citet{DopitaDiagnosis13} and later modified by \citet{BylerUVZ20} to better match observations in the low-metallicity regime (12 + log(O/H) < 8). The relationships are defined as 

\begin{equation}
    \rm log(C/O)= - 0.8 + 0.14 · (Z - 8.0 )+ log \left(1 + e^{\frac{Z-8}{0.2}}\right),
\end{equation}
\begin{equation}
    \rm log(N/O)= - 1.5 + log \left(1 + e^{\frac{Z-8.3}{0.1}}\right),
\end{equation}

\noindent
with Z = 12 + log(O/H).

Carbon is often referred to as a pseudosecondary element since the triple-$\alpha$ process is not directly dependent on metallicity. Instead, variations in the C/O ratio are typically linked to other metallicity-dependent processes, such as stellar winds or supernova explosions, rather than nucleosynthesis \citep{BergCO16,BylerUVZ20}. On the other hand, as the nitrogen forms as a secondary nucleosynthesis product at high metallicities, in our case it is expected to be constant \citep{DopitaDiagnosis13, BergCO16, BylerUVZ20}. Table \ref{tab:results} shows the C/O and N/O values for our sample with typical statistical errors of $\pm$0.14 dex \citep{BylerUVZ20}. The log(C/O) ratio ranges from -0.85 to -0.65, corresponding to -0.6 < [C/O] < -0.4.\footnote{We use the standard notation of [C/O], with the brackets indicating the log of the C/O ratio normalised to the Sun abundance, assuming log(C/O)$_{\odot}$ = -0.26 \citep{AsplundSUN21}.} Other rest-frame UV calibrations for deriving C/O abundances exist, such as the semi-empirical method presented by \citet{EnriqueCalUV17}, which used the \ciii, \civ, and \oiii doublets. They found that for galaxies at z > 2, the log(C/O) abundance ranges from -0.8 to -0.5 for 7 $\lesssim \rm 12 + log(O/H) \lesssim$ 7.5, which is in very good agreement with our results. In addition to  C/O, the log(N/O) abundance remains constant at $\approx-1.5$ ([N/O] = -0.42; \citealt{AsplundSUN21}), reinforcing the fact that all our \heii emitters  belong to a low-metallicity regime.

\subsection{Electron density}\label{sec:n_e}

The electron density (n$_e$) is defined as the number of free electrons per unit volume. By selecting two emission lines from the same ion that originate from nearly identical excitation energy levels, the relative excitation rate becomes primarily dependent on the collision strength \citep{OsterbrockBook06}. In such cases, if the collisional de-excitation rates of the two levels are different, the intensity ratio of these lines will depend mostly on the electron density. Traditionally, the optical \sii$\lambda$6716/$\lambda$6731 ratio has been used to estimate n$_e$. However, in the UV regime, the \ciii$\lambda$1907/$\lambda$1909 ratio exhibits similar characteristics, making it a suitable diagnostic for determining the electron density in this dataset \citep{KeenanNe92,OsterbrockBook06,ProxaufNe14}. We used {\sc PYNEB}, as described in Section \ref{sec:metal}, assuming an electron temperature of T$_e$ = 10$^{4}$K. The \ciii$\lambda$1907/$\lambda$1909 ratio is employed to estimate the electron density for those galaxies in which both \ciii lines are detected with a S/N greater than 3. For galaxies with ID 50, 22, and 1141, we instead used the optical \sii ratio from NIRSpec observations as the S/N of the \sii lines is higher than that of the \ciii doublet in these cases. The uncertainties associated with every value are obtained as the standard deviation of 10000 Monte Carlo runs. Table \ref{tab:results} summarises the values obtained for each galaxy. We find that the galaxies present  low to intermediate $n_e$ regimes, between 10$^2$-10$^3$ cm$^{-3}$, with only two cases with $n_e\approx10^4$cm$^{-3}$. However, it is important to note that the uncertainties associated with the derived values are significant (up to 50\%) due to the \ciii lines, which are intrinsically faint and therefore noisy.

\begin{figure}[t!]
\centering
    \includegraphics[width=\columnwidth]{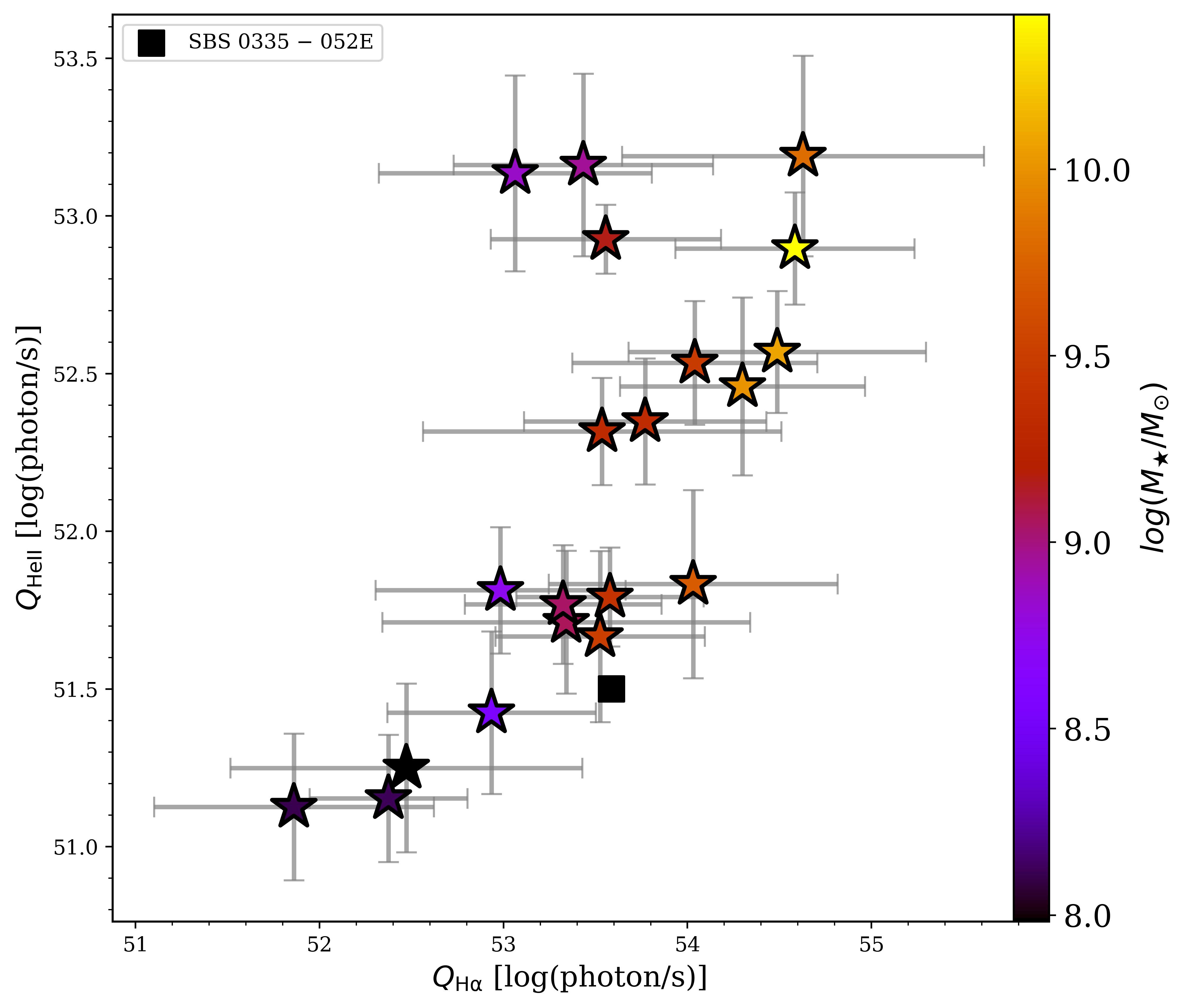}
\caption{He$^{++}$ vs. H$^{+}$ ionising photon production, derived from Equations \ref{eq:QHe} and \ref{eq:QH}. The colours indicate the stellar mass of each galaxy. The black square represents measurements for the extremely metal-poor star-forming galaxy SBS 0335-052E taken from \citep{KehrigSBS18}.}
\label{fig:QHe_QH_mass}     
\end{figure}

\subsection{Ionising photon production}\label{sec:photons}

\begin{figure*}[ht!]
\centering
    \includegraphics[width=0.49\textwidth]{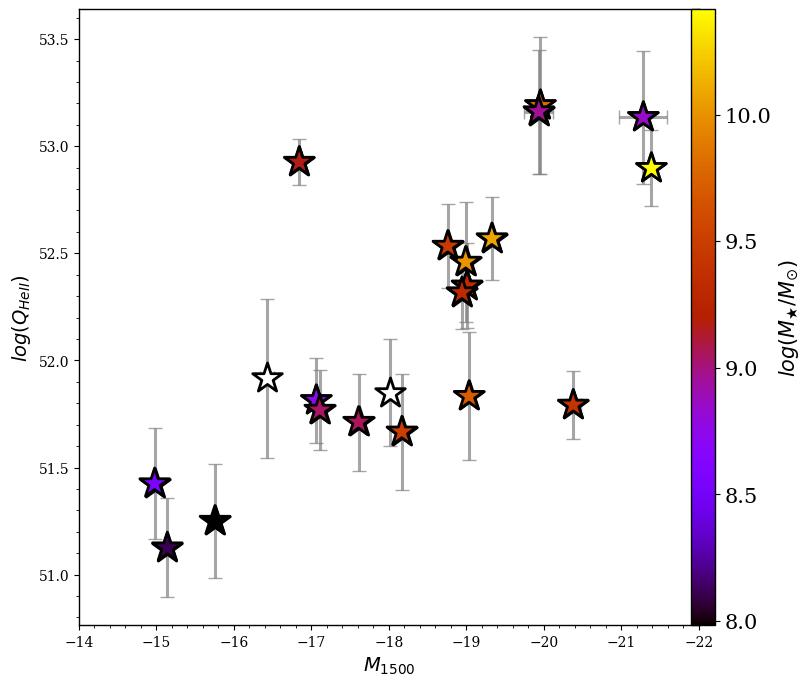}
    \includegraphics[width=0.49\textwidth]{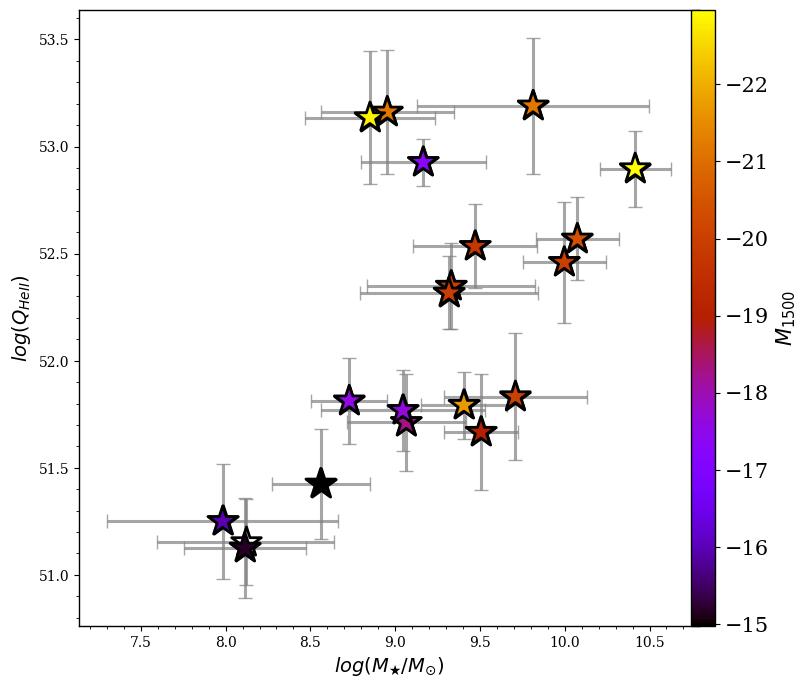}
\caption{Q$_{\rm HeII}$ in units of log(photons/s) as a function of UV magnitude (left panel) and of stellar mass (right panel). The colours indicate the stellar mass of each galaxy. The relationship between the two ionising photon production rates appears to be exponential, with both quantities increasing as a function of stellar mass.}
\label{fig:QHe_UV_mass}     
\end{figure*}

The presence of the nebular \heii 1640 \A line in at high redshift indicates the existence of hard ionising radiation, requiring photons with energies $h\nu>54.4$ eV to ionise the He$^{+}$ ion, which is characteristic of ongoing extreme star formation activity in very low-metallicity regimes \citep{KehrigM3311,KehrigIZw15,KehrigSBS18,CassataPopIII13}. 
\noindent
These photons  produce an ionising flux that is related to the \heii 1640 \A luminosity as follows \citep{OsterbrockBook06}:
\begin{equation}\label{eq:QHe}
    \rm Q_{\heii} = \frac{\rm L_{1640}}{\rm j_{1640}\alpha_B(\heii_{1640},T)} \quad [\rm phot/\rm s].
\end{equation}

\noindent
Here L$_{1640}$ is the \heii 1640 \A luminosity, reddening corrected using the A$_V$ derived from equation \ref{eq:Av_beta}; j$_{1640}$ is the \heii 1640 \A emissivity; and $\rm \alpha_B(\rm \heii_{1640},T)$ is the recombination coefficient for the \heii 1640 \A at a given temperature. 

Analogously, we can also calculate the hydrogen photon production rate using the \ha line as 

\begin{equation}\label{eq:QH}
    \rm Q_{\rm H\alpha} = \frac{\rm L_{\rm H\alpha}}{\rm j_{\rm H\alpha}\alpha_B({\rm H\alpha},T)} \quad [\rm phot/\rm s], 
\end{equation}

\noindent
where L$_{\rm H\alpha}$ is the \ha 6563 \A luminosity, j$_{\rm H\alpha}$ is the \ha emissivity, and $\alpha_B(\rm H\alpha,T)$ is the recombination coefficient for the \ha line at a given temperature. We computed both Q$_{\rm HeII}$ and Q$_{\rm H\alpha}$ using the Python module {\sc PYNEB} \citep{LuridianaPyneb15} assuming case B recombination, T = 10$^{4}$ K, and an electron density n$_{e}$ = 100 cm$^{-3}$.\footnote{We fix the assumed $n_e$ for all galaxies since the emissivity is only weakly dependent on the electron density, only around a 0.5\% between 10$^2$ and 10$^3$ and 1\% in the extreme cases of $n_e$ = 10$^4$ found in section \ref{sec:n_e}.} Since the \ha line is not included in the sample, we estimated $L_{H\alpha}$ using the SFR derived from the SED (table \ref{tab1:Sample}) and applying the SFR-L$_{H\alpha}$ relation \citep{KennicuttSFRHa94, MadauSFRHa98,KewleySFRHa02}: SFR$_{\rm H\alpha}(\rm M_{\odot}yr^{-1})=7.9\times10^{-42}\rm L_{\rm H\alpha}(erg/s)$.

Figure \ref{fig:QHe_QH_mass} shows the relation between He$^{++}$ versus H$^{+}$ ionising photon production. Given that the ionisation potential of He$^+$ (54.4 eV) is significantly higher than that of H$^0$ (13.6 eV), it is expected that Q$_{H\alpha}\geq$ Q$_{\rm HeII}$; if the incident SED is hard enough to doubly ionise helium, it is also sufficient to ionise hydrogen. In addition, the trend in figure \ref{fig:QHe_QH_mass} appears to be exponential, with the minimum value of Q$_{\rm H\alpha}$ (10$^{52.94}$) reached at the minimum value of Q$_{\rm HeII}$ (10$^{52.26}$), and the maximum value of Q$_{\rm H\alpha}$ (10$^{55.71}$) reached at the maximum value of Q$_{\rm HeII}$ (10$^{54.33}$; individual values can be found in Table \ref{tab:results}). Furthermore, both Q$_{\rm HeII}$ and Q$_{\rm H\alpha}$ appear to correlate with the stellar mass (see Figs. \ref{fig:QHe_QH_mass} and \ref{fig:QHe_UV_mass}). On the one hand, a correlation between $M_{\star}$ and Q$_{\rm H\alpha}$ is expected, as Q$_{\rm H\alpha}$ is proportional to the SFR, and the SFR and $M_{\star}$ follow a linear trend in the SFMS (figure \ref{fig:SFMS}).
On the other hand, Fig. \ref{fig:QHe_UV_mass} shows that Q$_{\rm HeII}$ values seem to increase with the absolute UV magnitude (M1500) and with the total stellar mass of our galaxy sample of nebular \heii emitters. These trends indicate that hot massive stars might be the main cause of  the \heii ionisation in our sample since such stars are expected to dominate the far-UV stellar continuum at 1500 \A and the stellar mass in metal-poor star-forming galaxies (e.g. \citealt{Guseva00,ShiraziWR12,KehrigIZw15,SzecsiTWUINS15,KehrigIZw16,KehrigSBS18,SenchynaXrays20,Berg21}). This is also expected since brighter galaxies have more star formation, and thus there can be more stars that can contribute to the ionising photon production.

\section{Discussion}\label{sec:4}

\subsection{UV diagnosis diagrams}\label{sec:discussion_models} 

\begin{figure*}[ht!]
\centering
    \includegraphics[width=0.49\textwidth]{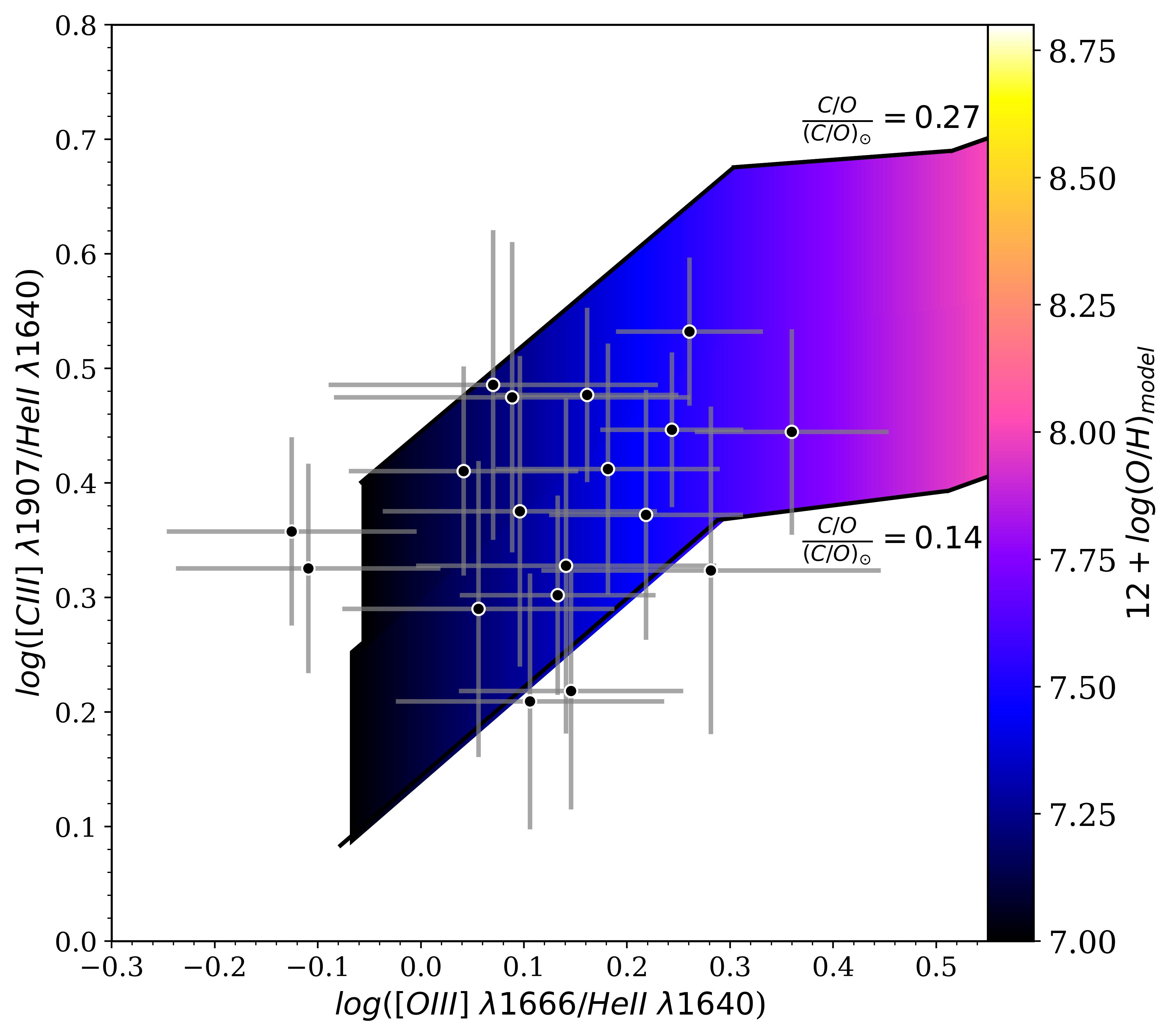}
    \includegraphics[width=0.5\textwidth]{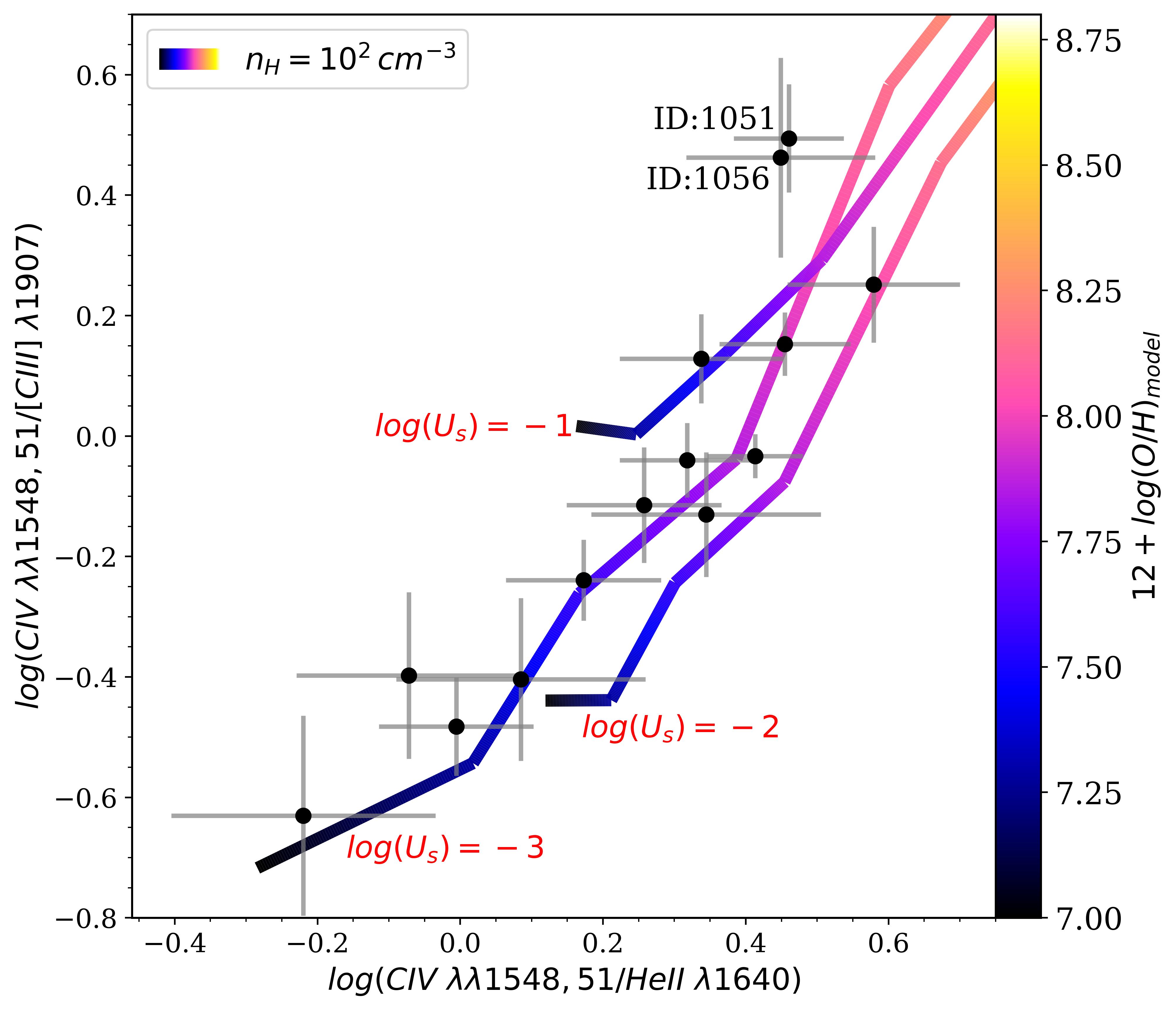}
\caption{Left: \ciii$\lambda1907$/\heii -- \oiii$\lambda1666$/\heii diagnosis. The \citet{GutkinModels16} models plotted correspond to a pure photoionisation scenario due to massive stars with log(U$_s$) = -1, $\xi_d$ = 0.1, $m_{\rm up}$ = 300 M$_\odot$, and 0.14 < (C/O)/(C/O)$_{\odot}$ < 0.27. The coloured area represents the region where the values of (C/O)/(C/O)$_{\odot}$ span vertically between 0.14 and 0.27 (both values marked by black lines), while the gas-phase metallicity varies horizontally between 12 + log(O/H) = 7 and 8.75. \\
Right: \civ$\lambda\lambda1548,51$/\ciii$\lambda1907$ -- \civ$\lambda\lambda1548,51$/\heii diagnosis for those galaxies where \civ$\lambda\lambda1548,51$ is detected. The rainbow lines correspond to photoionisation models with gas density n$_\mathrm{H}$ = 10$^{2}$ cm$^{-3}$, tracing a metallicity gradient from 12 + log(O/H) = 7 to 8.75. Three different paths are shown, each representing a different ionisation parameter log($U_s$), ranging from –3 to –1.  
All models assume a C/O abundance ratio of 0.1(C/O)$_\odot$. The observed scatter reflects a gradient in the ionisation parameter as \civ increases linearly relative to the other lines in a pure photoionisation scenario. In addition, the two AGNs (ID:1051 and ID:1056) present a significantly higher \civ/\ciii ratio, distinguishing them from the rest of the sample. All the points in the two diagnoses lie in the region of very low metallicity (12 + log(O/H) < 7.7), which are similar to the measured values.}

\label{fig:CIIIOIIIdiag}     
\end{figure*}

The characterisation of star formation and ISM conditions in star-forming galaxies and AGNs requires  a comprehensive interpretation of optical and/or UV emission lines. The computation of stellar population synthesis and photoionisation codes leads to the generation of various nebular emission models, which can be used to interpret the observed emission line fluxes. Furthermore, the \heii$\lambda$1640 line is commonly used as a standard reference in UV diagnostics, which is analogous to \hb in optical diagnostics, due to its strong dependence on both the metallicity and the ionisation parameter \citep{FeltreUVZ16, FeltreDiagnosis16}. In particular, \citet{GutkinModels16} present a set of UV models based on stellar population synthesis models \citep{BruzualModels03} and {\sc CLOUDY} photoionisation models \citep{FerlandCLOUDY13}. These models are suitable for studying the chemical evolution and ISM properties of galaxies at high redshift (2 $\lesssim$ z $\lesssim$ 9) based on UV emission features \citep[e.g. ][]{GutkinModels16,FeltreDiagnosis16,StarkDiagnosis16,NanayakkaraHeII19,Bunker23,Atek24}, and are parametrised using six values, which are more sensitive to UV lines than optical lines: the total ISM metallicity (Z$_{\rm ISM}$), the hydrogen gas density ($n_H$), the zero-age ionisation parameter at the Strömgren radius (U$_s$), the dust-to-metal ratio ($\xi_d$), the upper mass cut-off of the initial mass function (m$_{\rm up}$), and the C/O ratio.

In figure \ref{fig:CIIIOIIIdiag} (left) we show the \ciii$\lambda1907$/\heii versus \oiii$\lambda1666$/\heii diagnosis. The models plotted correspond to a pure photoionisation scenario due to massive stars, as pure AGN ionisation would produce \ciii/\heii ratios lower than 1 \citep{FeltreDiagnosis16,GutkinModels16}. These models, which best fit the data, correspond to the highest ionisation parameter in the set, with log(U$_s$) = -1, a low dust-to-metal ratio ($\xi_d$ = 0.1), and a top-heavy IMF with $m_{\rm up}$ = 300 M$_\odot$. The free parameters are the ISM metallicity (Z$_{\rm ISM}$), expressed in terms of 12 + log(O/H), and the C/O abundance (normalised to (C/O)$_{\odot}$ = 0.55; \citealt{AsplundSUN21}), represented in Figure \ref{fig:CIIIOIIIdiag} as an area with a width corresponding to Z$_{\rm ISM}$ and a height corresponding to C/O. All points fall within this area, with values of 12 + log(O/H)$_{\rm model}$ < 7.7, which are similar to the values found in Section \ref{sec:metal}. The C/O values range between 0.14(C/O)$_\odot$ and 0.27(C/O)$_\odot$, corresponding to -0.85 < [C/O] < -0.4, generally falling below the measured values shown in Section \ref{sec:metal} (-0.6 < [C/O] < -0.4). These variations in the C/O ratios when using UV lines are well documented in the literature and can fluctuate by more than 0.6 dex at low metallicities \citep{BergCO16, BergCO19, BylerUVZ20}. This is primarily due to the strong dependence of C/O on the star formation history (SFH) and supernova feedback, and thus the \ciii and \oiii lines alone are insufficient to accurately determine the gas-phase oxygen abundance \citep{BergCO19, BylerUVZ20}.

In addition, the diagnostic diagrams that best distinguish between star formation-driven photoionisation, AGN activity, and shocks are those involving the UV lines \civ${\lambda\lambda1548,51}$, \ciii${\lambda1907}$, and \heii$\lambda1640$, as AGNs and shocks produce markedly different line ratios in the UV compared to their optical counterparts \citep[e.g. ][]{Villar-Martin97CIV,AllenCIV98,GrovesCIV04,AllenModels08,FeltreUVZ16}. In addition, the ratios involving the lines mentioned above have  been found to be sensitive to metallicity, and  the \civ$\lambda\lambda1548,51$/\ciii$\lambda1907$ ratio also depends on the ionisation parameter \citep{Nagao06, FeltreDiagnosis16}. In figure \ref{fig:CIIIOIIIdiag} (right) we examine the \civ$\lambda\lambda1548,51$/\ciii$\lambda1907$ versus \civ$\lambda\lambda1548,51$/\heii diagnosis using the \citealt{GutkinModels16} models as in figure \ref{fig:CIIIOIIIdiag} (left). 

\noindent
The figure shows the mentioned line ratios for all our galaxies where the \civ lines are detected, including the two AGNs (ID:1051 and ID:1056), exhibiting a linear trend between the two ratios as the \civ$\lambda\lambda1548,51$ flux increases. This behaviour correspond with a photoionisation scenario driven by star formation in most galaxies. 
However, the trend breaks for the two AGNs, which show significantly higher \civ$\lambda\lambda1548,51$ emission relative to the other lines, which is an expected signature of AGN activity \citep{GutkinModels16,FeltreDiagnosis16}. The coloured lines represent models for pure star-forming photoionisation with the same gas density of n$_\mathrm{H}$ = 10$^{2}$ cm$^{-3}$ and three different ionisation parameters (from log(U$_s$) = -3 to -1). The rainbow-coded colours indicate the metallicity, ranging from 12 + log(O/H) = 7 to 8.75. An increase in ISM metallicity is also expected to lead to a higher abundance of cooling agents, resulting in an increased ratio of metal lines to \heii emission \citep{GutkinModels16}, and thus the increment in the \civ/\heii ratio of figure \ref{fig:CIIIOIIIdiag} (right) translates into a metallicity gradient. For similar metallicity values, \civ/\ciii increases with the ionisation parameter. For a fixed U$_s$, the models show that the increase in the \civ$\lambda\lambda1548,51$ emission relative to the collisionally excited \ciii line in our galaxies can be attributed to an increase in metallicity, which is compatible with our measured values of 12 + log(O/H) < 7.7. Since U$_s$ is proportional to  the H-ionising photon rate (Q$_H$) and to the gas density n$_H$, this parameter provides insight into the hardness of the ionising spectrum. A high U$_s$ can result from AGN activity, which tends to increase the gas density and ionising output, or from elevated ionising photon production \citep{FeltreUVZ16,FeltreDiagnosis16}. As discussed in section \ref{sec:photons}, Q$_H$ can be higher than 10$^{54}$ photons/s in star-forming galaxies, particularly in the most massive \heii emitters.

\subsection{Direct vs indirect abundance determination}\label{sec:discussion_metals}

The use of UV lines to estimate chemical abundances in high-redshift star-forming galaxies has been extensively discussed by several authors \citep[e.g. ][]{VillarTe04,YuanTe09,ErbTe10,ChristensenTe12,JamesTe14,BaylissTe14,SteidelTe16,KojimaTe17,NichollsTe20}. All these studies found that using the UV lines of the O$^{++}$ ion yields T$_e$ values similar to those obtained from optical lines, showing the importance of using UV lines to characterise physical conditions in high-redshift galaxies (see \citealt{NichollsTe20} for more details). In particular, UV \ciii and \oiii lines are optimal for this estimation since the emissivity of \oiii $\lambda$1666 \A and the strength of the \ciii $\lambda$1909 \A line are sensitive to $T_e$ and the oxygen gas abundance rather than the absolute carbon abundance. Thus, the combination of \ciii and \oiii emission lines serves as a tracer of the gas-phase oxygen abundance \citep{JaskotUVZ16,BylerUVZ18,BylerUVZ20}.

Galaxy 50 is the one with the broadest wavelength coverage when combining MUSE and NIRSpec data, allowing   a direct estimation of T$_e$ using the \oiii UV doublet 1660,66 and the optical \oiii 5007 \A line. \citet{NichollsTe20} presents a formula for obtaining T$_e$ by combining optical and UV O$^{++}$ ion lines, obtained by fitting several line ratios obtained from {\sc MAPPINGS v5.1} photoionisation models, using extensive collisional and radiative excitation/de-excitation data, as well as collisional cross-sections from \citet{Lennon94}. This fitting assumes a low-density limit and a T$_e$ range between 5000 and 32000 K. The results obtained by \citet{NichollsTe20} from this fitting are consistent with the results of the mentioned studies and is expressed as 
\begin{equation}
    log(T_e[\rm OIII])=\frac{5.0485 - 8.2350x + 0.6987x^2}{1 - 2.0120x + 0.2510x^2},
\end{equation}

\noindent
where
\begin{equation*}
    x=log\left(\frac{f_{\rm 1660}+f_{\rm 1666}}{f_{\rm 5007}}\right)
\end{equation*}

\noindent
is the ratio of the dust corrected fluxes of the UV to optical \oiii lines. 

For ID:50, we obtain a T$_e[\rm OIII]=(2.28\pm0.29)\times10^4$ K. This temperature can be used to derive the C/O using the \oiii 1666 \A and \ciii 1909 \A as  \citep{GarnettCO95}

\begin{equation} \label{eq:te}
    \frac{C^{++}}{O^{++}}=0.089e^{-1.09/t}\frac{f[\rm CIII]_{\lambda1909}}{f[\rm OIII]_{\lambda1666}}
\end{equation}

\noindent
and 

\begin{equation}
    \frac{C}{O}=ICF\times\left[\frac{C^{++}}{O^{++}} \right],
\end{equation}

\noindent
where
$t = T_e[\rm OIII]/10^4$ and ICF is the ionisation correction factor. The ICF is a challenging parameter to determine, particularly when no data are available to estimate the O$^+$ abundance. \citet{AmayoICF21} presented an extensive grid of photoionisation models to derive the ICF for various ionic species, including C$^{++}$/O$^{++}$. They found that at low metallicities (12 + log(O/H) < 7.7), the C$^{++}$/O$^{++}$ ICF tends to 1. Additionally, since O$^{++}$ and C$^{++}$ have similar ionisation potentials (54.9 eV and 47.9 eV, respectively), one can assume C$^{++}$/O$^{++} \approx$ C/O if the incident SED is  hard  enough to produce both species at low metallicity and the radiation field is not dominated by cooler stars. In the latter case, C$^{++}$ can exceed O$^{++}$ \citep{GarnettCO95}. In our case, since we are in a low-metallicity regime and the radiation field can indeed produce photons with energies higher than 54.9 eV (as helium is also ionised twice), we assume that ICF = 1.

\noindent
With these assumptions, and applying equation \ref{eq:te}, we found that the C/O abundance in ID:50 is 0.12$\pm$0.03 ($\sim21.3\%$ the solar abundance). This value is slightly lower compared to the obtained from the He2-O3C3 calibration (0.19$\pm$0.02, corresponding to $\sim35\%$ of the solar abundance), but still falls within the value of the calibration taking into account the uncertainties. These values are also similar to those obtained using the direct method and the semi-empirical calibration of \citet{EnriqueCalUV17}, which yields a value of $\sim$0.17 assuming the 12 + log(O/H) obtained for this galaxy (= 7.62; see Table \ref{tab:results}). Differences between these values can generally be expected, especially when combining optical and UV emission lines, as the uncertainties in dust extinction and de-reddening calculations are higher for UV line fluxes \citep{NichollsTe20}. Moreover, this correction does not necessarily need to be identical for optical and UV lines \citep{Cardelli89,CalzettiBeta94}, making it an intrinsic challenge when using UV emission lines \citep{NichollsTe20}. This is discussed in more detail in the next subsections.

\subsection{Optical versus UV dust absorption correction}

It is well known that interstellar extinction exhibits a strong wavelength dependence, particularly in the UV part of the spectrum. This becomes less pronounced towards the infrared, primarily due to the effects of scattering and absorption followed by re-emission, which are less significant at longer wavelengths \citep{Cardelli89,CalzettiBeta94,OsterbrockBook06}. The amount of stellar continuum absorbed by dust, typically estimated as the difference between the observed and intrinsic light \citep{CalzettiExtCurve00}, varies significantly from the UV to the IR. This absorption is commonly related to the visual extinction, A$_V$, and is described using an analytical expression known as the extinction law, which depends on wavelength, $k(\lambda)$ \citep[e.g. ][]{FitzpatrickExt90,CalzettiExtCurve00,FitzpatrickExt07,FitzpatrickExt09,ButlerExt21}. This wavelength dependence causes the stellar continuum to appear sloped, particularly at shorter wavelengths. The steepness of this slope can be parametrised to estimate the total extinction, as described in Section \ref{sec:data_UV} \citep{CalzettiBeta94,MeurerBeta99}. Nevertheless, we cannot assume that this steepness is uniform across the entire spectrum. As a result, the A$_V$ derived from the UV slope may differ from that obtained using the optical Balmer decrement or the near-infrared Paschen lines. Consequently, the dust-reddening correction can vary depending on the wavelength range we are working with, and can affect the results when we perform a multi-wavelength analysis, as mentioned in Sections \ref{sec:data_UV} and \ref{sec:discussion_metals}.

\begin{figure*}[ht!]
\centering
    \includegraphics[width=0.331\textwidth]{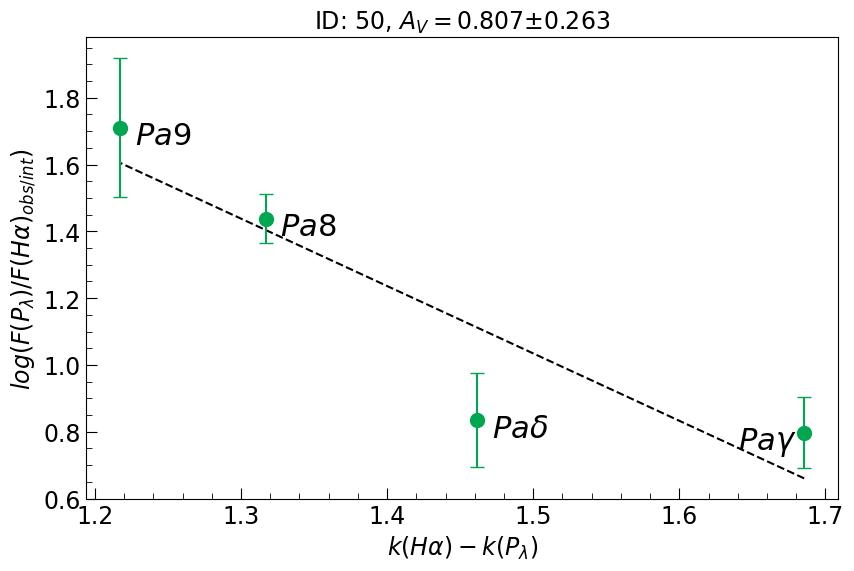}
    \includegraphics[width=0.33\textwidth]{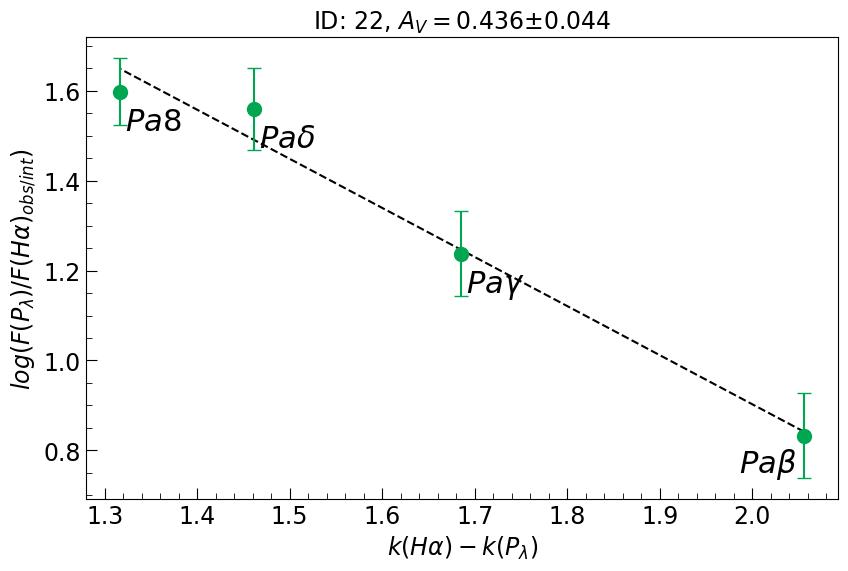}
    \includegraphics[width=0.33\textwidth]{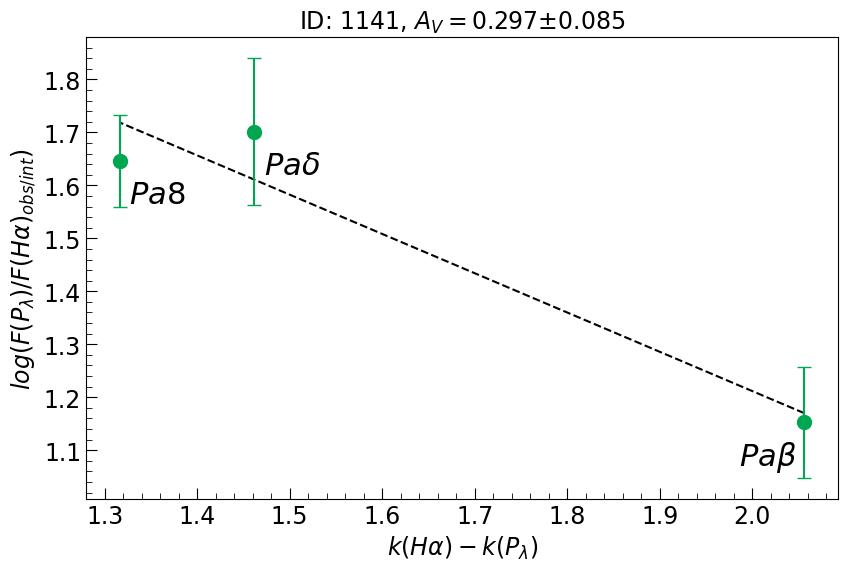}
\caption{Ratio between observed and theoretical Paschen to \ha flux versus the extinction curve, k(\rm \ha) - k(P\rm $_\lambda$), for the three galaxies. The black dashed line represents the linear regression fitting. Every plot is labelled with the A$_V$ obtained.}
\label{fig:Av_Pachen}     
\end{figure*}

In this section we check the possible discrepancies in the A$_V$ determination using the NIRspec sample described in section \ref{sec:JWST_data} for those galaxies where more than two Paschen lines are detected (ID:50, ID:22, and ID:1141).

\noindent
Only ID:50 presents at least two Balmer lines to obtain A$_V$ in the optical range (hereafter A$(V)_{\rm Balmer}$), as seen in Table \ref{tab:JWST_lines}. Taking the \ha and \hb values from the table, we computed A$(V)_{\rm Balmer}$ by using the \citet{CalzettiExtCurve00} extinction law as 

\begin{equation}
    \rm A(V)_{\rm Balmer} = \frac{2.5R_V}{k_{\rm H\beta}-k_{\rm H\alpha}} log\left(\frac{[F_{\rm H\alpha}/F_{\rm H\beta}]_{\rm obs}}{[F_{\rm H\alpha}/F_{\rm H\beta}]_{\rm int}}\right) ,
\end{equation}

\noindent
where $k_{\rm H\beta}$ and $k_{\rm H\alpha}$ are the values of the extinction law for \hb and \ha (= 3.61 and 2.53, respectively); $[F_{\rm H\alpha}/F_{\rm H\beta}]_{\rm obs}$ is the observed \ha/\hb ratio; $[F_{\rm H\alpha}/F_{\rm H\beta}]_{\rm int}$ is the intrinsic ratio, fixed at 2.86 (assuming case B recombination at low-density limit and T$_e$ = 10$^4$K \citealt{OsterbrockBook06}); and $R_V=4.05$ is the total V attenuation. 

The optical A$_V$ obtained for ID:50 is $A(V)_{\rm Balmer} = 0.798\pm0.118$, which is closely consistent with that obtained with the $\beta$ slope ($A(V)_{\beta} = 0.802\pm0.006$). In the near-IR, we can determine A$_V$ by using all the Paschen series lines detected (hereafter $\rm A(V)_{\rm Paschen}$). This can be done by considering the flux attenuation of a certain Paschen line (P$_{\lambda}$) 
$[\rm F(\rm P_\lambda)_{\rm obs}/F(\rm P_\lambda)_{\rm int}]=10^{-0.4A_V\cdot k(\rm P_\lambda)/R_V}$, and the \ha flux attenuation  $[F(\rm H\alpha)_{obs}/F(\rm H\alpha)_{int}]=10^{-0.4A_V\cdot k_{\rm H\alpha}/R_V}$. The ratio of the two Paschen values  to the \ha corrections is then
\begin{equation}\label{eq:Paschen_Av}
    \rm log\left(\frac{[\rm F(\rm P_\lambda)/F(\rm H\alpha)]_{\rm obs}}{[\rm F(\rm P_\lambda)/F(\rm H\alpha)]_{\rm int}}\right)=\rm -\frac{0.4}{R_V}A_V\left[k(\rm H\alpha)-k(\rm P_\lambda)\right].
\end{equation}

\noindent
$[\rm F(\rm P_\lambda)/F(\rm H\alpha)]_{\rm int}$ is the intrinsic ratio of the Paschen series to \ha, which can be obtained using {\sc PYNEB} and assuming case B recombination, low-density limit and T$_e$ = 10$^4$K \citep{OsterbrockBook06}. $[F(\rm P_\lambda)/F(\rm H\alpha)]_{\rm obs}$ is the observed ratio obtained from NIRSpec (Table \ref{tab:JWST_lines}). In addition, $k(\rm H\alpha)-k(\rm P_\lambda)$ is the difference in the \citet{CalzettiExtCurve00} extinction curve between the \ha and the Paschen series.  Equation \ref{eq:Paschen_Av} shows that, when representing the mentioned flux ratios and $k(\rm H\alpha)-k(\rm P_\lambda)$, its slope is proportional to $A(V)_{\rm Paschen}$ \citep{KehrigPachen06}. In Figure \ref{fig:Av_Pachen} we show the ratio of the observed and theoretical Paschen to \ha flux versus $k(\rm H\alpha)-k(\rm P_\lambda)$ for the three mentioned galaxies. 

In Table \ref{tab:Av_vs_Av} we present a summary of the results for the different A$_V$ values obtained using each method. In general, the A$_V$ values derived from the near-IR are lower than those obtained from the UV, as expected when the galaxies are  dustier. However, it is important to note that the number of data points used in the near-IR extinction estimation is limited, which can introduce a source of systematic error. A more robust and reliable approach to estimate $A(V)_{\rm Paschen}$ would involve using a larger set of Paschen series lines, particularly when the goal is to correct emission lines spanning the optical to near-IR range. The most remarkable case is ID:50, which shows consistent A$_V$ values across the near-IR, optical, and even UV, which is the most uncertain part of the extinction curve, indicating a uniform extinction across the spectrum for this source, clearly matching the assumed extinction curve.

\noindent
These discrepancies in the dust correction can be an important source of systematic error, especially when mixing UV and optical lines, as in the calculation of the electron temperature for ID:50 in Section \ref{sec:discussion_metals}. For the sake of consistency, we calculated the H$\alpha$ luminosity (L$_\mathrm{H\alpha}$) using the NIRSpec H$\alpha$ flux from Table \ref{tab:JWST_lines}, corrected using A(V)$_\mathrm{Balmer}$. We obtained L$_\mathrm{H\alpha} = (2.070 \pm 0.341) \times 10^{42}$ erg/s, which is in very good agreement with the value derived from the SFR estimated via SED fitting (Table \ref{tab1:Sample}) using the SFR–L$_\mathrm{H\alpha}$ relation \citep{KennicuttSFRHa94, MadauSFRHa98, KewleySFRHa02} applied in Section \ref{sec:photons}, which yields L$_\mathrm{H\alpha} = (2.797 \pm 0.397) \times 10^{42}$ erg/s.

\begin{table}[ht!]
\caption{A$_V$ values obtained using the three methods.}
\resizebox{0.9\columnwidth}{!}{%
\begin{tabular}{llll}
\hline
\hline
ID   & A(V)$_{\rm \beta}$ & A(V)$_{\rm Balmer}$ & A(V)$_{\rm Paschen}$ \\ \hline
50   & 0.802$\pm$0.006                   & 0.798$\pm$0.118    & 0.807$\pm$0.263      \\
22   & 1.262$\pm$0.012                  & -                  & 0.436$\pm$0.044      \\
1141 & 0.807$\pm$0.011                   & -                  & 0.297$\pm$0.085     \\ \hline
\end{tabular}%
}
\label{tab:Av_vs_Av}
\end{table}

\subsection{\heii ionisation budget}

\begin{figure}[t!]
\centering
    \includegraphics[width=\columnwidth]{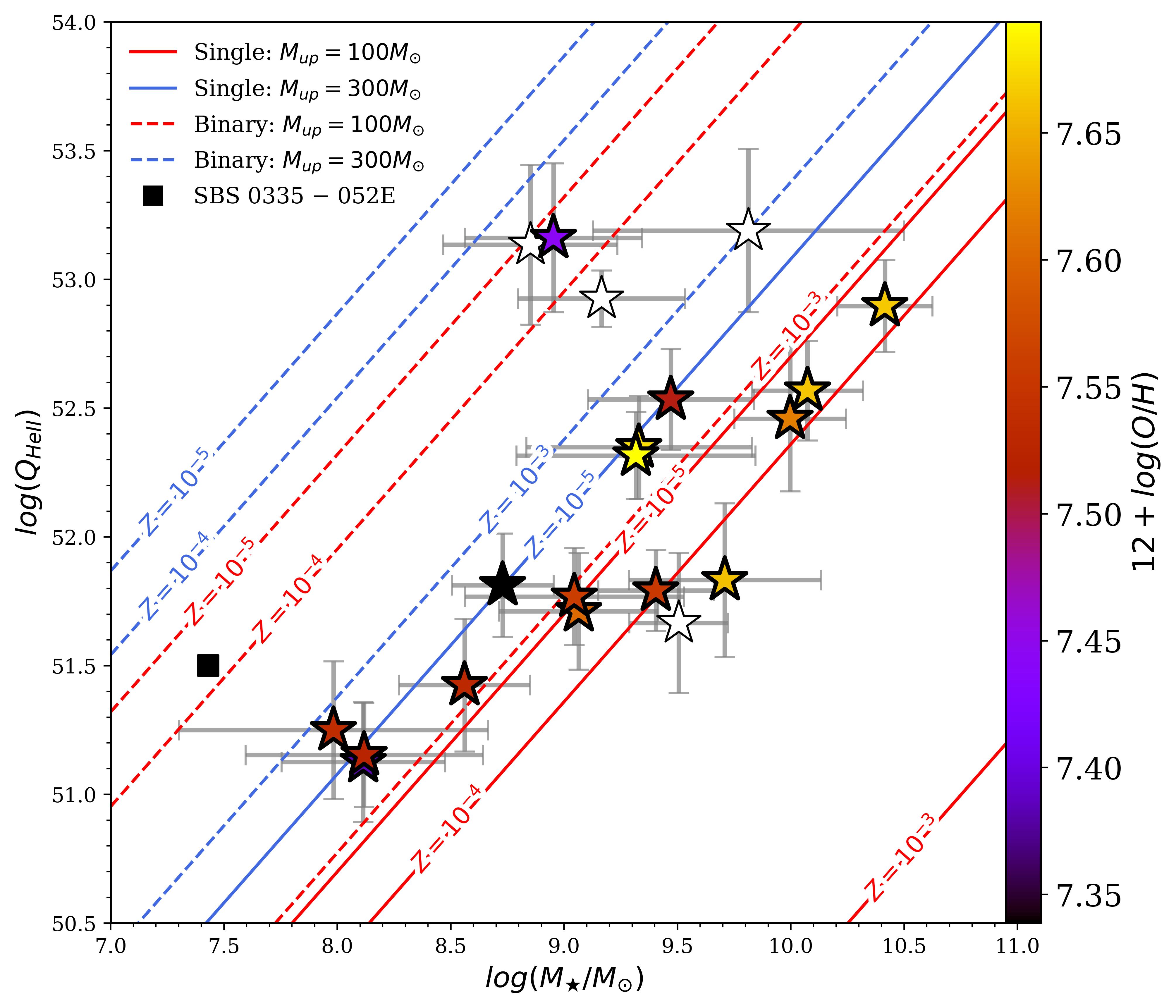}
\caption{Q$_{\rm HeII}$ vs. M$\star$ for our sample. The colours correspond to the gas-phase metallicity. The oblique lines represent the Q$_{\rm HeII}$ predictions from the BPASS models, scaled linearly with stellar mass. Models with interacting binaries are shown as dashed lines. Those without them are shown as solid lines. In all cases we show the metallicities of Z = 10$^{-3}$, 10$^{-4}$, and 10$^{-5}$. Additionally, we present models adopting a standard Salpeter IMF \citep{SalpeterIMF55} with an upper mass limit of M$_\mathrm{\rm up}$ = 100M$_\odot$ (red lines) and models assuming a top-heavy IMF with M$_\mathrm{up}$ = 300M$_\odot$ (blue lines). SBS 0335–052E \citep{KehrigSBS18} is included as a black square, which presents 12 + log(O/H) = 7.3.}
\label{fig:QHeII_BPASS_Z}     
\end{figure}

Different ionisation sources have been proposed to explain the origin of the \heii excitation. Among them, hot massive stars have been one of the most commonly suggested. In particular, some early-type WR stars with enough T$_{\rm eff}$ are often considered strong candidates, as they are capable of producing photons with energies exceeding 54.4 eV, sufficient to doubly ionise helium \citep[e.g.][]{SchaererWR96,SchaererWRPopII02,KehrigM3311,KehrigSBS18,NanayakkaraHeII19,MayyaWR23}. However, WR populations are not expected to be significant in very low-metallicity low-z galaxies \citep[e.g.][]{Guseva00, IzotovWR99, IzotovWR06, PapaderosWR06, KehrigSBS18}, suggesting the need for additional sources of ionisation such as interacting binaries \citep[e.g.][]{GotbergBinary17,SmithBinary18,KehrigSBS18,GotbergBinary1719}, shocks \citep[e.g.][]{SchaererXray19,Roy25}, or X-rays sources \citep[e.g.][]{SchaererXray19,SaxenaXrayVANDELS20,SenchynaXrays20}. Other types of hot massive stars may contribute more significantly to the global \heii ionisation budget in such environments, as WR stars alone cannot fully account for the observed nebular \heii emission \citep[e.g.][]{KehrigSBS18, Stanway19, Plat19, Roy25}. In particular, in extremely metal-poor galaxies such as IZw18 \citep{ThuanIZw1804, KehrigIZw15, KehrigIZw16} and SBS 0335-052E \citep{HuntSBSmass18, KehrigSBS18}, even after accounting for additional ionising sources, it has been necessary to invoke the presence of extremely metal-poor stars with metallicities lower than that of the surrounding gas in order to explain the observed \heii budget (extremely metal poor stars with Z $\sim10^{-5}$). These stars are peculiar, extremely hot, and nearly metal-free that can emit large amounts of He$^+$-ionising photons and may boost \heii emission in metal-poor galaxies at lower redshifts (e.g. \citealt{SchaererPOPIII08,YoonPopIIImodels12,CassataPopIII13,KehrigSBS18,EnriqueHeII20}). As Figures \ref{fig:QHe_UV_mass} and \ref{fig:CIIIOIIIdiag} suggest that massive stars could be the main ionising source, we assess whether the photon production from nearly metal-free stars and interacting binaries can reproduce the measured \heii photon production by considering the Binary Population and Spectral Synthesis models (BPASSv2.1; \citealt{EldridgeBinaries17}) models,\footnote{We use the v2.1 in order to use the "all binary" and "all single" star populations models.} which provide a comprehensive set of stellar SEDs that include both single and interacting binary systems (without including x-rays binaries) at very low metallicities (Z = 10$^{-3}$, 10$^{-4}$, and 10$^{-5}$). Q$_{\rm HeII}$ can be derived in these models for a given IMF by integrating the SEDs between 1 and 228 \A for instantaneous starburst models with an initial stellar mass of M$_\mathrm{ini}$ = 10$^6\,M_{\odot}$ at an age of 10$^{6.8}$ yr, which corresponds to the average age of stellar populations that predominantly produce photons with energies > 54.4 eV \citep[e.g. ][]{NanayakkaraHeII19,HawcroftSTARBURST25}. In Figure \ref{fig:QHeII_BPASS_Z}, we present the values of Q$_{\rm HeII}$ derived in Section \ref{sec:photons} for our \heii emitters, plotted against the stellar mass obtained from SED fitting (Table \ref{tab1:Sample}), and colour-coded by the gas-phase metallicity derived in Section \ref{sec:metal}. The oblique lines represent the Q$_{\rm HeII}$ predictions from the BPASS models, scaled linearly with stellar mass, i.e. if the predicted Q$_{\rm HeII}$ is 6.45$\times$10$^{48}$ photons/s for a star cluster of M = 10$^6$M$_{\odot}$, then a cluster with ten times the stellar mass is expected to produce ten times more ionising photons. We include both models, with interacting binaries (dashed lines) and without them (solid lines), for metallicities Z = 10$^{-3}$, 10$^{-4}$, and 10$^{-5}$. Additionally, we present models adopting a standard Salpeter IMF \citep{SalpeterIMF55} with an upper mass limit of M$_\mathrm{up}$ = 100M$_\odot$ (red lines), as well as models assuming a top-heavy IMF with M$_\mathrm{up}$ = 300M$_\odot$ (blue lines). 
\noindent
In general, most of the data points suggest an increase in Q$_{\rm HeII}$ as a function of stellar mass (M$_\star$), along with a trend on gas-phase metallicity.

The comparison with the predictions of BPASS indicate that, for the galaxies with higher gas-phase metallicity (7.7 > 12 +log(O/H) > 7.55) and stellar mass (log$(\rm M_{\star})\gtrsim 9\rm M_{\odot}$ ), the models including single stellar populations with a Salpeter IMF (M$_{\rm up}=100\rm M_{\odot}$ can reproduce the observed \heii ionising photon production when considering nearly metal-free stars (Z$_\odot$/2000 < Z < Z$_\odot$/200). This constraint on stellar metallicity can be relaxed when considering stellar clusters composed of interacting binaries with very low metallicities ($\rm Z_{\odot}/200 < Z < Z_{\odot}/20$). Thus, both scenarios—single, nearly metal-free stars and very low-metallicity interacting binaries (each with M$_{\rm up} = 100\rm M_{\odot}$)—can account for the observed ionising photon production. However, in the single-star scenario, the stellar metallicity is lower than the gas-phase metallicity. For the binary star models, the stellar metallicity closely matches the gas-phase metallicity only in the highest-metallicity galaxies (7.7 > 12 + log(O/H) > 7.6, which correspond to Z$_{\rm gas}\sim10^{-3}$).

On the other hand, for galaxies in our sample with lower stellar masses and lower gas-phase metallicities than in the previously discussed cases, the first two  scenarios cannot fully account for the observed ionising photon production. Instead, a top-heavy IMF with an upper mass limit of $M_{\text{up}} = 300M_{\odot}$ must be considered. In this context, both single nearly metal-free stars and very low-metallicity interacting binaries with a top-heavy IMF can reproduce the observed \heii-ionising photon production in these low-mass very low-metallicity galaxies. We note that hypothetical rapidly rotating chemically homogeneous evolution (CHE) stars could contribute to enhance UV emission, and therefore to the observed Q$_{\rm HeII}$ \citep{SzecsiCHE15,Szecsi25,LiuCHE25}. Analogously, quasi-chemically homogenous evolution (QHE) stars, which are hypothetical main sequence stars evolving to WR stars directly due to their mixing \citep[e.g. ][]{MartinsQHE13,EldridgeBinaries17}, can also contribute to the ionisation budget. Nevertheless, a detailed analysis of the contribution of such stars to the \heii ionising budget is beyond the scope of this work.

Four cases exhibit even more extreme behaviour (IDs 106, 3621, 8249, and 149), corresponding to the highest Q$_{\rm HeII}$ in our sample. The previously discussed scenarios cannot reproduce the observed Q$_{\rm HeII}$ values; only models that include nearly metal-free binaries or very low-metallicity binaries with a top-heavy IMF are able to account for the observed ionising photon production. One of the four mentioned   cases corresponds to the galaxy with the lowest measured O/H in our sample (ID:149 with 12 + log(O/H) = 7.45). The others lack sufficient signal-to-noise in the \ciii and \oiii lines to allow for a reliable O/H determination (see Section \ref{sec:metal}). These results are similar to those obtained for galaxies such as SBS 0335–052E (marked as a black square in Figure \ref{fig:QHeII_BPASS_Z}) and IZw18, both of which exhibit extremely low metallicities (12 + log(O/H) $\approx$ 7.2-7.3). In these cases, the observed Q$_{\rm HeII}$ can only be reproduced by models that include interacting binaries at Z = 10$^{-5}$ and a top-heavy IMF in the \heii ionisation budget. 

\noindent

It is important to note that the stellar metallicities in the BPASS models are defined in terms of Fe/H, not O/H. Therefore, any direct comparison between stellar and gas-phase metallicity must account for $\alpha$-enhancement, i.e. the overabundance of $\alpha$-elements such as oxygen relative to iron (O/Fe), which becomes more pronounced at lower values of 12 + log(O/H) \citep{SteidelTe16,Matteucci21Review,Byrne25}. Nevertheless, \citet{SteidelTe16} have shown that star-forming galaxies at z$\sim$2-3 exhibit emission-line properties indicative of photoionisation by young $\alpha$-enhanced massive stars, with O/Fe values up to two to five times the solar value. However, this level of enhancement remains insufficient to account for the most extreme cases in our sample, where the stellar metallicity required to reproduce the \heii ionisation rate falls in the range $Z_{\odot}/2000 < Z < Z_{\odot}/200$. In all cases, the gas-phase metallicity need to be lower than the metallicity of the stellar populations to reproduce the observed Q$_{\rm HeII}$, evidencing a decoupling between stellar and gas-phase abundances in high-redshift galaxies \citep{SteidelTe16, BylerUVZ20}, where the ionising stellar populations appear to be more metal-poor than the surrounding gas. This scenario can be attributed with the presence of nearly metal-free very massive stars, and may help explain the rapid enrichment of the ISM in $\alpha$-elements observed in galaxies with sufficiently high star formation rates. In addition, these findings are similar to previous studies, as the formation of the nebular \heii line continues to challenge current stellar models.

\section{Summary and conclusions}

In this work, we performed a physical and chemical characterisation of a sample of galaxies at z $\sim$ 2-4 exhibiting prominent nebular \heii$\lambda$1640 \A emission, commonly referred to as \heii emitters. Our sample was selected from the MUSE Hubble Ultra Deep Field surveys (MHUDF; \citealt{MHXDFBacon17, MHXDFBacon23}), and expanded upon the samples used by previous authors, which include three MUSE fields: the MUSE Extremely Deep Field (MXDF), the Ultra Deep Field survey (UDF-10), and the MOSAIC field. We used the publicly available AMUSED database to identify potential \heii emitters.

\noindent
From a total of 2221 objects in the MHUDF, we identified 25 galaxies exhibiting nebular \heii$\lambda$1640 \A emission with signal-to-noise ratios greater than 3, spanning a redshift range of 1.907 < z < 4.41. Among these, two are classified as AGNs. For the remaining 23 star-forming galaxies, all cases where the $\lambda=1214$\A falls within the MUSE range (ten sources excluding the AGNs) show a prominent Ly$\alpha$ emission line.

\noindent
To complement the MUSE rest-frame UV data, we also incorporated JWST/NIRSpec optical spectra by querying the public DJA database \citep{EugenioJADES25}. We found that five star-forming galaxies in our sample have publicly available and fully reduced NIRSpec observations.

\noindent
Using UV-based calibrations, we find that all \heii emitters in our sample exhibit very low gas-phase metallicities, with values in the range 7.3 < 12 + log(O/H) < 7.7. Additionally, these galaxies show sub-solar carbon-to-oxygen abundance ratios, with -0.6 < [C/O] < -0.4, similar to those found for other high-redshift \heii emitters. In contrast, the nitrogen-to-oxygen ratio remains approximately constant at [N/O] = -0.42, as expected given that nitrogen is primarily produced through secondary nucleosynthesis at higher metallicities.  Moreover, the theoretical photoionisation models of \citet{GutkinModels16} predict very low gas-phase metallicities and sub-solar C/O abundance ratios (-0.85 < [C/O] < -0.57. In addition, we derived electron densities using both UV and optical diagnostics, specifically the \ciii$\lambda$1907/$\lambda$1909 and \sii$\lambda$6716/$\lambda$6731 ratios, and found that all galaxies in the sample lie within a diverse low- to intermediate-density regime (10$^2$ < $n_e$ < 10$^3$ cm$^{-3}$).
\noindent
To assess the consistency of our calculations and assumptions, we compared dust extinction estimations derived from the UV (using MUSE data) with those obtained in the optical/near-IR (using JWST/NIRSpec data). We found that the A$_V$ values derived from the Paschen series in the near-IR are generally lower than those inferred from the UV $\beta$ slope. This is expected, particularly in galaxies with higher dust content, where the UV slope may overestimate extinction in the optical and near-IR lines. Therefore, in such cases, the $\beta$ slope should be used exclusively to correct UV fluxes, while the A$_V$ derived from the Paschen series is more appropriate for correcting emission lines in the optical/near-IR regime.

\noindent
A notable exception is object ID:50, which shows uniform extinction across the UV, optical, and near-IR wavelengths. For this galaxy, the extinction-corrected H$\alpha$ luminosity from NIRSpec is similar to the H$\alpha$ luminosity inferred from the SED fitting. Given that this object also exhibits the highest number of detected emission lines in the sample, we were able to compute its electron temperature and derive its C/O abundance ratio. The result from the direct method is in very good agreement with the value obtained from the empirical UV calibration.

\noindent
We computed the production rates of both H$^+$ and He$^{++}$ ionising photons and found that Q$_\mathrm{HeII}$ increases exponentially with Q$_{\rm H\alpha}$. Both Q$_\mathrm{HeII}$ and Q$_{\rm H\alpha}$ are higher in galaxies with higher stellar masses and greater UV luminosities.

\noindent
To investigate the origin of the observed \heii emission, we compared our results with stellar SED models from BPASS \citep{EldridgeBinaries17}. For our galaxies with higher gas-phase metallicities (7.7 > 12 + log(O/H) > 7.55) and higher stellar mass ($log(M_{\star})\gtrsim 9M_{\odot}$), we find that single-star populations with a Salpeter IMF can reproduce the observed Q$_{\rm HeII}$ emission, but only if the stars are nearly metal-free (Z$_\odot$/2000 < Z < Z$_\odot$/200). The condition of the metallicity can be relaxed to a very low stellar metallicity (Z$_\odot$/200 < Z < Z$_\odot$/20) when considering binary stars. For our galaxies with lower gas-phase metallicities (12 +log(O/H) $\lesssim$ 7.55) and lower stellar masses ($log(M_{\star})\lesssim 9M_{\odot}$), a top-heavy IMF with M$_\mathrm{\rm up}$ = 300M$_\odot$ is needed to account for the He$^+$ ionisation either assuming single nearly metal-free hot stars or binary stars with lower metallicity than the gas.

\noindent
In addition to the galaxies discussed above, four galaxies in our sample exhibit higher Q$_{\rm HeII}$ values relative to their stellar masses (IDs 106, 3621, 8249, and 149), resembling the well-known extremely metal-poor \heii emitters SBS 0335–052E and IZw18 \citep{KehrigIZw15, KehrigSBS18}, especially ID:149 with 12 + log(O/H) = 7.45. Their  high observed Q$_{\rm HeII}$ can be reproduced  either with models that include interacting binaries combined with a top-heavy IMF at Z$_\odot$/200 < Z < Z$_\odot$/20 or with binaries following a Salpeter IMF (M$_\mathrm{up}$ = 100M$_\odot$) at Z$_\odot$/2000 < Z < Z$_\odot$/200 (i.e. nearly metal-free). Notably, one of these four galaxies exhibits the lowest O/H ratio in our sample, further supporting the scenario observed in SBS 0335–052E and IZw18.

\noindent
In conclusion, the origin of the \heii$\lambda$1640 \A emission in this sample is influenced by a combination of factors, including stellar mass, IMF, stellar metallicity, and stellar multiplicity. Very hot massive stars can provide sufficient energy to efficiently ionise helium twice; however, in some cases the production of the He$^+$ ionising photons requires nearly metal-free hot stars, a condition that can be relaxed when considering binary stars. 

Our analysis, based on a larger and consistently selected sample of galaxies in the MHUDF at the cosmic noon and beyond, indicates that the challenge of reproducing nebular \heii emission seem to be a generalised problem in extremely metal poor galaxies.

A significantly larger sample is required to verify these findings and draw more robust conclusions. In future work, we will extend this analysis to a major sample and conduct a more in-depth investigation of Lyman continuum photon escape by examining Ly$\alpha$ line profiles and the two-dimensional spatially extended emission of Ly$\alpha$, \heii, and other detected lines. The galaxy with ID:50 will also be studied as a special case, given its rich set of emission features, spanning from Ly$\alpha$ at 1216\A to Pa$\gamma$ at 10938\A. This offers an extensive dataset that covers a broad spectral range, from the UV with MUSE to the near-infrared with JWST/NIRSpec.

\begin{acknowledgements}

The authors acknowledge financial support from the State Agency for Research of the Spanish MCIU through Center of Excellence Severo Ochoa’ award to the Instituto de Astrofísica de Andalucía CEX2021-001131-S funded by MCIN/AEI/10.13039/501100011033, and from the project PID2022-136598NB-C32 “Estallidos8”. R.G.D. acknowledge support by the project ref. AST22\_00001\_Subp\_15 funded from the EU – NextGenerationEU. I.M.C. acknowledges support from ANID programme FONDECYT Postdoctorado 3230653. This work is based in part on observations made with the NASA/ESA/CSA James Webb Space Telescope. The data were obtained from the Mikulski Archive for Space Telescopes at the Space Telescope Science Institute, which is operated by the Association of Universities for Research in Astronomy, Inc., under NASA contract NAS 5-03127 for JWST. The data products presented herein were retrieved from the Dawn JWST Archive (DJA) \url{https://dawn-cph.github.io/dja/}. DJA is an initiative of the Cosmic Dawn Center (DAWN), which is funded by the Danish National Research Foundation under grant DNRF140. The authors also acknowledge Roland Bacon for his valuable comments and suggestions, which helped to improve the quality of this manuscript. 

\end{acknowledgements}

\section*{Data availability}

The datasets used in this work were obtained from public sources. The MUSE data are available through the Advanced MUSE Data Products (AMUSED) public web interface, which provides access to MUSE datacubes for inspection and retrieval \url{https://amused.univ-lyon1.fr/}. The JWST/NIRSpec data were retrieved via DJA web interface \url{https://s3.amazonaws.com/msaexp-nirspec/extractions/nirspec_graded_v2.html}. Table \ref{tab:MUSE_lines} is only available in electronic form at the CDS via anonymous ftp to cdsarc.u-strasbg.fr (130.79.128.5) or via http://cdsweb.u-strasbg.fr/cgi-bin/qcat?J/A+A/.

\bibliographystyle{aa}
\bibliography{aa}

\begin{thebibliography}{}
\makeatletter
\relax
\def\mn@urlcharsother{\let\do\@makeother \do\$\do\&\do\#\do\^\do\_\do\%\do\~}
\def\mn@doi{\begingroup\mn@urlcharsother \@ifnextchar [ {\mn@doi@} {\mn@doi@[]}}
\def\mn@doi@[#1]#2{\def\@tempa{#1}\ifx\@tempa\@empty \href {http://dx.doi.org/#2} {doi:#2}\else \href {http://dx.doi.org/#2} {#1}\fi \endgroup}
\def\mn@eprint#1#2{\mn@eprint@#1:#2::\@nil}
\def\mn@eprint@arXiv#1{\href {http://arxiv.org/abs/#1} {{\tt arXiv:#1}}}
\def\mn@eprint@dblp#1{\href {http://dblp.uni-trier.de/rec/bibtex/#1.xml} {dblp:#1}}
\def\mn@eprint@#1:#2:#3:#4\@nil{\def\@tempa {#1}\def\@tempb {#2}\def\@tempc {#3}\ifx \@tempc \@empty \let \@tempc \@tempb \let \@tempb \@tempa \fi \ifx \@tempb \@empty \def\@tempb {arXiv}\fi \@ifundefined {mn@eprint@\@tempb}{\@tempb:\@tempc}{\expandafter \expandafter \csname mn@eprint@\@tempb\endcsname \expandafter{\@tempc}}}

\bibitem[\protect\citeauthoryear{{Allen}, {Wright}  \& {Goss}}{{Allen} et~al.}{1976}]{AllenWR76}
{Allen} D.~A.,  {Wright} A.~E.,   {Goss} W.~M.,  1976, \mn@doi [\mnras] {10.1093/mnras/177.1.91}, \href {https://ui.adsabs.harvard.edu/abs/1976MNRAS.177...91A} {177, 91}

\bibitem[\protect\citeauthoryear{{Allen}, {Dopita}  \& {Tsvetanov}}{{Allen} et~al.}{1998}]{AllenCIV98}
{Allen} M.~G.,  {Dopita} M.~A.,   {Tsvetanov} Z.~I.,  1998, \mn@doi [\apj] {10.1086/305145}, \href {https://ui.adsabs.harvard.edu/abs/1998ApJ...493..571A} {493, 571}

\bibitem[\protect\citeauthoryear{{Allen}, {Groves}, {Dopita}, {Sutherland}  \& {Kewley}}{{Allen} et~al.}{2008}]{AllenModels08}
{Allen} M.~G.,  {Groves} B.~A.,  {Dopita} M.~A.,  {Sutherland} R.~S.,   {Kewley} L.~J.,  2008, \mn@doi [\apjs] {10.1086/589652}, \href {https://ui.adsabs.harvard.edu/abs/2008ApJS..178...20A} {178, 20}

\bibitem[\protect\citeauthoryear{{Amayo}, {Delgado-Inglada}  \& {Stasi{\'n}ska}}{{Amayo} et~al.}{2021}]{AmayoICF21}
{Amayo} A.,  {Delgado-Inglada} G.,   {Stasi{\'n}ska} G.,  2021, \mn@doi [\mnras] {10.1093/mnras/stab1467}, \href {https://ui.adsabs.harvard.edu/abs/2021MNRAS.505.2361A} {505, 2361}

\bibitem[\protect\citeauthoryear{{Asplund}, {Amarsi}  \& {Grevesse}}{{Asplund} et~al.}{2021}]{AsplundSUN21}
{Asplund} M.,  {Amarsi} A.~M.,   {Grevesse} N.,  2021, \mn@doi [\aap] {10.1051/0004-6361/202140445}, \href {https://ui.adsabs.harvard.edu/abs/2021A&A...653A.141A} {653, A141}

\bibitem[\protect\citeauthoryear{{Atek} et~al.,}{{Atek} et~al.}{2024}]{Atek24}
{Atek} H.,  et~al., 2024, \mn@doi [\nat] {10.1038/s41586-024-07043-6}, \href {https://ui.adsabs.harvard.edu/abs/2024Natur.626..975A} {626, 975}

\bibitem[\protect\citeauthoryear{{Bacon} et~al.,}{{Bacon} et~al.}{2010}]{BaconMuse10}
{Bacon} R.,  et~al., 2010, in {McLean} I.~S.,  {Ramsay} S.~K.,   {Takami} H.,  eds,  Society of Photo-Optical Instrumentation Engineers (SPIE) Conference Series Vol. 7735, Ground-based and Airborne Instrumentation for Astronomy III. p. 773508 (\mn@eprint {arXiv} {2211.16795}), \mn@doi{10.1117/12.856027}

\bibitem[\protect\citeauthoryear{{Bacon} et~al.,}{{Bacon} et~al.}{2017}]{MHXDFBacon17}
{Bacon} R.,  et~al., 2017, \mn@doi [\aap] {10.1051/0004-6361/201730833}, \href {https://ui.adsabs.harvard.edu/abs/2017A&A...608A...1B} {608, A1}

\bibitem[\protect\citeauthoryear{{Bacon} et~al.,}{{Bacon} et~al.}{2023}]{MHXDFBacon23}
{Bacon} R.,  et~al., 2023, \mn@doi [\aap] {10.1051/0004-6361/202244187}, \href {https://ui.adsabs.harvard.edu/abs/2023A&A...670A...4B} {670, A4}

\bibitem[\protect\citeauthoryear{{Bayliss}, {Rigby}, {Sharon}, {Wuyts}, {Florian}, {Gladders}, {Johnson}  \& {Oguri}}{{Bayliss} et~al.}{2014}]{BaylissTe14}
{Bayliss} M.~B.,  {Rigby} J.~R.,  {Sharon} K.,  {Wuyts} E.,  {Florian} M.,  {Gladders} M.~D.,  {Johnson} T.,   {Oguri} M.,  2014, \mn@doi [\apj] {10.1088/0004-637X/790/2/144}, \href {https://ui.adsabs.harvard.edu/abs/2014ApJ...790..144B} {790, 144}

\bibitem[\protect\citeauthoryear{{Berg}, {Skillman}, {Henry}, {Erb}  \& {Carigi}}{{Berg} et~al.}{2016}]{BergCO16}
{Berg} D.~A.,  {Skillman} E.~D.,  {Henry} R. B.~C.,  {Erb} D.~K.,   {Carigi} L.,  2016, \mn@doi [\apj] {10.3847/0004-637X/827/2/126}, \href {https://ui.adsabs.harvard.edu/abs/2016ApJ...827..126B} {827, 126}

\bibitem[\protect\citeauthoryear{{Berg}, {Chisholm}, {Erb}, {Pogge}, {Henry}  \& {Olivier}}{{Berg} et~al.}{2019}]{BergCO19}
{Berg} D.~A.,  {Chisholm} J.,  {Erb} D.~K.,  {Pogge} R.,  {Henry} A.,   {Olivier} G.~M.,  2019, \mn@doi [\apjl] {10.3847/2041-8213/ab21dc}, \href {https://ui.adsabs.harvard.edu/abs/2019ApJ...878L...3B} {878, L3}

\bibitem[\protect\citeauthoryear{{Berg}, {Chisholm}, {Erb}, {Skillman}, {Pogge}  \& {Olivier}}{{Berg} et~al.}{2021}]{Berg21}
{Berg} D.~A.,  {Chisholm} J.,  {Erb} D.~K.,  {Skillman} E.~D.,  {Pogge} R.~W.,   {Olivier} G.~M.,  2021, \mn@doi [\apj] {10.3847/1538-4357/ac141b}, \href {https://ui.adsabs.harvard.edu/abs/2021ApJ...922..170B} {922, 170}

\bibitem[\protect\citeauthoryear{{Boogaard} et~al.,}{{Boogaard} et~al.}{2018}]{BoogaardSFMS18}
{Boogaard} L.~A.,  et~al., 2018, \mn@doi [\aap] {10.1051/0004-6361/201833136}, \href {https://ui.adsabs.harvard.edu/abs/2018A&A...619A..27B} {619, A27}

\bibitem[\protect\citeauthoryear{{Bromm}}{{Bromm}}{2013}]{BrommReview13}
{Bromm} V.,  2013, \mn@doi [Reports on Progress in Physics] {10.1088/0034-4885/76/11/112901}, \href {https://ui.adsabs.harvard.edu/abs/2013RPPh...76k2901B} {76, 112901}

\bibitem[\protect\citeauthoryear{{Bruzual} \& {Charlot}}{{Bruzual} \& {Charlot}}{2003}]{BruzualModels03}
{Bruzual} G.,  {Charlot} S.,  2003, \mn@doi [\mnras] {10.1046/j.1365-8711.2003.06897.x}, \href {https://ui.adsabs.harvard.edu/abs/2003MNRAS.344.1000B} {344, 1000}

\bibitem[\protect\citeauthoryear{{Bunker} et~al.,}{{Bunker} et~al.}{2023}]{Bunker23}
{Bunker} A.~J.,  et~al., 2023, \mn@doi [\aap] {10.1051/0004-6361/202346159}, \href {https://ui.adsabs.harvard.edu/abs/2023A&A...677A..88B} {677, A88}

\bibitem[\protect\citeauthoryear{{Butler} \& {Salim}}{{Butler} \& {Salim}}{2021}]{ButlerExt21}
{Butler} R.~E.,  {Salim} S.,  2021, \mn@doi [\apj] {10.3847/1538-4357/abe7e3}, \href {https://ui.adsabs.harvard.edu/abs/2021ApJ...911...40B} {911, 40}

\bibitem[\protect\citeauthoryear{{Byler}, {Dalcanton}, {Conroy}  \& {Johnson}}{{Byler} et~al.}{2017}]{BylerUVZ17}
{Byler} N.,  {Dalcanton} J.~J.,  {Conroy} C.,   {Johnson} B.~D.,  2017, \mn@doi [\apj] {10.3847/1538-4357/aa6c66}, \href {https://ui.adsabs.harvard.edu/abs/2017ApJ...840...44B} {840, 44}

\bibitem[\protect\citeauthoryear{{Byler}, {Dalcanton}, {Conroy}, {Johnson}, {Levesque}  \& {Berg}}{{Byler} et~al.}{2018}]{BylerUVZ18}
{Byler} N.,  {Dalcanton} J.~J.,  {Conroy} C.,  {Johnson} B.~D.,  {Levesque} E.~M.,   {Berg} D.~A.,  2018, \mn@doi [\apj] {10.3847/1538-4357/aacd50}, \href {https://ui.adsabs.harvard.edu/abs/2018ApJ...863...14B} {863, 14}

\bibitem[\protect\citeauthoryear{{Byler}, {Kewley}, {Rigby}, {Acharyya}, {Berg}, {Bayliss}  \& {Sharon}}{{Byler} et~al.}{2020}]{BylerUVZ20}
{Byler} N.,  {Kewley} L.~J.,  {Rigby} J.~R.,  {Acharyya} A.,  {Berg} D.~A.,  {Bayliss} M.,   {Sharon} K.,  2020, \mn@doi [\apj] {10.3847/1538-4357/ab7ea9}, \href {https://ui.adsabs.harvard.edu/abs/2020ApJ...893....1B} {893, 1}

\bibitem[\protect\citeauthoryear{{Byrne}, {Eldridge}  \& {Stanway}}{{Byrne} et~al.}{2025}]{Byrne25}
{Byrne} C.~M.,  {Eldridge} J.~J.,   {Stanway} E.~R.,  2025, \mn@doi [\mnras] {10.1093/mnras/staf178}, \href {https://ui.adsabs.harvard.edu/abs/2025MNRAS.537.2433B} {537, 2433}

\bibitem[\protect\citeauthoryear{{Cai} et~al.,}{{Cai} et~al.}{2011}]{CaiHST11}
{Cai} Z.,  et~al., 2011, \mn@doi [\apjl] {10.1088/2041-8205/736/2/L28}, \href {https://ui.adsabs.harvard.edu/abs/2011ApJ...736L..28C} {736, L28}

\bibitem[\protect\citeauthoryear{{Calzetti}, {Kinney}  \& {Storchi-Bergmann}}{{Calzetti} et~al.}{1994}]{CalzettiBeta94}
{Calzetti} D.,  {Kinney} A.~L.,   {Storchi-Bergmann} T.,  1994, \mn@doi [\apj] {10.1086/174346}, \href {https://ui.adsabs.harvard.edu/abs/1994ApJ...429..582C} {429, 582}

\bibitem[\protect\citeauthoryear{{Calzetti}, {Armus}, {Bohlin}, {Kinney}, {Koornneef}  \& {Storchi-Bergmann}}{{Calzetti} et~al.}{2000}]{CalzettiExtCurve00}
{Calzetti} D.,  {Armus} L.,  {Bohlin} R.~C.,  {Kinney} A.~L.,  {Koornneef} J.,   {Storchi-Bergmann} T.,  2000, \mn@doi [\apj] {10.1086/308692}, \href {https://ui.adsabs.harvard.edu/abs/2000ApJ...533..682C} {533, 682}

\bibitem[\protect\citeauthoryear{{Cardelli}, {Clayton}  \& {Mathis}}{{Cardelli} et~al.}{1989}]{Cardelli89}
{Cardelli} J.~A.,  {Clayton} G.~C.,   {Mathis} J.~S.,  1989, \mn@doi [\apj] {10.1086/167900}, \href {https://ui.adsabs.harvard.edu/abs/1989ApJ...345..245C} {345, 245}

\bibitem[\protect\citeauthoryear{{Cassata} et~al.,}{{Cassata} et~al.}{2013}]{CassataPopIII13}
{Cassata} P.,  et~al., 2013, \mn@doi [\aap] {10.1051/0004-6361/201220969}, \href {https://ui.adsabs.harvard.edu/abs/2013A&A...556A..68C} {556, A68}

\bibitem[\protect\citeauthoryear{{Christensen} et~al.,}{{Christensen} et~al.}{2012}]{ChristensenTe12}
{Christensen} L.,  et~al., 2012, \mn@doi [\mnras] {10.1111/j.1365-2966.2012.22007.x}, \href {https://ui.adsabs.harvard.edu/abs/2012MNRAS.427.1973C} {427, 1973}

\bibitem[\protect\citeauthoryear{{D'Eugenio} et~al.,}{{D'Eugenio} et~al.}{2025}]{EugenioJADES25}
{D'Eugenio} F.,  et~al., 2025, \mn@doi [\apjs] {10.3847/1538-4365/ada148}, \href {https://ui.adsabs.harvard.edu/abs/2025ApJS..277....4D} {277, 4}

\bibitem[\protect\citeauthoryear{{Dopita}, {Sutherland}, {Nicholls}, {Kewley}  \& {Vogt}}{{Dopita} et~al.}{2013}]{DopitaDiagnosis13}
{Dopita} M.~A.,  {Sutherland} R.~S.,  {Nicholls} D.~C.,  {Kewley} L.~J.,   {Vogt} F. P.~A.,  2013, \mn@doi [\apjs] {10.1088/0067-0049/208/1/10}, \href {https://ui.adsabs.harvard.edu/abs/2013ApJS..208...10D} {208, 10}

\bibitem[\protect\citeauthoryear{{Drake} et~al.,}{{Drake} et~al.}{2017}]{DrakeMHUDF17}
{Drake} A.~B.,  et~al., 2017, \mn@doi [\aap] {10.1051/0004-6361/201731431}, \href {https://ui.adsabs.harvard.edu/abs/2017A&A...608A...6D} {608, A6}

\bibitem[\protect\citeauthoryear{{Eldridge} \& {Stanway}}{{Eldridge} \& {Stanway}}{2012}]{EldridgeBinary12}
{Eldridge} J.~J.,  {Stanway} E.~R.,  2012, \mn@doi [\mnras] {10.1111/j.1365-2966.2011.19713.x}, \href {https://ui.adsabs.harvard.edu/abs/2012MNRAS.419..479E} {419, 479}

\bibitem[\protect\citeauthoryear{{Eldridge} \& {Stanway}}{{Eldridge} \& {Stanway}}{2022}]{EldridgeRev22}
{Eldridge} J.~J.,  {Stanway} E.~R.,  2022, \mn@doi [\araa] {10.1146/annurev-astro-052920-100646}, \href {https://ui.adsabs.harvard.edu/abs/2022ARA&A..60..455E} {60, 455}

\bibitem[\protect\citeauthoryear{{Eldridge}, {Stanway}, {Xiao}, {McClelland}, {Taylor}, {Ng}, {Greis}  \& {Bray}}{{Eldridge} et~al.}{2017}]{EldridgeBinaries17}
{Eldridge} J.~J.,  {Stanway} E.~R.,  {Xiao} L.,  {McClelland} L.~A.~S.,  {Taylor} G.,  {Ng} M.,  {Greis} S.~M.~L.,   {Bray} J.~C.,  2017, \mn@doi [\pasa] {10.1017/pasa.2017.51}, \href {https://ui.adsabs.harvard.edu/abs/2017PASA...34...58E} {34, e058}

\bibitem[\protect\citeauthoryear{{Erb}, {Pettini}, {Shapley}, {Steidel}, {Law}  \& {Reddy}}{{Erb} et~al.}{2010}]{ErbTe10}
{Erb} D.~K.,  {Pettini} M.,  {Shapley} A.~E.,  {Steidel} C.~C.,  {Law} D.~R.,   {Reddy} N.~A.,  2010, \mn@doi [\apj] {10.1088/0004-637X/719/2/1168}, \href {https://ui.adsabs.harvard.edu/abs/2010ApJ...719.1168E} {719, 1168}

\bibitem[\protect\citeauthoryear{{Feltre}, {Charlot}  \& {Gutkin}}{{Feltre} et~al.}{2016a}]{FeltreUVZ16}
{Feltre} A.,  {Charlot} S.,   {Gutkin} J.,  2016a, \mn@doi [\mnras] {10.1093/mnras/stv2794}, \href {https://ui.adsabs.harvard.edu/abs/2016MNRAS.456.3354F} {456, 3354}

\bibitem[\protect\citeauthoryear{{Feltre}, {Charlot}  \& {Gutkin}}{{Feltre} et~al.}{2016b}]{FeltreDiagnosis16}
{Feltre} A.,  {Charlot} S.,   {Gutkin} J.,  2016b, \mn@doi [\mnras] {10.1093/mnras/stv2794}, \href {https://ui.adsabs.harvard.edu/abs/2016MNRAS.456.3354F} {456, 3354}

\bibitem[\protect\citeauthoryear{{Feltre} et~al.,}{{Feltre} et~al.}{2018}]{FeltreMHUDF18}
{Feltre} A.,  et~al., 2018, \mn@doi [\aap] {10.1051/0004-6361/201833281}, \href {https://ui.adsabs.harvard.edu/abs/2018A&A...617A..62F} {617, A62}

\bibitem[\protect\citeauthoryear{{Ferland} et~al.,}{{Ferland} et~al.}{2013}]{FerlandCLOUDY13}
{Ferland} G.~J.,  et~al., 2013, \mn@doi [\rmxaa] {10.48550/arXiv.1302.4485}, \href {https://ui.adsabs.harvard.edu/abs/2013RMxAA..49..137F} {49, 137}

\bibitem[\protect\citeauthoryear{{Finkelstein} et~al.,}{{Finkelstein} et~al.}{2019}]{Finkelstein19}
{Finkelstein} S.~L.,  et~al., 2019, \mn@doi [\apj] {10.3847/1538-4357/ab1ea8}, \href {https://ui.adsabs.harvard.edu/abs/2019ApJ...879...36F} {879, 36}

\bibitem[\protect\citeauthoryear{{Fitzpatrick} \& {Massa}}{{Fitzpatrick} \& {Massa}}{1990}]{FitzpatrickExt90}
{Fitzpatrick} E.~L.,  {Massa} D.,  1990, \mn@doi [\apjs] {10.1086/191413}, \href {https://ui.adsabs.harvard.edu/abs/1990ApJS...72..163F} {72, 163}

\bibitem[\protect\citeauthoryear{{Fitzpatrick} \& {Massa}}{{Fitzpatrick} \& {Massa}}{2007}]{FitzpatrickExt07}
{Fitzpatrick} E.~L.,  {Massa} D.,  2007, \mn@doi [\apj] {10.1086/518158}, \href {https://ui.adsabs.harvard.edu/abs/2007ApJ...663..320F} {663, 320}

\bibitem[\protect\citeauthoryear{{Fitzpatrick} \& {Massa}}{{Fitzpatrick} \& {Massa}}{2009}]{FitzpatrickExt09}
{Fitzpatrick} E.~L.,  {Massa} D.,  2009, \mn@doi [\apj] {10.1088/0004-637X/699/2/1209}, \href {https://ui.adsabs.harvard.edu/abs/2009ApJ...699.1209F} {699, 1209}

\bibitem[\protect\citeauthoryear{{Garnett}, {Kennicutt}, {Chu}  \& {Skillman}}{{Garnett} et~al.}{1991}]{GarnettWR91}
{Garnett} D.~R.,  {Kennicutt} Jr. R.~C.,  {Chu} Y.-H.,   {Skillman} E.~D.,  1991, \mn@doi [\apj] {10.1086/170065}, \href {https://ui.adsabs.harvard.edu/abs/1991ApJ...373..458G} {373, 458}

\bibitem[\protect\citeauthoryear{{Garnett}, {Skillman}, {Dufour}, {Peimbert}, {Torres-Peimbert}, {Terlevich}, {Terlevich}  \& {Shields}}{{Garnett} et~al.}{1995}]{GarnettCO95}
{Garnett} D.~R.,  {Skillman} E.~D.,  {Dufour} R.~J.,  {Peimbert} M.,  {Torres-Peimbert} S.,  {Terlevich} R.,  {Terlevich} E.,   {Shields} G.~A.,  1995, \mn@doi [\apj] {10.1086/175503}, \href {https://ui.adsabs.harvard.edu/abs/1995ApJ...443...64G} {443, 64}

\bibitem[\protect\citeauthoryear{{Gonz{\'a}lez-D{\'\i}az} et~al.,}{{Gonz{\'a}lez-D{\'\i}az} et~al.}{2024}]{BETISI}
{Gonz{\'a}lez-D{\'\i}az} R.,  et~al., 2024, \mn@doi [\aap] {10.1051/0004-6361/202348453}, \href {https://ui.adsabs.harvard.edu/abs/2024A&A...687A..20G} {687, A20}

\bibitem[\protect\citeauthoryear{{G{\"o}tberg}, {de Mink}  \& {Groh}}{{G{\"o}tberg} et~al.}{2017}]{GotbergBinary17}
{G{\"o}tberg} Y.,  {de Mink} S.~E.,   {Groh} J.~H.,  2017, \mn@doi [\aap] {10.1051/0004-6361/201730472}, \href {https://ui.adsabs.harvard.edu/abs/2017A&A...608A..11G} {608, A11}

\bibitem[\protect\citeauthoryear{{G{\"o}tberg}, {de Mink}, {Groh}, {Leitherer}  \& {Norman}}{{G{\"o}tberg} et~al.}{2019}]{GotbergBinary1719}
{G{\"o}tberg} Y.,  {de Mink} S.~E.,  {Groh} J.~H.,  {Leitherer} C.,   {Norman} C.,  2019, \mn@doi [\aap] {10.1051/0004-6361/201834525}, \href {https://ui.adsabs.harvard.edu/abs/2019A&A...629A.134G} {629, A134}

\bibitem[\protect\citeauthoryear{{G{\"o}tberg} et~al.,}{{G{\"o}tberg} et~al.}{2023}]{GotbergStrip23}
{G{\"o}tberg} Y.,  et~al., 2023, \mn@doi [\apj] {10.3847/1538-4357/ace5a3}, \href {https://ui.adsabs.harvard.edu/abs/2023ApJ...959..125G} {959, 125}

\bibitem[\protect\citeauthoryear{{Groves}, {Dopita}  \& {Sutherland}}{{Groves} et~al.}{2004}]{GrovesCIV04}
{Groves} B.~A.,  {Dopita} M.~A.,   {Sutherland} R.~S.,  2004, \mn@doi [\apjs] {10.1086/421114}, \href {https://ui.adsabs.harvard.edu/abs/2004ApJS..153...75G} {153, 75}

\bibitem[\protect\citeauthoryear{{Guseva}, {Izotov}  \& {Thuan}}{{Guseva} et~al.}{2000}]{Guseva00}
{Guseva} N.~G.,  {Izotov} Y.~I.,   {Thuan} T.~X.,  2000, \mn@doi [\apj] {10.1086/308489}, \href {https://ui.adsabs.harvard.edu/abs/2000ApJ...531..776G} {531, 776}

\bibitem[\protect\citeauthoryear{{Gutkin}, {Charlot}  \& {Bruzual}}{{Gutkin} et~al.}{2016}]{GutkinModels16}
{Gutkin} J.,  {Charlot} S.,   {Bruzual} G.,  2016, \mn@doi [\mnras] {10.1093/mnras/stw1716}, \href {https://ui.adsabs.harvard.edu/abs/2016MNRAS.462.1757G} {462, 1757}

\bibitem[\protect\citeauthoryear{{Hawcroft} et~al.,}{{Hawcroft} et~al.}{2025}]{HawcroftSTARBURST25}
{Hawcroft} C.,  et~al., 2025, \mn@doi [arXiv e-prints] {10.48550/arXiv.2505.24841}, \href {https://ui.adsabs.harvard.edu/abs/2025arXiv250524841H} {p. arXiv:2505.24841}

\bibitem[\protect\citeauthoryear{{Heintz} et~al.,}{{Heintz} et~al.}{2024}]{HeintzMSAEXP24}
{Heintz} K.~E.,  et~al., 2024, \mn@doi [Science] {10.1126/science.adj0343}, \href {https://ui.adsabs.harvard.edu/abs/2024Sci...384..890H} {384, 890}

\bibitem[\protect\citeauthoryear{{Hunt} et~al.,}{{Hunt} et~al.}{2014}]{HuntSBSmass18}
{Hunt} L.~K.,  et~al., 2014, \mn@doi [\aap] {10.1051/0004-6361/201322739}, \href {https://ui.adsabs.harvard.edu/abs/2014A&A...561A..49H} {561, A49}

\bibitem[\protect\citeauthoryear{{Iani} et~al.,}{{Iani} et~al.}{2023}]{IaniUVZ23}
{Iani} E.,  et~al., 2023, \mn@doi [\mnras] {10.1093/mnras/stac3198}, \href {https://ui.adsabs.harvard.edu/abs/2023MNRAS.518.5018I} {518, 5018}

\bibitem[\protect\citeauthoryear{{Izotov} \& {Thuan}}{{Izotov} \& {Thuan}}{2004}]{IzotovWR04}
{Izotov} Y.~I.,  {Thuan} T.~X.,  2004, \mn@doi [\apj] {10.1086/380830}, \href {https://ui.adsabs.harvard.edu/abs/2004ApJ...602..200I} {602, 200}

\bibitem[\protect\citeauthoryear{{Izotov}, {Chaffee}, {Foltz}, {Green}, {Guseva}  \& {Thuan}}{{Izotov} et~al.}{1999}]{IzotovWR99}
{Izotov} Y.~I.,  {Chaffee} F.~H.,  {Foltz} C.~B.,  {Green} R.~F.,  {Guseva} N.~G.,   {Thuan} T.~X.,  1999, \mn@doi [\apj] {10.1086/308119}, \href {https://ui.adsabs.harvard.edu/abs/1999ApJ...527..757I} {527, 757}

\bibitem[\protect\citeauthoryear{{Izotov}, {Schaerer}, {Blecha}, {Royer}, {Guseva}  \& {North}}{{Izotov} et~al.}{2006}]{IzotovWR06}
{Izotov} Y.~I.,  {Schaerer} D.,  {Blecha} A.,  {Royer} F.,  {Guseva} N.~G.,   {North} P.,  2006, \mn@doi [\aap] {10.1051/0004-6361:20065622}, \href {https://ui.adsabs.harvard.edu/abs/2006A&A...459...71I} {459, 71}

\bibitem[\protect\citeauthoryear{{James} et~al.,}{{James} et~al.}{2014}]{JamesTe14}
{James} B.~L.,  et~al., 2014, \mn@doi [\mnras] {10.1093/mnras/stu287}, \href {https://ui.adsabs.harvard.edu/abs/2014MNRAS.440.1794J} {440, 1794}

\bibitem[\protect\citeauthoryear{{Jaskot} \& {Ravindranath}}{{Jaskot} \& {Ravindranath}}{2016}]{JaskotUVZ16}
{Jaskot} A.~E.,  {Ravindranath} S.,  2016, \mn@doi [\apj] {10.3847/1538-4357/833/2/136}, \href {https://ui.adsabs.harvard.edu/abs/2016ApJ...833..136J} {833, 136}

\bibitem[\protect\citeauthoryear{{Johnson}, {Leja}, {Conroy}  \& {Speagle}}{{Johnson} et~al.}{2021}]{prospectorJohnson21}
{Johnson} B.~D.,  {Leja} J.,  {Conroy} C.,   {Speagle} J.~S.,  2021, \mn@doi [\apjs] {10.3847/1538-4365/abef67}, \href {https://ui.adsabs.harvard.edu/abs/2021ApJS..254...22J} {254, 22}

\bibitem[\protect\citeauthoryear{{Keenan}, {Feibelman}  \& {Berrington}}{{Keenan} et~al.}{1992}]{KeenanNe92}
{Keenan} F.~P.,  {Feibelman} W.~A.,   {Berrington} K.~A.,  1992, \mn@doi [\apj] {10.1086/171220}, \href {https://ui.adsabs.harvard.edu/abs/1992ApJ...389..443K} {389, 443}

\bibitem[\protect\citeauthoryear{{Kehrig}, {V{\'\i}lchez}, {Telles}, {Cuisinier}  \& {P{\'e}rez-Montero}}{{Kehrig} et~al.}{2006}]{KehrigPachen06}
{Kehrig} C.,  {V{\'\i}lchez} J.~M.,  {Telles} E.,  {Cuisinier} F.,   {P{\'e}rez-Montero} E.,  2006, \mn@doi [\aap] {10.1051/0004-6361:20054488}, \href {https://ui.adsabs.harvard.edu/abs/2006A&A...457..477K} {457, 477}

\bibitem[\protect\citeauthoryear{{Kehrig} et~al.,}{{Kehrig} et~al.}{2011}]{KehrigM3311}
{Kehrig} C.,  et~al., 2011, \mn@doi [\aap] {10.1051/0004-6361/201015493}, \href {https://ui.adsabs.harvard.edu/abs/2011A&A...526A.128K} {526, A128}

\bibitem[\protect\citeauthoryear{{Kehrig}, {V{\'\i}lchez}, {P{\'e}rez-Montero}, {Iglesias-P{\'a}ramo}, {Brinchmann}, {Kunth}, {Durret}  \& {Bayo}}{{Kehrig} et~al.}{2015}]{KehrigIZw15}
{Kehrig} C.,  {V{\'\i}lchez} J.~M.,  {P{\'e}rez-Montero} E.,  {Iglesias-P{\'a}ramo} J.,  {Brinchmann} J.,  {Kunth} D.,  {Durret} F.,   {Bayo} F.~M.,  2015, \mn@doi [\apjl] {10.1088/2041-8205/801/2/L28}, \href {https://ui.adsabs.harvard.edu/abs/2015ApJ...801L..28K} {801, L28}

\bibitem[\protect\citeauthoryear{{Kehrig} et~al.,}{{Kehrig} et~al.}{2016}]{KehrigIZw16}
{Kehrig} C.,  et~al., 2016, \mn@doi [\mnras] {10.1093/mnras/stw806}, \href {https://ui.adsabs.harvard.edu/abs/2016MNRAS.459.2992K} {459, 2992}

\bibitem[\protect\citeauthoryear{{Kehrig}, {V{\'\i}lchez}, {Guerrero}, {Iglesias-P{\'a}ramo}, {Hunt}, {Duarte-Puertas}  \& {Ramos-Larios}}{{Kehrig} et~al.}{2018}]{KehrigSBS18}
{Kehrig} C.,  {V{\'\i}lchez} J.~M.,  {Guerrero} M.~A.,  {Iglesias-P{\'a}ramo} J.,  {Hunt} L.~K.,  {Duarte-Puertas} S.,   {Ramos-Larios} G.,  2018, \mn@doi [\mnras] {10.1093/mnras/sty1920}, \href {https://ui.adsabs.harvard.edu/abs/2018MNRAS.480.1081K} {480, 1081}

\bibitem[\protect\citeauthoryear{{Kehrig}, {Guerrero}, {V{\'\i}lchez}  \& {Ramos-Larios}}{{Kehrig} et~al.}{2021}]{KehrigXRays21}
{Kehrig} C.,  {Guerrero} M.~A.,  {V{\'\i}lchez} J.~M.,   {Ramos-Larios} G.,  2021, \mn@doi [\apjl] {10.3847/2041-8213/abe41b}, \href {https://ui.adsabs.harvard.edu/abs/2021ApJ...908L..54K} {908, L54}

\bibitem[\protect\citeauthoryear{{Kennicutt}, {Tamblyn}  \& {Congdon}}{{Kennicutt} et~al.}{1994}]{KennicuttSFRHa94}
{Kennicutt} Jr. R.~C.,  {Tamblyn} P.,   {Congdon} C.~E.,  1994, \mn@doi [\apj] {10.1086/174790}, \href {https://ui.adsabs.harvard.edu/abs/1994ApJ...435...22K} {435, 22}

\bibitem[\protect\citeauthoryear{{Kewley} \& {Dopita}}{{Kewley} \& {Dopita}}{2002}]{KewleySFRHa02}
{Kewley} L.~J.,  {Dopita} M.~A.,  2002, \mn@doi [\apjs] {10.1086/341326}, \href {https://ui.adsabs.harvard.edu/abs/2002ApJS..142...35K} {142, 35}

\bibitem[\protect\citeauthoryear{{Kojima}, {Ouchi}, {Nakajima}, {Shibuya}, {Harikane}  \& {Ono}}{{Kojima} et~al.}{2017}]{KojimaTe17}
{Kojima} T.,  {Ouchi} M.,  {Nakajima} K.,  {Shibuya} T.,  {Harikane} Y.,   {Ono} Y.,  2017, \mn@doi [\pasj] {10.1093/pasj/psx017}, \href {https://ui.adsabs.harvard.edu/abs/2017PASJ...69...44K} {69, 44}

\bibitem[\protect\citeauthoryear{{Le F{\`e}vre} et~al.,}{{Le F{\`e}vre} et~al.}{2003}]{LeFevreVIMOS03}
{Le F{\`e}vre} O.,  et~al., 2003, The Messenger, \href {https://ui.adsabs.harvard.edu/abs/2003Msngr.111...18L} {111, 18}

\bibitem[\protect\citeauthoryear{{Lennon} \& {Burke}}{{Lennon} \& {Burke}}{1994}]{Lennon94}
{Lennon} D.~J.,  {Burke} V.~M.,  1994, \aaps, \href {https://ui.adsabs.harvard.edu/abs/1994A&AS..103..273L} {103, 273}

\bibitem[\protect\citeauthoryear{{Lilly}, {Tresse}, {Hammer}, {Crampton}  \& {Le Fevre}}{{Lilly} et~al.}{1995}]{LillyCosmo95}
{Lilly} S.~J.,  {Tresse} L.,  {Hammer} F.,  {Crampton} D.,   {Le Fevre} O.,  1995, \mn@doi [\apj] {10.1086/176560}, \href {https://ui.adsabs.harvard.edu/abs/1995ApJ...455..108L} {455, 108}

\bibitem[\protect\citeauthoryear{{Liu}, {Sibony}, {Meynet}  \& {Bromm}}{{Liu} et~al.}{2025}]{LiuCHE25}
{Liu} B.,  {Sibony} Y.,  {Meynet} G.,   {Bromm} V.,  2025, \mn@doi [\apjl] {10.3847/2041-8213/adb151}, \href {https://ui.adsabs.harvard.edu/abs/2025ApJ...980L..30L} {980, L30}

\bibitem[\protect\citeauthoryear{{Luridiana}, {Morisset}  \& {Shaw}}{{Luridiana} et~al.}{2015}]{LuridianaPyneb15}
{Luridiana} V.,  {Morisset} C.,   {Shaw} R.~A.,  2015, \mn@doi [\aap] {10.1051/0004-6361/201323152}, \href {https://ui.adsabs.harvard.edu/abs/2015A&A...573A..42L} {573, A42}

\bibitem[\protect\citeauthoryear{{Madau}, {Ferguson}, {Dickinson}, {Giavalisco}, {Steidel}  \& {Fruchter}}{{Madau} et~al.}{1996}]{MadauCosmo96}
{Madau} P.,  {Ferguson} H.~C.,  {Dickinson} M.~E.,  {Giavalisco} M.,  {Steidel} C.~C.,   {Fruchter} A.,  1996, \mn@doi [\mnras] {10.1093/mnras/283.4.1388}, \href {https://ui.adsabs.harvard.edu/abs/1996MNRAS.283.1388M} {283, 1388}

\bibitem[\protect\citeauthoryear{{Madau}, {Pozzetti}  \& {Dickinson}}{{Madau} et~al.}{1998}]{MadauSFRHa98}
{Madau} P.,  {Pozzetti} L.,   {Dickinson} M.,  1998, \mn@doi [\apj] {10.1086/305523}, \href {https://ui.adsabs.harvard.edu/abs/1998ApJ...498..106M} {498, 106}

\bibitem[\protect\citeauthoryear{{Martins}, {Depagne}, {Russeil}  \& {Mahy}}{{Martins} et~al.}{2013}]{MartinsQHE13}
{Martins} F.,  {Depagne} E.,  {Russeil} D.,   {Mahy} L.,  2013, \mn@doi [\aap] {10.1051/0004-6361/201321282}, \href {https://ui.adsabs.harvard.edu/abs/2013A&A...554A..23M} {554, A23}

\bibitem[\protect\citeauthoryear{{Mary}, {Bacon}, {Conseil}, {Piqueras}  \& {Schutz}}{{Mary} et~al.}{2020}]{MaryORIGIN20}
{Mary} D.,  {Bacon} R.,  {Conseil} S.,  {Piqueras} L.,   {Schutz} A.,  2020, \mn@doi [\aap] {10.1051/0004-6361/201937001}, \href {https://ui.adsabs.harvard.edu/abs/2020A&A...635A.194M} {635, A194}

\bibitem[\protect\citeauthoryear{{Matteucci}}{{Matteucci}}{2021}]{Matteucci21Review}
{Matteucci} F.,  2021, \mn@doi [\aapr] {10.1007/s00159-021-00133-8}, \href {https://ui.adsabs.harvard.edu/abs/2021A&ARv..29....5M} {29, 5}

\bibitem[\protect\citeauthoryear{{Mayya} et~al.,}{{Mayya} et~al.}{2020}]{Mayya20}
{Mayya} Y.~D.,  et~al., 2020, \mn@doi [\mnras] {10.1093/mnras/staa2335}, \href {https://ui.adsabs.harvard.edu/abs/2020MNRAS.498.1496M} {498, 1496}

\bibitem[\protect\citeauthoryear{{Mayya}, {Plat}, {G{\'o}mez-Gonz{\'a}lez}, {Zaragoza-Cardiel}, {Charlot}  \& {Bruzual}}{{Mayya} et~al.}{2023}]{MayyaWR23}
{Mayya} Y.~D.,  {Plat} A.,  {G{\'o}mez-Gonz{\'a}lez} V.~M.~A.,  {Zaragoza-Cardiel} J.,  {Charlot} S.,   {Bruzual} G.,  2023, \mn@doi [\mnras] {10.1093/mnras/stad017}, \href {https://ui.adsabs.harvard.edu/abs/2023MNRAS.519.5492M} {519, 5492}

\bibitem[\protect\citeauthoryear{{Meurer}, {Heckman}  \& {Calzetti}}{{Meurer} et~al.}{1999}]{MeurerBeta99}
{Meurer} G.~R.,  {Heckman} T.~M.,   {Calzetti} D.,  1999, \mn@doi [\apj] {10.1086/307523}, \href {https://ui.adsabs.harvard.edu/abs/1999ApJ...521...64M} {521, 64}

\bibitem[\protect\citeauthoryear{{Mondal} et~al.,}{{Mondal} et~al.}{2025}]{MondalJWST2025}
{Mondal} C.,  et~al., 2025, \mn@doi [\apj] {10.3847/1538-4357/ade2cd}, \href {https://ui.adsabs.harvard.edu/abs/2025ApJ...988..171M} {988, 171}

\bibitem[\protect\citeauthoryear{{Nagao}, {Maiolino}  \& {Marconi}}{{Nagao} et~al.}{2006}]{Nagao06}
{Nagao} T.,  {Maiolino} R.,   {Marconi} A.,  2006, \mn@doi [\aap] {10.1051/0004-6361:20054127}, \href {https://ui.adsabs.harvard.edu/abs/2006A&A...447..863N} {447, 863}

\bibitem[\protect\citeauthoryear{{Nakajima} \& {Maiolino}}{{Nakajima} \& {Maiolino}}{2022}]{NakajimaPopIII22}
{Nakajima} K.,  {Maiolino} R.,  2022, \mn@doi [\mnras] {10.1093/mnras/stac1242}, \href {https://ui.adsabs.harvard.edu/abs/2022MNRAS.513.5134N} {513, 5134}

\bibitem[\protect\citeauthoryear{{Nanayakkara} et~al.,}{{Nanayakkara} et~al.}{2019}]{NanayakkaraHeII19}
{Nanayakkara} T.,  et~al., 2019, \mn@doi [\aap] {10.1051/0004-6361/201834565}, \href {https://ui.adsabs.harvard.edu/abs/2019A&A...624A..89N} {624, A89}

\bibitem[\protect\citeauthoryear{{Nicholls}, {Kewley}  \& {Sutherland}}{{Nicholls} et~al.}{2020}]{NichollsTe20}
{Nicholls} D.~C.,  {Kewley} L.~J.,   {Sutherland} R.~S.,  2020, \mn@doi [\pasp] {10.1088/1538-3873/ab6818}, \href {https://ui.adsabs.harvard.edu/abs/2020PASP..132c3001N} {132, 033001}

\bibitem[\protect\citeauthoryear{{Osterbrock} \& {Ferland}}{{Osterbrock} \& {Ferland}}{2006}]{OsterbrockBook06}
{Osterbrock} D.~E.,  {Ferland} G.~J.,  2006, {Astrophysics of gaseous nebulae and active galactic nuclei}

\bibitem[\protect\citeauthoryear{{Papaderos}, {Izotov}, {Guseva}, {Thuan}  \& {Fricke}}{{Papaderos} et~al.}{2006}]{PapaderosWR06}
{Papaderos} P.,  {Izotov} Y.~I.,  {Guseva} N.~G.,  {Thuan} T.~X.,   {Fricke} K.~J.,  2006, \mn@doi [\aap] {10.1051/0004-6361:20065110}, \href {https://ui.adsabs.harvard.edu/abs/2006A&A...454..119P} {454, 119}

\bibitem[\protect\citeauthoryear{{P{\'e}rez-Montero} \& {Amor{\'\i}n}}{{P{\'e}rez-Montero} \& {Amor{\'\i}n}}{2017}]{EnriqueCalUV17}
{P{\'e}rez-Montero} E.,  {Amor{\'\i}n} R.,  2017, \mn@doi [\mnras] {10.1093/mnras/stx186}, \href {https://ui.adsabs.harvard.edu/abs/2017MNRAS.467.1287P} {467, 1287}

\bibitem[\protect\citeauthoryear{{P{\'e}rez-Montero}, {Kehrig}, {V{\'\i}lchez}, {Garc{\'\i}a-Benito}, {Duarte Puertas}  \& {Iglesias-P{\'a}ramo}}{{P{\'e}rez-Montero} et~al.}{2020}]{EnriqueHeII20}
{P{\'e}rez-Montero} E.,  {Kehrig} C.,  {V{\'\i}lchez} J.~M.,  {Garc{\'\i}a-Benito} R.,  {Duarte Puertas} S.,   {Iglesias-P{\'a}ramo} J.,  2020, \mn@doi [\aap] {10.1051/0004-6361/202038509}, \href {https://ui.adsabs.harvard.edu/abs/2020A&A...643A..80P} {643, A80}

\bibitem[\protect\citeauthoryear{{Plat}, {Charlot}, {Bruzual}, {Feltre}, {Vidal-Garc{\'\i}a}, {Morisset}, {Chevallard}  \& {Todt}}{{Plat} et~al.}{2019}]{Plat19}
{Plat} A.,  {Charlot} S.,  {Bruzual} G.,  {Feltre} A.,  {Vidal-Garc{\'\i}a} A.,  {Morisset} C.,  {Chevallard} J.,   {Todt} H.,  2019, \mn@doi [\mnras] {10.1093/mnras/stz2616}, \href {https://ui.adsabs.harvard.edu/abs/2019MNRAS.490..978P} {490, 978}

\bibitem[\protect\citeauthoryear{{Proxauf}, {{\"O}ttl}  \& {Kimeswenger}}{{Proxauf} et~al.}{2014}]{ProxaufNe14}
{Proxauf} B.,  {{\"O}ttl} S.,   {Kimeswenger} S.,  2014, \mn@doi [\aap] {10.1051/0004-6361/201322581}, \href {https://ui.adsabs.harvard.edu/abs/2014A&A...561A..10P} {561, A10}

\bibitem[\protect\citeauthoryear{{Roy}, {Sutherland}, {Krumholz}, {Heger}  \& {Dopita}}{{Roy} et~al.}{2020}]{RoyWR20}
{Roy} A.,  {Sutherland} R.~S.,  {Krumholz} M.~R.,  {Heger} A.,   {Dopita} M.~A.,  2020, \mn@doi [\mnras] {10.1093/mnras/staa781}, \href {https://ui.adsabs.harvard.edu/abs/2020MNRAS.494.3861R} {494, 3861}

\bibitem[\protect\citeauthoryear{{Roy} et~al.,}{{Roy} et~al.}{2025}]{Roy25}
{Roy} A.,  et~al., 2025, \mn@doi [\aap] {10.1051/0004-6361/202553697}, \href {https://ui.adsabs.harvard.edu/abs/2025A&A...696A..29R} {696, A29}

\bibitem[\protect\citeauthoryear{{Salpeter}}{{Salpeter}}{1955}]{SalpeterIMF55}
{Salpeter} E.~E.,  1955, \mn@doi [\apj] {10.1086/145971}, \href {https://ui.adsabs.harvard.edu/abs/1955ApJ...121..161S} {121, 161}

\bibitem[\protect\citeauthoryear{{Saxena} et~al.,}{{Saxena} et~al.}{2020}]{SaxenaXrayVANDELS20}
{Saxena} A.,  et~al., 2020, \mn@doi [\aap] {10.1051/0004-6361/201937170}, \href {https://ui.adsabs.harvard.edu/abs/2020A&A...636A..47S} {636, A47}

\bibitem[\protect\citeauthoryear{{Schaerer}}{{Schaerer}}{1996}]{SchaererWR96}
{Schaerer} D.,  1996, \mn@doi [\apjl] {10.1086/310193}, \href {https://ui.adsabs.harvard.edu/abs/1996ApJ...467L..17S} {467, L17}

\bibitem[\protect\citeauthoryear{{Schaerer}}{{Schaerer}}{2002}]{SchaererWRPopII02}
{Schaerer} D.,  2002, \mn@doi [\aap] {10.1051/0004-6361:20011619}, \href {https://ui.adsabs.harvard.edu/abs/2002A&A...382...28S} {382, 28}

\bibitem[\protect\citeauthoryear{{Schaerer}}{{Schaerer}}{2008}]{SchaererPOPIII08}
{Schaerer} D.,  2008, in {Hunt} L.~K.,  {Madden} S.~C.,   {Schneider} R.,  eds,  IAU Symposium Vol. 255, Low-Metallicity Star Formation: From the First Stars to Dwarf Galaxies. pp 66--74 (\mn@eprint {arXiv} {0809.1988}), \mn@doi{10.1017/S1743921308024599}

\bibitem[\protect\citeauthoryear{{Schaerer}, {Fragos}  \& {Izotov}}{{Schaerer} et~al.}{2019}]{SchaererXray19}
{Schaerer} D.,  {Fragos} T.,   {Izotov} Y.~I.,  2019, \mn@doi [\aap] {10.1051/0004-6361/201935005}, \href {https://ui.adsabs.harvard.edu/abs/2019A&A...622L..10S} {622, L10}

\bibitem[\protect\citeauthoryear{{Senchyna}, {Stark}, {Mirocha}, {Reines}, {Charlot}, {Jones}  \& {Mulchaey}}{{Senchyna} et~al.}{2020}]{SenchynaXrays20}
{Senchyna} P.,  {Stark} D.~P.,  {Mirocha} J.,  {Reines} A.~E.,  {Charlot} S.,  {Jones} T.,   {Mulchaey} J.~S.,  2020, \mn@doi [\mnras] {10.1093/mnras/staa586}, \href {https://ui.adsabs.harvard.edu/abs/2020MNRAS.494..941S} {494, 941}

\bibitem[\protect\citeauthoryear{{Shirazi} \& {Brinchmann}}{{Shirazi} \& {Brinchmann}}{2012}]{ShiraziWR12}
{Shirazi} M.,  {Brinchmann} J.,  2012, \mn@doi [\mnras] {10.1111/j.1365-2966.2012.20439.x}, \href {https://ui.adsabs.harvard.edu/abs/2012MNRAS.421.1043S} {421, 1043}

\bibitem[\protect\citeauthoryear{{Smith}, {G{\"o}tberg}  \& {de Mink}}{{Smith} et~al.}{2018}]{SmithBinary18}
{Smith} N.,  {G{\"o}tberg} Y.,   {de Mink} S.~E.,  2018, \mn@doi [\mnras] {10.1093/mnras/stx3181}, \href {https://ui.adsabs.harvard.edu/abs/2018MNRAS.475..772S} {475, 772}

\bibitem[\protect\citeauthoryear{{Soto}, {Lilly}, {Bacon}, {Richard}  \& {Conseil}}{{Soto} et~al.}{2016}]{SotoZap16}
{Soto} K.~T.,  {Lilly} S.~J.,  {Bacon} R.,  {Richard} J.,   {Conseil} S.,  2016, \mn@doi [\mnras] {10.1093/mnras/stw474}, \href {https://ui.adsabs.harvard.edu/abs/2016MNRAS.458.3210S} {458, 3210}

\bibitem[\protect\citeauthoryear{{Stanway} \& {Eldridge}}{{Stanway} \& {Eldridge}}{2019}]{Stanway19}
{Stanway} E.~R.,  {Eldridge} J.~J.,  2019, \mn@doi [\aap] {10.1051/0004-6361/201834359}, \href {https://ui.adsabs.harvard.edu/abs/2019A&A...621A.105S} {621, A105}

\bibitem[\protect\citeauthoryear{{Stark} et~al.,}{{Stark} et~al.}{2017}]{StarkDiagnosis16}
{Stark} D.~P.,  et~al., 2017, \mn@doi [\mnras] {10.1093/mnras/stw2233}, \href {https://ui.adsabs.harvard.edu/abs/2017MNRAS.464..469S} {464, 469}

\bibitem[\protect\citeauthoryear{{Steidel}, {Strom}, {Pettini}, {Rudie}, {Reddy}  \& {Trainor}}{{Steidel} et~al.}{2016}]{SteidelTe16}
{Steidel} C.~C.,  {Strom} A.~L.,  {Pettini} M.,  {Rudie} G.~C.,  {Reddy} N.~A.,   {Trainor} R.~F.,  2016, \mn@doi [\apj] {10.3847/0004-637X/826/2/159}, \href {https://ui.adsabs.harvard.edu/abs/2016ApJ...826..159S} {826, 159}

\bibitem[\protect\citeauthoryear{{Sz{\'e}csi}, {Langer}, {Yoon}, {Sanyal}, {de Mink}, {Evans}  \& {Dermine}}{{Sz{\'e}csi} et~al.}{2015a}]{SzecsiTWUINS15}
{Sz{\'e}csi} D.,  {Langer} N.,  {Yoon} S.-C.,  {Sanyal} D.,  {de Mink} S.,  {Evans} C.~J.,   {Dermine} T.,  2015a, \mn@doi [\aap] {10.1051/0004-6361/201526617}, \href {https://ui.adsabs.harvard.edu/abs/2015A&A...581A..15S} {581, A15}

\bibitem[\protect\citeauthoryear{{Sz{\'e}csi}, {Langer}, {Yoon}, {Sanyal}, {de Mink}, {Evans}  \& {Dermine}}{{Sz{\'e}csi} et~al.}{2015b}]{SzecsiCHE15}
{Sz{\'e}csi} D.,  {Langer} N.,  {Yoon} S.-C.,  {Sanyal} D.,  {de Mink} S.,  {Evans} C.~J.,   {Dermine} T.,  2015b, \mn@doi [\aap] {10.1051/0004-6361/201526617}, \href {https://ui.adsabs.harvard.edu/abs/2015A&A...581A..15S} {581, A15}

\bibitem[\protect\citeauthoryear{{Sz{\'e}csi}, {Tramper}, {Kub{\'a}tov{\'a}}, {Kehrig}, {Kub{\'a}t}, {Krti{\v{c}}ka}, {Sander}  \& {Garcia}}{{Sz{\'e}csi} et~al.}{2025}]{Szecsi25}
{Sz{\'e}csi} D.,  {Tramper} F.,  {Kub{\'a}tov{\'a}} B.,  {Kehrig} C.,  {Kub{\'a}t} J.,  {Krti{\v{c}}ka} J.,  {Sander} A. A.~C.,   {Garcia} M.,  2025, \mn@doi [arXiv e-prints] {10.48550/arXiv.2506.21442}, \href {https://ui.adsabs.harvard.edu/abs/2025arXiv250621442S} {p. arXiv:2506.21442}

\bibitem[\protect\citeauthoryear{{Thuan}, {Bauer}, {Papaderos}  \& {Izotov}}{{Thuan} et~al.}{2004}]{ThuanIZw1804}
{Thuan} T.~X.,  {Bauer} F.~E.,  {Papaderos} P.,   {Izotov} Y.~I.,  2004, \mn@doi [\apj] {10.1086/382949}, \href {https://ui.adsabs.harvard.edu/abs/2004ApJ...606..213T} {606, 213}

\bibitem[\protect\citeauthoryear{{Topping} et~al.,}{{Topping} et~al.}{2024}]{ToppingJWSTz624}
{Topping} M.~W.,  et~al., 2024, \mn@doi [\mnras] {10.1093/mnras/stae682}, \href {https://ui.adsabs.harvard.edu/abs/2024MNRAS.529.3301T} {529, 3301}

\bibitem[\protect\citeauthoryear{{Tumlinson} \& {Shull}}{{Tumlinson} \& {Shull}}{2000}]{TumlinsonZeroAge00}
{Tumlinson} J.,  {Shull} J.~M.,  2000, \mn@doi [\apjl] {10.1086/312432}, \href {https://ui.adsabs.harvard.edu/abs/2000ApJ...528L..65T} {528, L65}

\bibitem[\protect\citeauthoryear{{Valentino} et~al.,}{{Valentino} et~al.}{2023}]{ValentinoGrizli23}
{Valentino} F.,  et~al., 2023, \mn@doi [\apj] {10.3847/1538-4357/acbefa}, \href {https://ui.adsabs.harvard.edu/abs/2023ApJ...947...20V} {947, 20}

\bibitem[\protect\citeauthoryear{{Venditti}, {Bromm}, {Finkelstein}, {Calabr{\`o}}, {Napolitano}, {Graziani}  \& {Schneider}}{{Venditti} et~al.}{2024}]{VendittiJWSTz624}
{Venditti} A.,  {Bromm} V.,  {Finkelstein} S.~L.,  {Calabr{\`o}} A.,  {Napolitano} L.,  {Graziani} L.,   {Schneider} R.,  2024, \mn@doi [\apjl] {10.3847/2041-8213/ad7387}, \href {https://ui.adsabs.harvard.edu/abs/2024ApJ...973L..12V} {973, L12}

\bibitem[\protect\citeauthoryear{{Villar-Martin}, {Tadhunter}  \& {Clark}}{{Villar-Martin} et~al.}{1997}]{Villar-Martin97CIV}
{Villar-Martin} M.,  {Tadhunter} C.,   {Clark} N.,  1997, \mn@doi [\aap] {10.48550/arXiv.astro-ph/9701016}, \href {https://ui.adsabs.harvard.edu/abs/1997A&A...323...21V} {323, 21}

\bibitem[\protect\citeauthoryear{{Villar-Mart{\'\i}n}, {Cervi{\~n}o}  \& {Gonz{\'a}lez Delgado}}{{Villar-Mart{\'\i}n} et~al.}{2004}]{VillarTe04}
{Villar-Mart{\'\i}n} M.,  {Cervi{\~n}o} M.,   {Gonz{\'a}lez Delgado} R.~M.,  2004, \mn@doi [\mnras] {10.1111/j.1365-2966.2004.08395.x}, \href {https://ui.adsabs.harvard.edu/abs/2004MNRAS.355.1132V} {355, 1132}

\bibitem[\protect\citeauthoryear{{Wise}, {Turk}, {Norman}  \& {Abel}}{{Wise} et~al.}{2012}]{Wise12}
{Wise} J.~H.,  {Turk} M.~J.,  {Norman} M.~L.,   {Abel} T.,  2012, \mn@doi [\apj] {10.1088/0004-637X/745/1/50}, \href {https://ui.adsabs.harvard.edu/abs/2012ApJ...745...50W} {745, 50}

\bibitem[\protect\citeauthoryear{{Wise}, {Demchenko}, {Halicek}, {Norman}, {Turk}, {Abel}  \& {Smith}}{{Wise} et~al.}{2014}]{Wise14}
{Wise} J.~H.,  {Demchenko} V.~G.,  {Halicek} M.~T.,  {Norman} M.~L.,  {Turk} M.~J.,  {Abel} T.,   {Smith} B.~D.,  2014, \mn@doi [\mnras] {10.1093/mnras/stu979}, \href {https://ui.adsabs.harvard.edu/abs/2014MNRAS.442.2560W} {442, 2560}

\bibitem[\protect\citeauthoryear{{Xiao}, {Stanway}  \& {Eldridge}}{{Xiao} et~al.}{2018}]{XiaoBPASS18}
{Xiao} L.,  {Stanway} E.~R.,   {Eldridge} J.~J.,  2018, \mn@doi [\mnras] {10.1093/mnras/sty646}, \href {https://ui.adsabs.harvard.edu/abs/2018MNRAS.477..904X} {477, 904}

\bibitem[\protect\citeauthoryear{{Yoon}, {Dierks}  \& {Langer}}{{Yoon} et~al.}{2012}]{YoonPopIIImodels12}
{Yoon} S.~C.,  {Dierks} A.,   {Langer} N.,  2012, \mn@doi [\aap] {10.1051/0004-6361/201117769}, \href {https://ui.adsabs.harvard.edu/abs/2012A&A...542A.113Y} {542, A113}

\bibitem[\protect\citeauthoryear{{Yuan} \& {Kewley}}{{Yuan} \& {Kewley}}{2009}]{YuanTe09}
{Yuan} T.~T.,  {Kewley} L.~J.,  2009, \mn@doi [\apjl] {10.1088/0004-637X/699/2/L161}, \href {https://ui.adsabs.harvard.edu/abs/2009ApJ...699L.161Y} {699, L161}

\bibitem[\protect\citeauthoryear{{Zhou}, {Shi}, {Zhang}  \& {Wang}}{{Zhou} et~al.}{2021}]{ZhouIWZ18mass21}
{Zhou} L.,  {Shi} Y.,  {Zhang} Z.-Y.,   {Wang} J.,  2021, \mn@doi [\aap] {10.1051/0004-6361/202039033}, \href {https://ui.adsabs.harvard.edu/abs/2021A&A...653L..10Z} {653, L10}

\bibitem[\protect\citeauthoryear{{de Graaff} et~al.,}{{de Graaff} et~al.}{2025}]{GraaffMSAEXP24}
{de Graaff} A.,  et~al., 2025, \mn@doi [\aap] {10.1051/0004-6361/202452186}, \href {https://ui.adsabs.harvard.edu/abs/2025A&A...697A.189D} {697, A189}

\bibitem[\protect\citeauthoryear{{del Moral-Castro}, {V{\'\i}lchez}, {Iglesias-P{\'a}ramo}  \& {Arroyo-Polonio}}{{del Moral-Castro} et~al.}{2024}]{IgnacioMHUDF24}
{del Moral-Castro} I.,  {V{\'\i}lchez} J.~M.,  {Iglesias-P{\'a}ramo} J.,   {Arroyo-Polonio} A.,  2024, \mn@doi [\aap] {10.1051/0004-6361/202348185}, \href {https://ui.adsabs.harvard.edu/abs/2024A&A...688A..28D} {688, A28}

\makeatother
\end{thebibliography}

\appendix

\section{\heii emitters spectra}

\begin{center}
\centering
    \includegraphics[width=\textwidth]{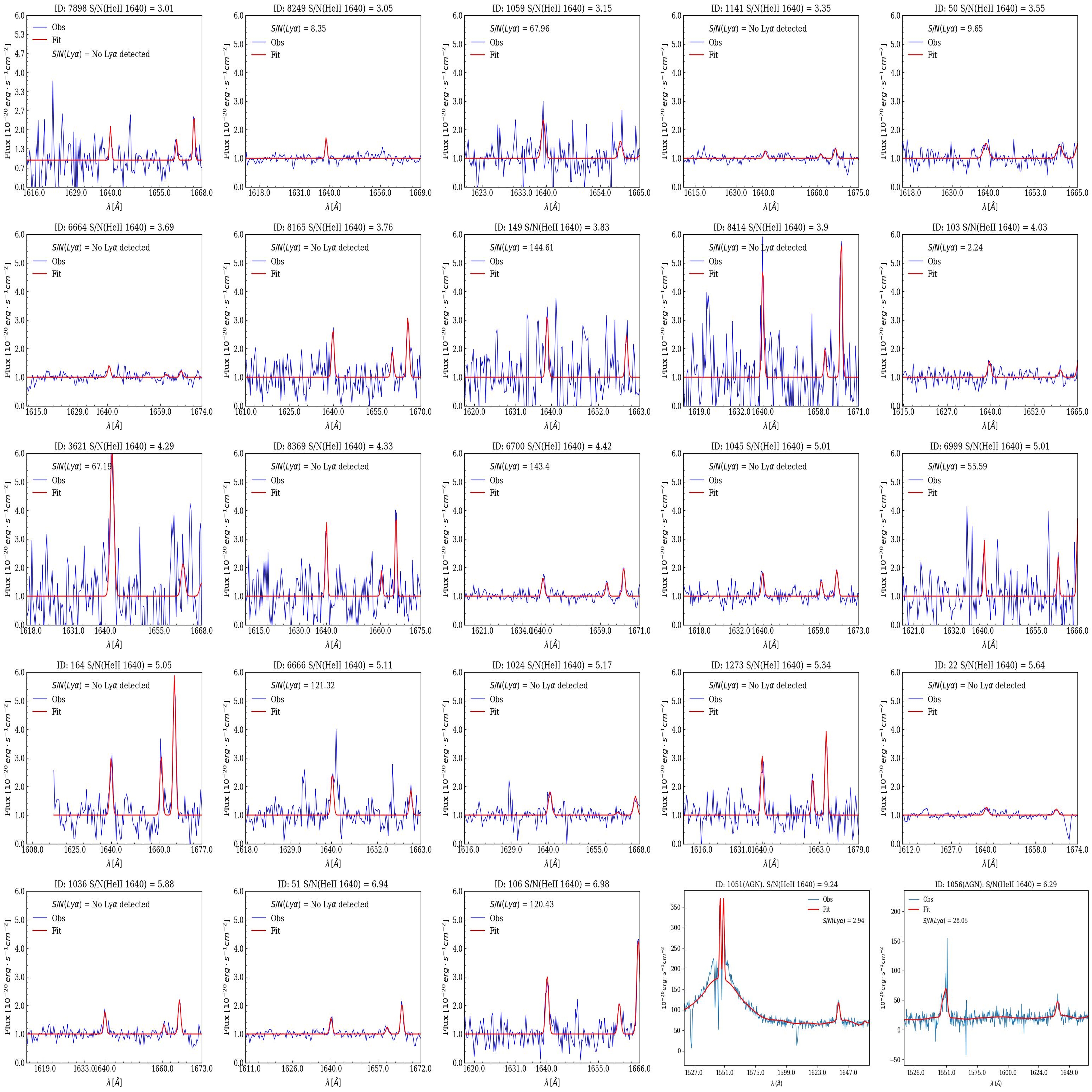}
\makebox[\textwidth][l]{%
  \parbox{\textwidth}{%
\captionof{figure}{\heii emitters sample ordered by S/N(\heii). The blue and red spectra represent the reference spectrum and the model continuum + fit, respectively. The S/N of \heii and Ly$\alpha$ (when detected) are included in the labels. The last two panels correspond to the two AGNs identified in the sample, exhibiting broad \heii and \civ 1548,51 \A lines.}
\label{fig:collage}     
}
}
\end{center}

\clearpage

\section{Additional tables}\label{A3:tables}

\makeatletter
\renewcommand{\thetable}{B\@arabic\c@table}
\makeatother
\setcounter{table}{-1}

\begin{center}
\makebox[\textwidth][c]{%
  \parbox{\textwidth}{%
  \refstepcounter{table}
    \captionof{table}{General characteristics of the \heii emitters sample.}
    \label{tab1:Sample}
    }
    }
\resizebox{\textwidth}{!}{%
\begin{tabular}{lllcclcccc}
\hline
\hline
ID & RA & DEC & EW(Ly$\alpha$) & EW(\heii) & z & M$_{\star}$ & SFR & A$_V$ & M$_{1500}$ \\ 
& J2000 & J2000 & \r{A} & \r{A} & & $\rm log(M/M_{\odot})$ & $\rm log(M_{\odot}\cdot yr^{-1})$ & & \\ \hline
7898 & 03:32:34.778 & -27:47:59.447 &  & 12.4$\pm$2.9 & 2.867 &  &  & 0.351$\pm$0.058 & -16.426$\pm$0.086 \\
8249 & 03:32:37.931 & -27:47:39.686 & 12.4$\pm$2.0 & 1.9$\pm$0.5 & 2.955 &  &  & 1.259$\pm$0.017 & -22.968$\pm$0.025 \\
1059 & 03:32:36.826 & -27:45:58.029 & 223.1$\pm$44.8 & 7.9$\pm$1.3 & 3.806 & $9.813^{0.685}_{-0.459}$ & $1.466^{+0.483}_{-0.427}$ & 0.877$\pm$0.008 & -19.955$\pm$0.012 \\
1141* & 03:32:39.617 & -27:47:37.179 &  & 1.8$\pm$0.2 & 2.344 & $9.709^{0.423}_{-0.343}$ & $0.869^{+0.287}_{-0.341}$ & 0.807$\pm$0.011 & -19.034$\pm$0.017 \\
50* & 03:32:39.083 & -27:46:17.855 & 24.9$\pm$13.7 & 3.4$\pm$0.5 & 3.325 & $9.997^{0.246}_{-0.327}$ & $1.137^{+0.286}_{-0.289}$ & 0.802$\pm$0.006 & -18.99$\pm$0.009 \\
6664 & 03:32:38.962 & -27:47:03.994 &  & 2.2$\pm$0.2 & 2.391 & $9.506^{0.218}_{-0.236}$ & $0.363^{+0.589}_{-0.247}$ & 0.618$\pm$0.009 & -18.167$\pm$0.014 \\
8165 & 03:32:39.698 & -27:46:48.250 &  & 5.7$\pm$1.3 & 2.338 & $7.983^{0.682}_{-0.372}$ & $-0.69^{+0.744}_{-0.415}$ & 0.31$\pm$0.028 & -15.755$\pm$0.042 \\
149 & 03:32:40.287 & -27:46:43.113 & 469.3$\pm$24.6 & 12.1$\pm$1.3 & 3.721 & $8.953^{0.392}_{-0.315}$ & $0.272^{+0.438}_{-0.306}$ & 1.598$\pm$0.128 & -19.934$\pm$0.189 \\
8414 & 03:32:41.061 & -27:47:20.897 &  & 11.0$\pm$2.5 & 2.844 & $8.561^{0.289}_{-0.281}$ & $-0.227^{+0.497}_{-0.246}$ & 0.319$\pm$0.026 & -14.977$\pm$0.039 \\
103 & 03:32:39.919 & -27:46:47.418 & 8.0$\pm$4.1 & 1.4$\pm$0.7 & 2.994 &  &  & 0.622$\pm$0.012 & -18.014$\pm$0.018 \\
3621 & 03:32:39.525 & -27:48:53.532 & 882.4$\pm$10.9 & 31.5$\pm$2.8 & 3.062 & $8.851^{0.384}_{-0.471}$ & $-0.098^{+1.246}_{-0.322}$ & 0.722$\pm$0.211 & -21.282$\pm$0.31 \\
8369 & 03:32:40.774 & -27:47:27.606 &  & 5.6$\pm$1.3 & 2.344 & $8.114^{0.361}_{-0.358}$ & $-1.301^{+2.397}_{-0.33}$ & 0.137$\pm$0.027 & -15.141$\pm$0.04 \\
6700 & 03:32:40.387 & -27:46:51.802 & 136.7$\pm$0.9 & 3.2$\pm$0.5 & 2.994 & $9.065^{0.351}_{-0.785}$ & $0.179^{+0.778}_{-0.434}$ & 0.342$\pm$0.005 & -17.608$\pm$0.007 \\
1045 & 03:32:33.780 & -27:48:14.354 &  & 3.6$\pm$0.5 & 2.616 & $9.33^{0.497}_{-0.299}$ & $0.608^{+0.822}_{-0.286}$ & 0.454$\pm$0.008 & -19.007$\pm$0.012 \\
6999 & 03:32:38.297 & -27:47:28.514 & 227.2$\pm$15.7 & 5.6$\pm$1.5 & 3.427 & $8.729^{0.225}_{-0.326}$ & $-0.178^{+0.438}_{-0.295}$ & 0.758$\pm$0.019 & -17.062$\pm$0.029 \\
164 & 03:32:40.643 & -27:47:05.944 &  & 7.7$\pm$1.2 & 1.907 & $8.118^{0.523}_{-0.285}$ & $-0.787^{+0.76}_{-0.186}$ &  &  \\
6666 & 03:32:38.290 & -27:46:36.267 & 221.8$\pm$2.2 & 15.8$\pm$1.7 & 3.434 & $9.471^{0.366}_{-0.43}$ & $0.878^{+0.312}_{-0.289}$ & 0.762$\pm$0.008 & -18.76$\pm$0.012 \\
1024 & 03:32:31.454 & -27:47:25.116 &  & 4.7$\pm$0.6 & 2.868 & $10.074^{0.244}_{-0.519}$ & $1.326^{+0.372}_{-0.351}$ & 0.619$\pm$0.007 & -19.328$\pm$0.01 \\
1273 & 03:32:35.475 & -27:46:16.915 &  & 10.2$\pm$0.7 & 2.172 & $9.045^{0.483}_{-0.463}$ & $0.162^{+0.762}_{-0.232}$ &  & -17.107$\pm$0.03 \\
22* & 03:32:37.071 & -27:46:17.176 &  & 2.6$\pm$0.2 & 2.226 & $10.415^{0.21}_{-0.236}$ & $1.422^{+0.458}_{-0.282}$ & 1.262$\pm$0.012 & -21.387$\pm$0.017 \\
1036 & 03:32:43.387 & -27:47:10.541 &  & 4.4$\pm$0.6 & 2.69 & $9.317^{0.527}_{-0.493}$ & $0.374^{+1.459}_{-0.423}$ & 0.401$\pm$0.007 & -18.942$\pm$0.01 \\
\textbf{1056} & 03:32:39.670 & -27:48:50.598 & 105.5$\pm$58.2 & 12.5$\pm$0.6 & 3.063 & $11.631^{0.142}_{-0.498}$ & $1.356^{+4.193}_{-0.379}$ & 2.577$\pm$0.015 & -23.591$\pm$0.022 \\
51 & 03:32:39.643 & -27:46:53.809 &  & 5.1$\pm$0.2 & 2.229 & $9.405^{0.254}_{-0.267}$ & $0.417^{+0.476}_{-0.221}$ & 0.608$\pm$0.066 & -20.378$\pm$0.098 \\
106 & 03:32:39.294 & -27:46:44.671 & 215.2$\pm$16.1 & 10.0$\pm$1.5 & 3.276 & $9.165^{0.368}_{-0.398}$ & $0.394^{+0.557}_{-0.272}$ & 0.334$\pm$0.015 & -16.843$\pm$0.022 \\
\textbf{1051} & 03:32:42.837 & -27:47:02.532 & 16.6$\pm$4.8 & 6.3$\pm$0.2 & 3.19 & $10.807^{0.276}_{-0.258}$ & $2.52^{+0.256}_{-0.354}$ & 0.57$\pm$0.045 & -20.243$\pm$0.066 \\
\hline
\end{tabular}%
}

\vspace{0.5em}
\noindent
\makebox[\textwidth][c]{%
  \parbox{\textwidth}{%
    \small {\textbf{Notes. }}ID, RA, DEC, z, $M_{\star}$ and SFR values were taken from the AMUSED database \citep{MHXDFBacon23}. For galaxies with ID: 7898, 8249 and 103, the SED was not performed due to the lack of HST photometric data. Consequently, the $M_{\star}$ and SFR values were not calculated and are left as blank spaces in the table. The $A_V$ from the $\beta$ slope, $M_{1500}$ and rest-frame EWs were calculated in this work (see section \ref{sec:data_UV}). The A$_V$ obtained using the Paschen series for the galaxies marked by an asterisk can be found in table \ref{tab:Av_vs_Av}. The two AGNs (ID:1056 and ID:1051) which present broad \heii, \ciii and \civ features are marked in bold face.}}
\end{center}

\begin{center}
\makebox[\textwidth][c]{%
  \parbox{\textwidth}{%
    \captionof{table}{Sample UV lines fluxes measured with MUSE.}
    \label{tab:MUSE_lines}
    }
    }
\resizebox{\textwidth}{!}{%
\begin{tabular}{lrrrrrrrrrrrrrrrrrrrrl}
                 
\cline{2-4} \cline{6-8} \cline{10-12} \cline{14-16} \cline{18-20} \cline{22-22}
\multicolumn{1}{c}{} & \multicolumn{3}{c}{3621} &  & \multicolumn{3}{c}{8369} &  & \multicolumn{3}{c}{6700} &  & \multicolumn{3}{c}{1045} &  & \multicolumn{3}{c}{6999} &  & \multicolumn{1}{c}{...} \\ \hline Line & Flux & Error & FWHM &  & Flux & Error & FWHM &  & Flux & Error & FWHM &  & Flux & Error & FWHM &  & Flux & Error & FWHM &  & \\ &
\multicolumn{2}{r}{$10^{-20}erg/s/cm^{2}$} & \A &  &
\multicolumn{2}{r}{$10^{-20}erg/s/cm^{2}$} & \A &  &
\multicolumn{2}{r}{$10^{-20}erg/s/cm^{2}$} & \A &  &
\multicolumn{2}{r}{$10^{-20}erg/s/cm^{2}$} & \A &  &
\multicolumn{2}{r}{$10^{-20}erg/s/cm^{2}$} & \A & \\ \hline
Ly$\alpha$ & 10087.44 & 150.14 & 6.92 &  &  &  &  &  & 2500.30 & 17.44 & 4.31 &  &  &  &  &  & 414.33 & 7.45 & 5.85 &  & \\
\sitwo1527* &  &  &  &  &  &  &  &  & -25.87 & 7.02 & 4.70 &  &  &  &  &  &  &  &  &  & \multicolumn{1}{c}{} \\
\civ1548 &  &  &  &  & 32.34 & 7.14 & 2.70 &  & 21.15 & 5.54 & 2.74 &  & -334.32 & 63.44 & 9.66 &  & 40.15 & 3.49 & 4.86 &  & \multicolumn{1}{c}{} \\
\civ1549b & 1048.72 & 205.92 & 6.76 &  & 56.16 & 9.45 & 2.70 &  & -94.09 & 9.76 & 4.75 &  & 97.89 & 36.85 & 2.52 &  & 62.45 & 5.13 & 4.86 &  & \multicolumn{1}{c}{} \\
\civ1550 & 825.62 & 179.74 & 6.76 &  & 23.83 & 6.19 & 2.70 &  & -46.12 & 6.72 & 4.75 &  & 84.47 & 33.34 & 2.52 &  & 22.30 & 5.66 & 4.87 &  & \multicolumn{1}{c}{} \\
Fe II 1608* &  &  &  &  &  &  &  &  & -20.43 & 6.62 & 4.87 &  &  &  &  &  &  &  &  &  &  \\
Fe II 1610b* &  &  &  &  &  &  &  &  & -23.51 & 7.43 & 4.87 &  &  &  &  &  &  &  &  &  &  \\
\heii1640 & 373.03 & 86.95 & 5.02 &  & 25.80 & 5.96 & 2.69 &  & 34.78 & 7.87 & 4.03 &  & 162.38 & 32.38 & 3.96 &  & 12.52 & 2.50 & 2.53 &  & ... \\
\oiii1660 &  &  &  &  &  &  &  &  & 25.37 & 6.23 & 4.06 &  & 103.89 & 31.28 & 3.98 &  & 9.31 & 2.63 & 2.53 &  &  \\
\oiii1663b &  &  &  &  & 42.45 & 7.73 & 2.68 &  & 78.17 & 7.67 & 4.07 &  & 282.46 & 40.07 & 3.98 &  & 26.81 & 3.59 & 2.53 &  &  \\
\oiii1666 &  &  &  &  & 32.92 & 6.29 & 2.68 &  & 52.80 & 5.65 & 4.07 &  & 178.57 & 28.72 & 3.99 &  & 17.50 & 2.66 & 2.53 &  &  \\
Al II 1671* &  &  &  &  &  &  &  &  & -17.01 & 5.43 & 5.00 &  &  &  &  &  &  &  &  &  &  \\
\sithree 1883 &  &  &  &  &  &  &  &  & 39.10 & 6.66 & 4.40 &  & 135.56 & 19.78 & 4.26 &  & 8.15 & 2.54 & 2.61 &  &  \\
\sithree1886b &  &  &  &  & 22.42 & 6.56 & 2.61 &  & 57.47 & 8.99 & 4.41 &  & 213.81 & 27.74 & 4.27 &  & 8.15 & 2.54 & 2.61 &  &  \\
\sithree1892 &  &  &  &  & 12.65 & 4.77 & 2.61 &  & 18.37 & 7.29 & 4.42 &  & 78.25 & 22.21 & 4.28 &  &  &  &  &  &  \\
\ciii1907 & 361.71 & 118.31 & 5.65 &  & 41.79 & 4.73 & 2.60 &  & 89.85 & 10.04 & 4.44 &  & 417.93 & 27.79 & 4.30 &  & 20.70 & 2.69 & 2.62 &  &  \\
\ciii1909b & 361.71 & 118.31 & 5.65 &  & 67.11 & 6.59 & 2.60 &  & 149.74 & 13.88 & 4.45 &  & 690.33 & 38.02 & 4.30 &  & 35.23 & 3.90 & 2.62 &  &  \\
\ciii1909 &  &  &  &  & 25.32 & 5.00 & 2.60 &  & 59.89 & 9.94 & 4.45 &  & 272.40 & 26.77 & 4.30 &  & 14.53 & 3.08 & 2.62 &  &  \\
... &  &  &  &  &  &  &  &  &  &  &  &  &  &  &  &  &  &  &  &  & ... \\ \cline{1-4} \cline{6-8} \cline{10-12} \cline{14-16} \cline{18-20} \cline{22-22} 
\end{tabular}%
}

\vspace{0.5em}

\noindent
\makebox[\textwidth][c]{%
  \parbox{\textwidth}{%
    \small {\textbf{Notes. }}The fluxes, errors, and FWHM are   from the AMUSED database \citep{MHXDFBacon23}. We excluded those with a S/N lower than 3, left as a blank space in the table. The full table (available at the CDS) includes all the emission and absorption features measured with MUSE, which ranges from Ly$\alpha$-1216 \r{A} to MgI-2853 \r{A} (54 lines). Absorption features are marked with an asterisk, and the flux of combined doublets are marked with a 'b'.
    }%
}
\end{center}

\clearpage

\setcounter{table}{1}

\begin{center}
\makebox[\textwidth][c]{%
  \parbox{\textwidth}{%
  \refstepcounter{table}
    \captionof{table}{\heii emitters NIRSpec sample.}
    \label{tab:JWST_lines}
    }
    }
\resizebox{\textwidth}{!}{%
\begin{tabular}{lrrrrrrrrrrrrrr}
\cline{2-3} \cline{5-6} \cline{8-9} \cline{11-12} \cline{14-15}
\multicolumn{1}{c}{} & \multicolumn{2}{c}{50} &  & \multicolumn{2}{c}{22} &  & \multicolumn{2}{c}{1141} &  & \multicolumn{2}{c}{6666} &  & \multicolumn{2}{c}{6700} \\ \hline
Line & Flux & Error &  & Flux & Error &  & Flux & Error &  & Flux & Error &  & Flux & Error \\
 & \multicolumn{2}{r}{$10^{-20}erg/s/cm^{2}$} &  & \multicolumn{2}{r}{$10^{-20}erg/s/cm^{2}$} &  & \multicolumn{2}{r}{$10^{-20}erg/s/cm^{2}$} &  & \multicolumn{2}{r}{$10^{-20}erg/s/cm^{2}$} &  & \multicolumn{2}{r}{$10^{-20}erg/s/cm^{2}$} \\ \hline
\oii3729 &  &  &  & 295.60 & 20.58 &  & 629.06 & 41.80 &  &  &  &  & 29.57 & 11.91 \\
H$\delta$ 4102 & 93.09 & 30.03 &  &  &  &  &  &  &  &  &  &  & 53.26 & 20.87 \\
H$\beta$ 4861 & 284.66 & 23.37 &  &  &  &  &  &  &  &  &  &  &  &  \\
\oiii5007 & 1823.99 & 148.41 &  &  &  &  &  &  &  &  &  &  &  &  \\
\hei5877 & 81.96 & 29.86 &  &  &  &  &  &  &  &  &  &  &  &  \\
\ha6563 & 1140.54 & 46.12 &  & 767.34 & 55.18 &  & 1417.88 & 73.86 &  &  &  &  & 228.58 & 13.57 \\
\sii6717 & 252.61 & 63.41 &  & 500.95 & 85.26 &  & 970.52 & 219.01 &  &  &  &  &  &  \\
\sii6731 & 232.53 & 55.74 &  & 498.55 & 80.88 &  & 892.97 & 172.20 &  &  &  &  &  &  \\
\hei7065 &  &  &  &  &  &  &  &  &  & 49.13 & 14.17 &  &  &  \\
$\left[\rm Ar III\right]$7138 & 55.36 & 22.17 &  &  &  &  &  &  &  &  &  &  &  &  \\
\hei8446 &  &  &  &  &  &  &  &  &  & 32.42 & 10.26 &  &  &  \\
\siii9068 & 87.46 & 23.43 &  & 443.55 & 109.40 &  & 722.92 & 178.28 &  & 28.18 & 6.35 &  &  &  \\
Pa9 9229 & 312.25 & 147.74 &  &  &  &  &  &  &  &  &  &  &  &  \\
Pa8 9546 & 239.83 & 37.29 &  & 346.16 & 54.54 &  & 387.04 & 72.64 &  & 32.45 & 5.23 &  & 31.26 & 17.42 \\
C I 9850 &  &  &  &  &  &  &  &  &  &  &  &  &  &  \\
Pa$\delta$ 10049 & 90.76 & 29.06 &  & 480.58 & 95.55 &  & 666.61 & 208.11 &  & 22.89 & 9.36 &  & 28.74 & 23.72 \\
\hei10830 & 304.88 & 38.61 &  & 277.28 & 55.93 &  & 335.23 & 66.09 &  & 91.70 & 6.95 &  &  &  \\
Pa$\gamma$ 10938 & 135.76 & 32.20 &  & 373.92 & 76.72 &  &  &  &  & 21.89 & 5.88 &  &  &  \\
Fe II 11128 &  &  &  &  &  &  &  &  &  & 16.94 & 13.07 &  &  &  \\
Pa$\beta$ 12818 &  &  &  & 265.46 & 54.86 &  & 554.66 & 129.48 &  &  &  &  &  &  \\ \cline{1-3} \cline{5-6} \cline{8-9} \cline{11-12} \cline{14-15} 
\end{tabular}%
}

\vspace{0.5em}

\noindent
\makebox[\textwidth][c]{%
  \parbox{\textwidth}{%
    \small {\textbf{Notes. }}NIRSpec fluxes, are measured as described in section \ref{sec:JWST_data}.
    }
    }
\end{center}

\setcounter{table}{2}

\begin{center}
\centering
\refstepcounter{table}
\makebox[0.8\textwidth][c]{%
  \parbox{0.8\textwidth}{%
    \captionof{table}{\heii emitters main results.}
    \label{tab:results}
    }
    }
\resizebox{0.8\textwidth}{!}{%

\begin{tabular}{lllllll}

\cline{1-6}
\hline
\hline
ID   & Q$_H$($\times10^{54}$) & Q$_{\rm HeII}$($\times10^{52}$) & 12+log(O/H)     & log(C/O)    & log(N/O)      & n$_e$($\times10^{3}$)            \\ 
 & photons/s & photons/s & & & & cm$^{-3}$
\\
\hline 
7898 &                     & 0.823$\pm$0.307        & 7.339$\pm$0.241 & -0.857 & -1.49993 &                 \\
8249 &                     & 14.173$\pm$4.659       &                 &        &          &                 \\
1059 & 4.243$\pm$4.172     & 15.497$\pm$4.923       &                 &        &          &                 \\
1141 & 1.073$\pm$0.842     & 0.680$\pm$0.203        & 7.663$\pm$0.164 & -0.677 & -1.49830 & 0.043$\pm$0.034 \\
50   & 1.989$\pm$1.323     & 2.880$\pm$0.812        & 7.621$\pm$0.165 & -0.713 & -1.49887 & 0.661$\pm$0.142 \\
6664 & 0.334$\pm$0.190     & 0.464$\pm$0.125        &                 &        &          & 3.123$\pm$1.811 \\
8165 & 0.029$\pm$0.028     & 0.177$\pm$0.047        & 7.545$\pm$0.16  & -0.766 & -1.49948 & 6.833$\pm$4.185 \\
149  & 0.271$\pm$0.191     & 14.51$\pm$4.196        & 7.456$\pm$0.189 & -0.813 & -1.49978 & 29.03$\pm$12.56 \\
8414 & 0.086$\pm$0.048     & 0.266$\pm$0.068        & 7.539$\pm$0.146 & -0.770 & -1.49951 & 4.087$\pm$2.581 \\
103  &                     & 0.707$\pm$0.175        & 7.668$\pm$0.146 & -0.672 & -1.49820 & 15.30$\pm$7.673 \\
3621 & 0.115$\pm$0.085     & 13.65$\pm$4.238        &                 &        &          &                 \\
8369 & 0.007$\pm$0.005     & 0.133$\pm$0.031        & 7.399$\pm$0.145 & -0.836 & -1.49988 & 2.965$\pm$1.903 \\
6700 & 0.219$\pm$0.218     & 0.515$\pm$0.116        & 7.606$\pm$0.127 & -0.724 & -1.49903 & 4.302$\pm$2.748 \\
1045 & 0.588$\pm$0.387     & 2.231$\pm$0.445        & 7.678$\pm$0.107 & -0.663 & -1.49802 & 2.058$\pm$1.068 \\
6999 & 0.096$\pm$0.065     & 0.650$\pm$0.130        & 7.356$\pm$0.133 & -0.851 & -1.49992 & 6.235$\pm$4.298 \\
164  & 0.023$\pm$0.010     & 0.142$\pm$0.028        & 7.532$\pm$0.11  & -0.774 & -1.49954 & 0.213$\pm$0.157 \\
6666 & 1.095$\pm$0.729     & 3.420$\pm$0.670        & 7.521$\pm$0.123 & -0.780 & -1.49959 & 6.755$\pm$4.665 \\
1024 & 3.074$\pm$2.484     & 3.706$\pm$0.717        & 7.665$\pm$0.109 & -0.675 & -1.49826 & 3.182$\pm$1.899 \\
1273 & 0.210$\pm$0.112     & 0.586$\pm$0.110        & 7.567$\pm$0.103 & -0.752 & -1.49934 & 6.245$\pm$2.705 \\
22   & 3.834$\pm$2.489     & 7.889$\pm$1.402        & 7.666$\pm$0.1   & -0.674 & -1.49823 & 3.328$\pm$2.034 \\
1036 & 0.343$\pm$0.334     & 2.073$\pm$0.352        & 7.694$\pm$0.09  & -0.647 & -1.49768 & 1.791$\pm$1.157 \\
51   & 0.379$\pm$0.192     & 0.619$\pm$0.097        & 7.562$\pm$0.081 & -0.756 & -1.49938 &                 \\
106  & 0.359$\pm$0.225     & 8.435$\pm$0.921        &                 &        &          & 2.802$\pm$0.155            \\ \hline    
\end{tabular}%
}

\vspace{0.5em}

\noindent
\makebox[0.8\textwidth][c]{%
  \parbox{0.8\textwidth}{%
    \small {\textbf{Notes. }}Q$_H$ and Q$_{\rm HeII}$ are in units of $10^{54}$ and $10^{52}$ photons/s,  respectively. The electron densities are in units of $10^{3}cm^{-3}$. Both AGNs are excluded from the results.
    }
    }
\end{center}

\clearpage

\section{Flux  comparison}\label{A2:fluxes_compar}

\begin{figure}[ht!]
\centering
    \includegraphics[width=\columnwidth]{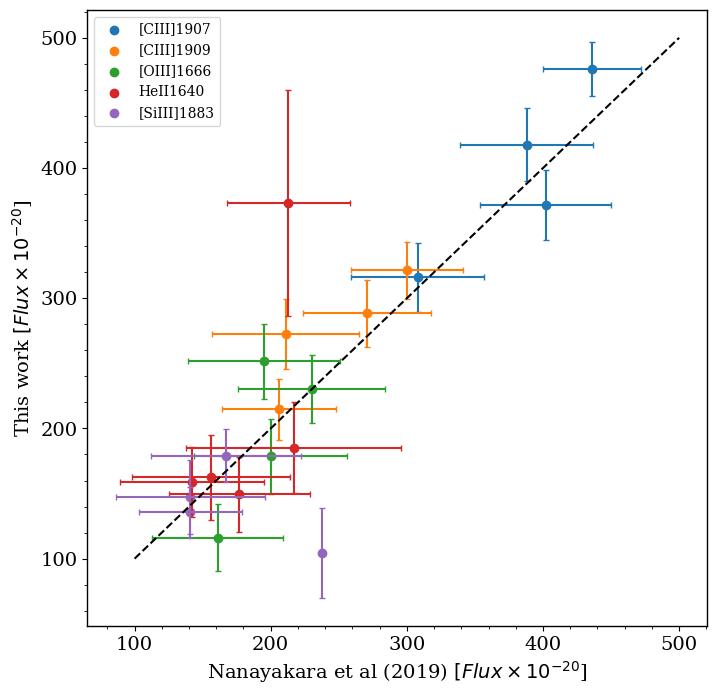}
\captionof{figure}{Comparison between the fluxes used in this work (when the target S/N is reached) and the fluxes obtained by \citet{NanayakkaraHeII19} in their work for the same objects. The diagonal dashed line indicates the 1:1 relation, showing that our fluxes are in excellent agreement.}
\label{fig:flux_compar}     
\end{figure}

As a sanity check, we compared our measured fluxes for the five \heii emitters in common with \citet{NanayakkaraHeII19} (IDs: 1024, 1036, 1045, 1273, and 3621). In Figure \ref{fig:flux_compar} we show this comparison for the matched detections, considering only the emission lines with a S/N greater than 3. In general, our fluxes are in good agreement. 

\clearpage

\end{document}